\documentclass[12pt]{article}
\usepackage{amsmath}
\usepackage{graphicx,psfrag,epsf}
\usepackage{enumerate}
\usepackage{natbib}
\usepackage{url} 
\usepackage{times}

\usepackage[top = 1 in, bottom = 1in, left = 0.92in, right=0.92in]{geometry}

\usepackage{color,xcolor}
\usepackage{float}
\usepackage[subrefformat=parens,labelformat=parens]{subcaption}
\usepackage{booktabs}
\usepackage{tikz}
\usepackage{multirow}
\usepackage{amsfonts}
\usepackage{amsthm}
\usepackage{bm}
\usepackage[inline]{enumitem}

\usetikzlibrary{positioning}
\usetikzlibrary{arrows.meta, matrix}
\usetikzlibrary{shapes,arrows,chains}
\usetikzlibrary{arrows,positioning,calc,topaths}

\usepackage{titlesec}
\titleformat*{\section}{\large\bfseries}
\titleformat*{\subsection}{\normalsize\bfseries}
\titleformat{\subsubsection}[runin]
{\normalfont\normalsize\bfseries}{\thesubsubsection}{1em}{}
\titleformat*{\paragraph}{\normalsize\bfseries}
\titleformat*{\subparagraph}{\normalsize\bfseries}

\usepackage[toc,page,header]{appendix}
\usepackage{minitoc}

\RequirePackage[citecolor=blue,colorlinks=true,linkcolor=blue,urlcolor=blue]{hyperref}
\usepackage[capitalise,noabbrev]{cleveref}
\crefformat{equation}{(#2#1#3)}

\theoremstyle{definition}
\newtheorem{theorem}{Theorem}
\newtheorem{proposition}{Proposition}

\newtheorem{assumption}{Assumption}
\newtheorem{remark}{Remark}

\newcommand{\E}{\mathbb{E}}
\newcommand{\Var}{\mathrm{Var}}
\newcommand{\indep}{\perp \!\!\! \perp}
\newcommand\numberthis{\addtocounter{equation}{1}\tag{\theequation}}

\usepackage{xargs, xstring}
\newcommandx{\tor}[1][1=r]{{(#1)}}

\newcommand{\cng}[1]{\textcolor{black}{#1}}

\allowdisplaybreaks

\begin{document}

\def\spacingset#1{\renewcommand{\baselinestretch}%
{#1}\small\normalsize} \spacingset{1}

\title{\bf Spatial causal inference in the presence of unmeasured confounding and interference}
\author{Georgia Papadogeorgou\thanks{The author was supported by the National Science Foundation under Grant No 2124124.} \\ 
{\small Department of Statistics, University of Florida, Email: \href{mailto:gpapadogeorgou@ufl.edu}{gpapadogeorgou@ufl.edu}} \\ and \\  \quad Srijata Samanta \\ {\small Global Biometrics and Data Sciences, Bristol Myers Squibb}}
\date{}
\maketitle
\vspace{-25pt}

\bigskip

\begin{abstract}
This manuscript unites causal inference and spatial statistics, presenting novel insights for causal inference in spatial data analysis, and drawing from tools in spatial statistics to estimate causal effects. We introduce spatial causal graphs to highlight that spatial confounding and interference can be entangled, in that investigating the presence of one can lead to wrongful conclusions in the presence of the other. Moreover, we show that spatial dependence in the exposure variable can render standard analyses invalid. To remedy these issues, we propose a Bayesian parametric approach based on tools commonly-used in spatial statistics. This approach simultaneously accounts for interference and mitigates bias from local and neighborhood unmeasured spatial confounding. From a Bayesian perspective, we show that incorporating an exposure model is necessary. Under a specific model formulation, we prove that all parameters are identifiable including the causal effects, even in the presence of unmeasured confounding. We illustrate the approach with a simulation study. \cng{We evaluate the effect of local and neighboring sulfur dioxide emissions from power plants on county-level cardiovascular mortality from observational spatial data in the United States, where unmeasured spatial confounding and interference might be present simultaneously.}
\end{abstract}

\noindent%



\section{Introduction}

\cng{Air pollution is a leading environmental determinant of cardiovascular morbidity and mortality, and quantifying its health impacts has been a focus of environmental epidemiology for decades. It is well-established that higher exposures to ambient fine particulate matter (PM$_{2.5}$) cause increased risks of cardiovascular mortality and morbidity \citep{dockery1993association, pope1995particulate, samet2000fine, dominici2014particulate, di2017air}.
Sulfur dioxide (SO$_2$) emissions from power plants contribute to PM$_{2.5}$ formation. Emissions are transported through space following atmospheric and meteorological processes, leading to potentially far-reaching effects \citep{henneman2019accountability, henneman2023mortality, zigler2025bipartite}. At the same time, key determinants of cardiovascular health and potential confounders of the relationship between emissions, air pollution and health often vary smoothly over space.}

\cng{We study the effect of SO$_2$ power plant emissions on cardiovascular mortality in the United States. This setting motivates the development of statistical methods that accommodate spatial dependence, spillover effects, and unmeasured confounding by spatial variables.}

Methodology for drawing causal inferences from observational data with complex interdependencies remains sparse. 
We study the challenges and opportunities in causal inference with spatial data that pertain specifically to the observations' statistical and causal dependence. We investigate the implications that arise due to spatial interference, unmeasured spatial confounding, and the variables' inherent spatial dependence. We aim to unite the spatial and causal inference literatures and address the gaps that arise when a research question is investigated under the lens of one, but not both. On one hand, we draw from the causal inference literature to provide new insights for the analysis of spatial data. On the other hand, we draw from the spatial statistics literature to develop tools for estimating causal effects. 

In spatial causal inference, a common challenge is that the outcome in one location might be driven by exposures in other locations, often referred to as {\it spillover effects} or {\it interference}. When interference is present, the interpretation of estimates from estimators that ignore it is complicated \citep{savje2021average, Tchetgen2012}.
Interference has attracted a lot of attention \citep[e.g.][]{Sobel2006, Hudgens2008, manski2013identification, aronow2017estimating, Tchetgen2017auto, ogburn2024causal} with some studies focusing on how interference manifests across space \citep{Verbitsky-savitz2012, wang2020design, zigler2025bipartite, papadogeorgou2022causal, giffin2022generalized, shin2023spatial, antonelli2023heterogeneous}.

Spatial data also presents opportunities for causal inference. The term ``spatial confounding'' has been used to represent drastically different notions in the spatial and causal literatures \citep[see][for relevant discussion]{reich2021review, gilbert2021causal, papadogeorgou2022discussion}. In spatial statistics, it is used to describe collinearity between covariates and spatial random effects in regression models \citep{reich2006effects, Hodges2010, Paciorek2010, Hanks2015, Prates2019alleviating}. We adopt the notion of spatial confounding encountered in causal inference \citep{papadogeorgou2019adjusting, gilbert2021causal}, where the measured variables do not suffice for confounding adjustment, but the missing confounders exhibit a spatial structure.
Recent developments use that nearby observations have similar values for unmeasured spatial confounders to mitigate unmeasured confounding bias \citep{thaden2018structural, papadogeorgou2019adjusting, keller2020selecting, schnell2020mitigating, gilbert2021causal, dupont2022spatial, christiansen2022toward, Guan2022spectral}. Therefore, the data's inherent dependence structure provides an opportunity to mitigate bias due to violations of the no unmeasured confounding assumption in causal inference.

There is limited literature investigating interference and spatial confounding simultaneously. 
\cite{graham2013quantifying} adopted a modelling approach which included spatial predictors, spatial random effects, and functions of the neighboring areas' exposure in a Poisson regression. However, causal quantities of interest are not clearly stated, their approach does not allow for confounding from unmeasured variables, and it is susceptible to biases due to spatial random effects.
\cite{giffin2021instrumental} introduced an instrumental-variable approach for spatial data with interference. 
Their approach provides a promising direction forward, though it requires access to a valid instrument.


Our work achieves the following goals.
\begin{enumerate*}[label=(\alph*)]
\item 
We introduce causal diagrams for spatial data.
\item We illustrate theoretically and practically that spatial confounding and spatial spillover effects can manifest as one another: If spatial confounding is present and not accounted, investigators might misinterpret the spatial structures induced by the confounder as interference, which would lead to wrongful conclusions about the causal effect of a potential intervention. In reverse, if interference is present and not accounted, researchers might mis-attribute spatial dependencies induced by interference to spatial confounding.\label{item:interference_confounding}
\item We demonstrate that statistical dependence in the exposure variable can render standard analyses for estimating causal effects invalid.\label{item:spatial_dependence}
These results establish that drawing causal inferences in spatial settings is faced with unique challenges compared to settings with independent observations.
%
%
%
\item 
We show that spatial causal inference must simultaneously account for interference, and local and neighborhood 
confounding in order to avoid the aforementioned pitfalls, an important guidance for practitioners.
\item We introduce a Bayesian causal inference framework for spatial data which 
\begin{enumerate*}[label=\roman*)]
\item is based on tools amenable to spatial statisticians,
\item incorporates interference, 
\item mitigates bias in effect estimation from local and neighborhood unmeasured confounding,
\item provides straightforward uncertainty quantification, and
\item establishes that modelling both the exposure and the outcome process is necessary in the presence of unmeasured spatial confounding.
\end{enumerate*}
\item In a special case, we show theoretically that all model parameters are identifiable, even in the presence of interference and unmeasured spatial confounding.
\item Across a variety of dependence structures, we illustrate  in simulations that this approach reduces bias in the estimation of local and interference effects from unmeasured spatial confounding or inherent statistical dependencies.
\end{enumerate*}


\section{Evaluating the effect of SO$_2$ emissions from power plants on cardiovascular mortality}
\label{sec:data}

\begin{figure}
\centering
\begin{subfigure}[t]{185pt}
\centering
\includegraphics[width = \textwidth,trim=40 0 40 40, clip]{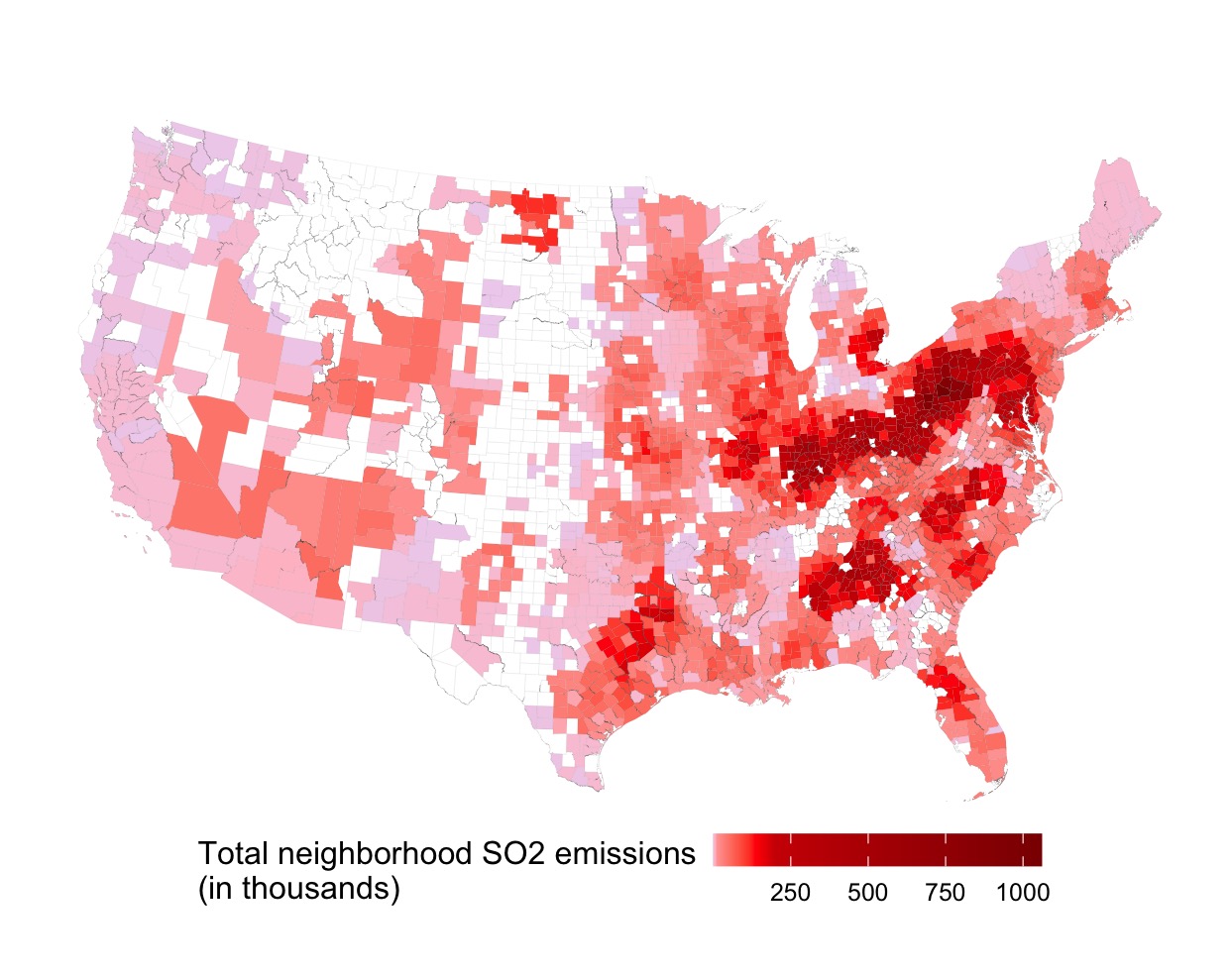} \caption{Total SO$_2$ emissions from power plants in counties within 50 kilometers during 2004.}
\label{fig:app_neighborhood_exposure}
\end{subfigure}
\hspace{30pt}
\begin{subfigure}[t]{185pt}
\centering
\includegraphics[width = \textwidth,trim=40 0 40 40, clip]{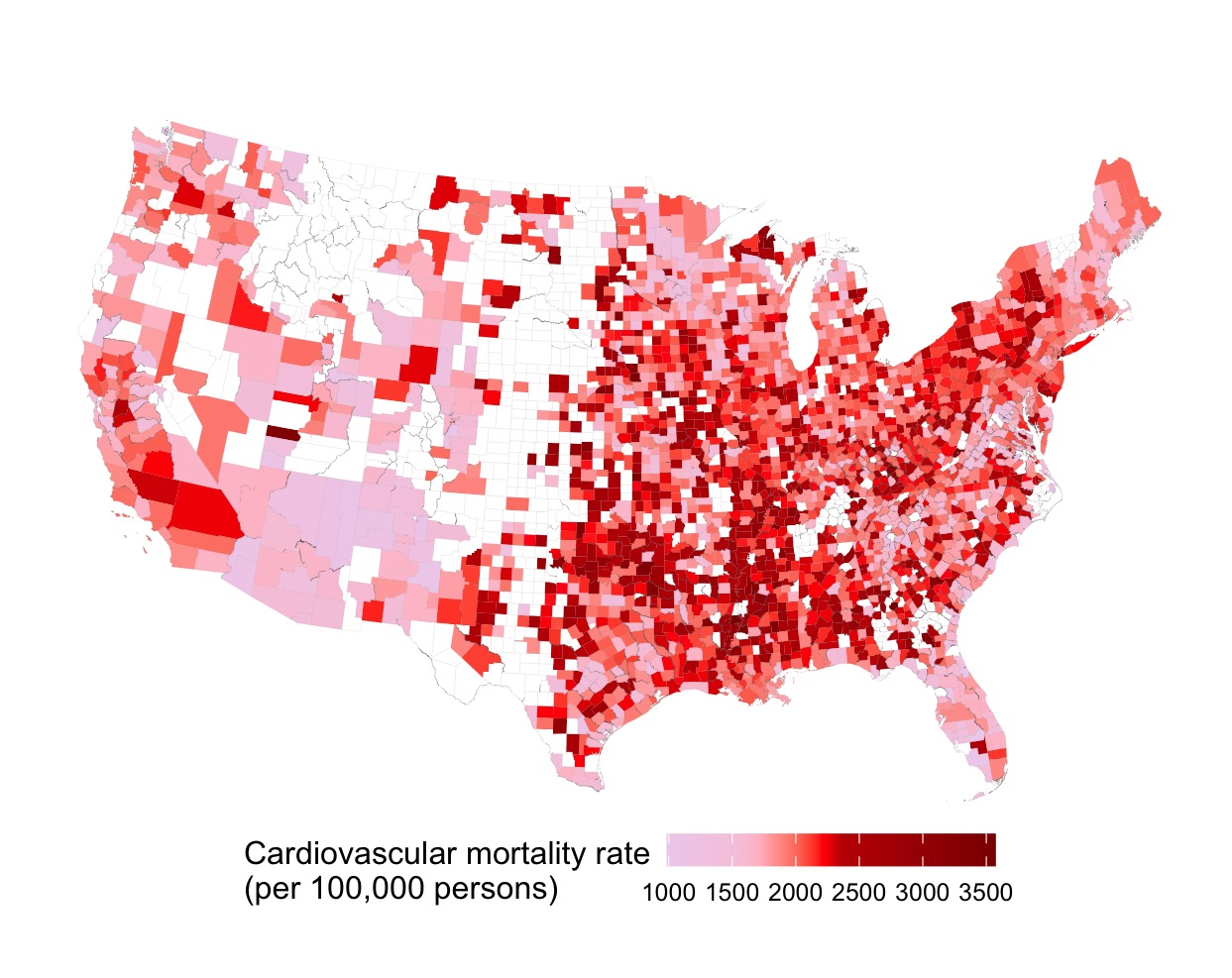}
\caption{Cardiovascular mortality rate per 100,000 residents in 2005.}
\label{fig:app_outcome}
\end{subfigure}
\caption{County-level neighborhood exposure and outcome in the evaluation of local and interference effects of SO$_2$ emissions from power plants on cardiovascular mortality. Local values of SO$_2$ emissions for each county are shown in the Supplement.}
\label{fig:app_data}
\end{figure}

\cng{In our motivating application, we investigate the relationship between sulfur dioxide (SO$_2$) emissions from power plants and cardiovascular mortality, while addressing the challenges that arise due to spatial dependence, interference, and unmeasured confounding by spatial variables.
We use data on counties (or county-equivalent units) in the 48 contiguous states and the District of Columbia.
Since emissions generated in one location may influence ambient pollution concentrations and health outcomes in other locations, we distinguish between \emph{local} and \emph{neighborhood} SO$_2$ emissions, with the latter defined as the total amount of emissions originating from power plants in counties within 50 kilometers. 
Our data include 2,440 counties, 476 of which experience both local and neighborhood exposure, and the remaining 1,964 have only neighborhood exposure. We study cardiovascular mortality among adults aged 65 years and older, defined using ICD-10 codes I00--I99 and obtained from the CDC WONDER system \citep{cdcwonder}. In \cref{fig:app_data} we illustrate the neighborhood SO$_2$ emissions and mortality outcome, which both indicate spatial patterns.
Additional information on the data compilation is included in Section
H
of the Supplementary Material.}

\cng{We consider 23 measured covariates as potential confounders of the effect of local and neighborhood SO$_2$ emissions on cardiovascular mortality. Each covariate is considered at the local and the neighborhood level, for a total of 46 variables in our adjustment set. They include power plant information (9 covariates) such as their size and operation capacity, demographic variables (11 covariates) from the 2000 Census and CDC Behavioral Risk Factor Surveillance System such as information on urbanicity, poverty and smoking rates, and weather information (3 covariates) on precipitation and temperature.}

\cng{Despite the large number of covariates in our adjustment set, unmeasured confounding due to completely missing, mis-measured, or mis-adjusted variables is possible when using observational data. Furthermore, in air pollution data, units tend to exhibit spatially correlated exposures and covariates. Therefore, explicitly accounting for unmeasured confounding from spatially structured variables and for spatial dependence in the exposure variable across observations can improve the credibility of causal conclusions.
By formulating and estimating causal effects in the presence of interference, spatial data, and unmeasured confounding, this analysis demonstrates its relevance in large-scale environmental health applications.}

\section{Causal diagrams and identifiability of estimands with paired spatial data}
\label{sec:pairs}

We introduce causal diagrams for spatial data in the presence of interference, spatial confounding, and inherent spatial dependence. We illustrate that formal causal reasoning is crucial for identifying and addressing the challenges in causal inference with spatial observational data. 
To establish our key notions, we first consider a simplified setting with a binary treatment and paired data for which dependencies exist within a pair but not across them. 
Spatial data on an interconnected network, which is our main focus, is discussed in \cref{sec:one_network}.

\subsection{Causal estimands for paired spatial data with a binary treatment}
\label{subsec:pairs_estimands}

Consider a sample of pairs of spatial observations, where there is a natural ordering within each pair that allows us to name them Unit 1 and Unit 2. For example, in air pollution studies, Unit 1 might be located downwind of Unit 2. A treatment is applied to or withheld from each of the units. {\it Spatial interference} reflects the situation where the treatment status in one location might affect the outcome at other locations. Then, the units have potential outcomes $Y_1(z_1, z_2)$ and $Y_2(z_2, z_1)$ representing the outcome that would occur for Units 1 and 2 if their treatment was set to $z_1, z_2$, respectively.
In our notation, the subscript reflects the unit whose potential outcome we focus on, we drop the notation that corresponds to the pair, and we write the individual's own treatment first.

\cng{For simplicity, we focus on a binary treatment and estimands reflecting causal effects on Unit 1's outcome for changes in Unit 1's or Unit 2's treatment status.} We refer to these estimands as {\it local} and {\it interference} effects, and we use $\lambda$ and $\iota$ to denote them, respectively. Specifically, we define the local effect for Unit $1$ when fixing the treatment of Unit $2$ to $z$ as
\begin{equation}
\begin{aligned}
\lambda_1(z) 
&= \E[Y_1(z_1 = 1, z_2 = z) - Y_1(z_1 = 0, z_2= z)] 
\equiv \E[ Y_1(1, z) - Y_1(0, z)],
\end{aligned}
\label{eq:local_effect}
\end{equation}
and the interference effect for Unit $1$ when setting its own treatment level at $z$ as
\begin{equation}
\begin{aligned}
\iota_1(z) &= \E[Y_1(z_1 = z, z_2 = 1) - Y_1(z_1 = z, z_2= 0)] 
\equiv \E[Y_1(z, 1) - Y_1(z, 0)].
\end{aligned}
\label{eq:interference_effect}
\end{equation}
\cng{The observed pairs are considered as drawn independently from a superpopulation of pairs.}
\cng{The local and interference effects reflect expected changes in the potential outcomes of Unit 1 in a pair drawn from the superpopulation. For example, if Unit 1 is the downwind unit in a pair, the interference effect $\iota_1$ reflects the expected change in a unit's outcome for a change in the treatment status of its upwind unit.}

Effects for Unit 2 can be defined similarly. Alternative definitions of local and interference effects, and the case with blocks of units of larger or varying sizes
are discussed in Section
A
of the Supplementary Material. 

\subsection{The observed pair data}

Let $\bm Z = (Z_1, Z_2)$ and $\bm Y = (Y_1, Y_2)$ denote the pair-level observed treatment and outcome. The observed outcomes are equal to the potential outcomes under the observed treatment, $Y_1 = Y_1(Z_1, Z_2)$ and $Y_2 = Y_2(Z_2, Z_1)$. We assume that ignorability holds conditional on a covariate $\bm U= (U_1, U_2)$. We denote the covariate with the letter ``U'' because it will be considered unmeasured later in the manuscript. We refrain from considering additional covariates until \cref{sec:one_network} for ease of exposition.
\begin{assumption}
[Pair ignorability]
It holds that \cng{$\bm Z \indep \left\{Y_1(z_1, z_2), Y_2(z_1, z_2) \right\}_{z_1, z_2 \in \{0, 1\}}  \mid \bm U$.}
For $\bm u = (u_1, u_2)$ with $P_{\bm U}(\bm u) > 0$, we have that $P_{\bm Z\mid \bm u}(\bm z \mid \bm U = \bm u) > 0$ for $\bm z = (z_1, z_2) \in \{0, 1\}^2$ where $P_{\bm U}, P_{\bm Z \mid \bm u}$ denote the corresponding distribution in the superpopulation of pairs.
\label{ass:paired_ignorability}
\end{assumption}
\vspace{-8pt} \noindent
It is known that this assumption suffices to write the local and interference effects in \cref{subsec:pairs_estimands}, which are defined in terms of unobserved potential outcomes, as functions of observable quantities. (In principle, for identifiability of the causal effects on Unit 1, the conditional independence statement suffices to hold for the potential outcomes of that unit only.)
Crucially, this nonparametric identifiability result establishes that causal effects can be estimated based on data $(\bm U, \bm Z, \bm Y)$, where {\it the pair is the unit of analysis}. Such analysis is possible when the population is a collection of pairs, but it is infeasible with data on an interconnected spatial network as the one in \cref{sec:one_network}.
Instead, in spatial settings, estimation techniques are at the level of the {\it unit}, where a unit's outcome is regressed on treatments and covariates. 
Therefore, we must uncover the complications in identifying causal effects, 
and provide guidance on how statistical and causal dependencies can be addressed within the context of spatial unit-level analysis.

\subsection{Identifiability of estimands in the presence of statistical and causal dependencies}
\label{subsec:pairs_graphs}


To do so, we introduce expanded causal diagrams for paired spatial data, with the units within a pair depicted separately. We consider cases with spatial confounding, interference, and inherent spatial statistical dependence in the treatment variable. The interpretation of spatial confounding and inherent dependencies is discussed below.

\begin{figure}[!t]
\centering
\begin{subfigure}[t]{.81\linewidth}
\centering
\resizebox{0.3\linewidth}{!}{
\begin{tikzpicture}
\node (1) {$\bm U$};
\node[right=of 1] (2) {$\bm Z$};
\node[right=of 2] (3) {$\bm Y$};
\draw[->] (1) to (2);
\draw[->] (1) to [out=30,in=150] (3);
\draw[->] (2) to (3);
\end{tikzpicture}
}
\caption{Causal graph at the pair level. For the two units, $\bm Z$ is the treatment vector, $\bm Y$ is the outcome vector, and $\bm U$ is the vector of the spatial confounder.}
\label{fig:dag_compact}
\end{subfigure}
\\[-10pt]
\begin{subfigure}[t]{.31\linewidth}
\centering
\resizebox{\linewidth}{!}{
\begin{tikzpicture}
        \node (1) {$U_1$};
		\node[ right= of 1] (2) {$Z_1$};
		\node[ right= of 2] (3) {$Y_1$};
		\node[below = of 1] (12) {$U_2$};
		\node[ right= of 12] (22) {$Z_2$};
		\node[ right= of 22] (32) {$Y_2$};
		\draw[->] (1) to (2); 
		\draw[->] (12) to (2); 
		\draw[->] (1) to (22); 
		\draw[->] (12) to (22); 
		\draw[->] (2) to (3); 
		\draw[->] (2) to (32);
		\draw[->] (22) to (3);
		\draw[->] (12) to [out=330,in=210] (32);
		\draw[->] (1) to [out=45,in=45,looseness=1.35] (32);
\phantom{\draw[->] (32) to [out=60,in=120,looseness=1.15] (1);}
		\draw[->] (12) to [out=315,in=315,looseness=1.35] (3);
		\draw[->] (1) to [out=30,in=150] (3);
		\draw[->] (22) to (32);
		\draw[<->] (1) to (12);
		\draw[<->] (2) to (22);
		\draw[<->] (3) to (32);
	\end{tikzpicture}
}
\vspace{-25pt}
\caption{The two units depicted separately with all possible causal and statistical relationships.}
\label{fig:dag_expanded}
\end{subfigure}
\hfill
\begin{subfigure}[t]{.31\linewidth}
\centering
\resizebox{\linewidth}{!}{
\begin{tikzpicture}
        \node (1) {$U_1$};
		\node[ right= of 1] (2) {$Z_1$};
		\node[ right= of 2] (3) {$Y_1$};
		\node[below = of 1] (12) {$U_2$};
		\node[ right= of 12] (22) {$Z_2$};
		\node[ right= of 22] (32) {$Y_2$};
		\draw[->] (1) to (2); 
		\draw[->] (12) to (22); 
		\draw[->] (2) to (3); 
		\draw[->] (2) to (32);
		\draw[->] (22) to (3);
		\draw[->] (12) to [out=330,in=210] (32);
		\draw[->] (1) to [out=45,in=45,looseness=1.35] (32);
\phantom{\draw[->] (32) to [out=60,in=120,looseness=1.15] (1);}
		\draw[->] (12) to [out=315,in=315,looseness=1.35] (3);
		\draw[->] (1) to [out=30,in=150] (3);
		\draw[->] (22) to (32);
		\draw[<->] (1) to (12);
		\draw[<->] (2) to (22);
	\end{tikzpicture}
}
\vspace{-25pt}
\caption{The two units with the relationships we consider in this work.}
\label{fig:dag_expanded_considered}
\end{subfigure}
\hfill
\begin{subfigure}[t]{.33\linewidth}
\centering
\resizebox{\linewidth}{!}{
\begin{tikzpicture}
        \node (1) {$U_1$};
        \node [below left =0.25 and 0.25 of 1] (10) {$U^u$};
        \node [below left =0.25 and 0.25 of 2] (20) {$Z^u$};
		\node[ right= of 1] (2) {$Z_1$};
		\node[ right= of 2] (3) {$Y_1$};
		\node[below = of 1] (12) {$U_2$};
		\node[ right= of 12] (22) {$Z_2$};
		\node[ right= of 22] (32) {$Y_2$};
		\draw[->] (1) to (2); 
		\draw[->] (12) to (22); 
		\draw[->] (2) to (3); 
		\draw[->] (2) to (32);
		\draw[->] (22) to (3);
		\draw[->] (12) to [out=330,in=210] (32);
		\draw[->] (1) to [out=45,in=45,looseness=1.35] (32);
		\draw[->] (12) to [out=315,in=315,looseness=1.35] (3);
		\draw[->] (1) to [out=30,in=150] (3);
		\draw[->] (22) to (32);
		\draw[->] (10) to (1);
		\draw[->] (10) to (12); 
		\draw[->] (20) to (2);
		\draw[->] (20) to (22);
\end{tikzpicture}
}
\vspace{-25pt}
\caption{Statistical dependencies occur due to an unobservable underlying common trend.}
\label{fig:dag_underlying}
\end{subfigure}
\caption{Causal and statistical dependencies depicted at the pair- and at the unit-level. 
}
\label{fig:dag}
\end{figure}

We first discuss some basics from causal graph theory. In a causal directed acyclic graph (DAG), nodes represent random variables, and they are connected with arrows. 
An arrow indicates potential causation from the tail variable to the head variable. No arrow signifies the absence of a causal relationship, but the presence of an arrow does not guarantee the occurrence of the depicted relationship. A DAG for the pair spatial data is shown in \cref{fig:dag_compact}. A backdoor path from $\bm Z$ to $\bm Y$ starts with an arrow pointing into $\bm Z$ and ends with an arrow pointing into $\bm Y$. If the path is open, conditioning on a variable on this path blocks it. 
Colliders are nodes where arrows on either side converge. These nodes block paths when left unconditioned, but conditioning on a collider opens the path.
If all backdoor paths from $\bm Z$ to $\bm Y$ are blocked, the variables are called d-separated, and the conditional independence statement in \cref{ass:paired_ignorability} holds. In \cref{fig:dag_compact}, $\bm U$ blocks the backdoor path from $\bm Z$ to $\bm Y$.
%
%

The graph in \cref{fig:dag_compact} compacts the variables at the pair-level which masks the underlying dependencies that are important for investigating identifiability of causal contrasts based on unit-level analyses. The two units are depicted separately in \cref{fig:dag_expanded}.
The scenarios we consider are depicted in \cref{fig:dag_expanded_considered}: the outcome is not inherently spatial ($Y_1, \leftrightarrow Y_2$ missing), and $U$ in one location does not predict $Z$ in a different location.
The double-headed arrows between $U_1, U_2$ and between $Z_1, Z_2$ represent {\it inherent spatial statistical dependencies} due to an underlying common trend that drives both variables.
A DAG that represents this structure is shown in \cref{fig:dag_underlying}: $U^u$ induces correlation between $U_1$ and $U_2$, and similarly for $Z^u, Z_1$, and $Z_2.$ 
%
%
The superscript $^u$ is used to stand for {\it u}nderlying variables that drive the spatial structure.
\cng{In our study, such inherent spatial dependence in the treatment variable can occur if different air pollution regulations apply 
to different geographical areas.}
It can also arise by exogenous processes that are possible to measure {\it in theory} but not in practice, such as the intricate pollution transport processes that dictate 
ambient air pollution levels.
Therefore, the underlying $U^u$ and $Z^u$ describe the inherent spatial structure in $\bm U$ and $\bm Z$ which cannot be ``adjusted away'' by conditioning on more covariates.
Finally, the term {\it spatial confounding} is used to represent the situation where an inherently spatial variable ($U_1 \leftrightarrow U_2$) exists on open backdoor paths from the treatment to the outcome.

In \cref{fig:graphs} we present causal diagrams with spatial confounding, interference, and inherent spatial dependence that correspond to subgraphs of \cref{fig:dag_expanded_considered} with different arrows missing. A researcher that is interested in drawing causal inferences does not know which of these graphs describes the dependencies in their data. We use these graphs to illustrate the biases in estimating causal effects from unit-level analyses that manifest {\it due to} the spatial dependence in the covariate and the treatment.
Conditional independence and identifiability statements are based on viewing the inherent spatial dependencies within the realm of the underlying DAG in \cref{fig:dag_underlying}.
The identifiability results discussed in this section 
are stated and proven in Section 
B
of the Supplementary Material, some of which are based on well-known theory of graphical models \citep{spirtes1993causation, pearl1995causal, pearl2000models}.

\begin{figure*}[!t]
\centering
\begin{subfigure}[t]{.31\linewidth}
\centering
\resizebox{\linewidth}{!}{
\begin{tikzpicture}
		\node (1) {$U_1$};
		\node[ right= of 1] (2) {$Z_1$};
		\node[ right= of 2] (3) {$Y_1$};
		\node[below = of 1] (12) {$U_2$};
		\node[ right= of 12] (22) {$Z_2$};
		\node[ right= of 22] (32) {$Y_2$};
		\draw[->] (1) to (2);
		\draw[->] (1) to [out=30,in=150] (3);
		\draw[->] (2) to (3); 
		\draw[->] (12) to (22);
		\draw[->] (12) to [out=330,in=210] (32);
		\draw[->] (22) to (32);
		\draw[<->] (1) to (12);
		\draw[<->] (2) to (22);
            \phantom{
		\clip (0,-2.7) rectangle (4.3,1);
		\draw[->] (1) to [out=45,in=45,looseness=1.35] (32);
		\draw[->] (12) to [out=315,in=315,looseness=1.35] (3);
		}
	\end{tikzpicture}
 }
 \vspace{-15pt}
\caption{{\bf Direct Spatial Confounding.}
The covariate predicts the exposure and the outcome only locally. 
}
\label{fig:direct}
\end{subfigure}
\hfill
%
\begin{subfigure}[t]{.31\linewidth}
\centering
\resizebox{\linewidth}{!}{
\begin{tikzpicture}
		\node (1) {$U_1$};
		\node[ right= of 1] (2) {$Z_1$};
		\node[ right= of 2] (3) {$Y_1$};
		\node[below = of 1] (12) {$U_2$};
		\node[ right= of 12] (22) {$Z_2$};
		\node[ right= of 22] (32) {$Y_2$};
		\draw[->] (1) to (2);
		\draw[->] (1) to [out=30,in=150] (3);
		\draw[->] (2) to (3); 
		\draw[->] (12) to (22);
		\draw[->] (12) to [out=330,in=210] (32);
		\draw[->] (22) to (32);
		\draw[<->] (1) to (12);
		\draw[<->] (2) to (22);
	    \clip (0,-2.7) rectangle (4.3,1);
		\draw[->] (1) to [out=45,in=45,looseness=1.35] (32);
		\draw[->] (12) to [out=315,in=315,looseness=1.35] (3);
	\end{tikzpicture}
} \vspace{-15pt}
\caption{{\bf Direct and Indirect Spatial Confounding.}
The covariate predicts the exposure, and the local and neighboring outcome.
}
\label{fig:general_spatial_conf}
\end{subfigure}
\hfill
%
\begin{subfigure}[t]{.32\linewidth}
\centering
\resizebox{\linewidth}{!}{
\begin{tikzpicture}
		\node (1) {$U_1$};
		\node[ right= of 1] (2) {$Z_1$};
		\node[ right= of 2] (3) {$Y_1$};
		\node[below = of 1] (12) {$U_2$};
		\node[ right= of 12] (22) {$Z_2$};
		\node[ right= of 22] (32) {$Y_2$};
		\draw[->] (2) to (3); 
		\draw[->] (2) to (32);
		\draw[->] (22) to (3);
		\phantom{\draw[->] (12) to [out=330,in=210] (32);}
		\draw[->] (22) to (32);
		\draw[<->] (1) to (12);
		\draw[<->] (2) to (22);
             \phantom{
		\clip (0,-2.7) rectangle (4.3,1);
		\draw[->] (1) to [out=45,in=45,looseness=1.35] (32);
		\draw[->] (12) to [out=315,in=315,looseness=1.35] (3);
		}
	\end{tikzpicture}
 }
 \vspace{-15pt}
\caption{{\bf Spatial interference.}
One unit's treatment affects the other unit's outcome.
Spatial covariates are not confounders.}
\label{fig:interference}
\end{subfigure}
$\ $ \\[-2pt]
%
\begin{subfigure}[t]{.31\linewidth}
\centering
\resizebox{\linewidth}{!}{
\begin{tikzpicture}
		\node (1) {$U_1$};
		\node[ right= of 1] (2) {$Z_1$};
		\node[ right= of 2] (3) {$Y_1$};
		\node[below = of 1] (12) {$U_2$};
		\node[ right= of 12] (22) {$Z_2$};
		\node[ right= of 22] (32) {$Y_2$};
		\draw[->] (1) to (2);
		\draw[->] (2) to (32);
		\draw[->] (22) to (3);
		\draw[->] (2) to (3); 
		\draw[->] (12) to (22);
		\draw[->] (22) to (32);
		\draw[<->] (1) to (12);
		\draw[<->] (2) to (22);
		\phantom{
		\clip (0,-2.7) rectangle (4.3,1);
		\draw[->] (1) to [out=45,in=45,looseness=1.35] (32);
		\draw[->] (12) to [out=315,in=315,looseness=1.35] (3);
		}
	\end{tikzpicture}
}
\vspace{-15pt}
\caption{{\bf Interference and a spatial predictor of the exposure.}
}
\label{fig:predictor_interference}
\end{subfigure}
\hfill
%
\begin{subfigure}[t]{.31\linewidth}
\centering
\resizebox{\linewidth}{!}{
\begin{tikzpicture}
		\node (1) {$U_1$};
		\node[ right= of 1] (2) {$Z_1$};
		\node[ right= of 2] (3) {$Y_1$};
		\node[below = of 1] (12) {$U_2$};
		\node[ right= of 12] (22) {$Z_2$};
		\node[ right= of 22] (32) {$Y_2$};
		\draw[->] (1) to (2);
		\draw[->] (2) to (32);
		\draw[->] (22) to (3);
		\draw[->] (1) to [out=30,in=150] (3);
		\draw[->] (2) to (3); 
		\draw[->] (12) to (22);
		\draw[->] (12) to [out=330,in=210] (32);
		\draw[->] (22) to (32);
		\draw[<->] (1) to (12);
		\draw[<->] (2) to (22);
		\phantom{
		\clip (0,-2.7) rectangle (4.3,1);
		\draw[->] (1) to [out=45,in=45,looseness=1.35] (32);
		\draw[->] (12) to [out=315,in=315,looseness=1.35] (3);
		}
	\end{tikzpicture}
 }
 \vspace{-15pt}
\caption{{\bf Direct Spatial Confounding and Interference.} 
}
\label{fig:direct_interference}
\end{subfigure}
\hfill
%
\begin{subfigure}[t]{.31\linewidth}
\centering
\resizebox{\linewidth}{!}{
\begin{tikzpicture}
		\node (1) {$U_1$};
		\node[ right= of 1] (2) {$Z_1$};
		\node[ right= of 2] (3) {$Y_1$};
		\node[below = of 1] (12) {$U_2$};
		\node[ right= of 12] (22) {$Z_2$};
		\node[ right= of 22] (32) {$Y_2$};
		\draw[->] (1) to (2);
		\draw[->] (1) to [out=30,in=150] (3);
		\draw[->] (2) to (3); 
		\draw[->] (12) to (22);
		\draw[->] (12) to [out=330,in=210] (32);
		\draw[->] (22) to (32);
		\draw[->] (2) to (32);
		\draw[->] (22) to (3);
		\draw[<->] (1) to (12);
		\draw[<->] (2) to (22);
	    \clip (0,-2.7) rectangle (4.3,1);
		\draw[->] (1) to [out=45,in=45,looseness=1.35] (32);
		\draw[->] (12) to [out=315,in=315,looseness=1.35] (3);
	\end{tikzpicture}
}
\vspace{-15pt}
\caption{{\bf Direct, Indirect Spatial Confounding, and Interference.}
}
\label{fig:general_interference}
\end{subfigure}
\caption{Graphical representation of spatial confounding and interference with a spatially correlated covariate $\bm U = (U_1, U_2)$, a spatial exposure $\bm Z = (Z_1, Z_2)$, and outcome $\bm Y = (Y_1, Y_2)$.}
\label{fig:graphs}
\end{figure*}

The graphs in Figures \ref{fig:direct} and \ref{fig:general_spatial_conf} correspond to scenarios with spatial confounding and no interference.
Scenario \ref{fig:direct} represents the case of {\it direct} spatial confounding where it is only the {\it local} value of $U$ that drives the local value for $Y$. In this scenario there exists {\it statistical}, but not {\it causal}, dependence across units.
Scenario \ref{fig:general_spatial_conf} allows also for {\it indirect} spatial confounding, which describes a causal dependence across units in that a local spatial predictor of the exposure ($U_j \rightarrow Z_j$) predicts the outcome in a different location ($U_j \rightarrow Y_i$).

Under scenario \ref{fig:direct}, to identify the local causal effect it suffices to control for the local value of the confounder.
If $\bm U$ and $\bm Z$ were {\it not} spatial, the interference effect could be identified without any adjustments.
With spatial data, this analysis would attribute spatial statistical dependence to interference, and mis-identify the presence of spatial interference.
Instead, local and interference effects should be investigated simultaneously adjusting for the local value of the confounder. 
However, in the presence of indirect spatial confounding in Scenario \ref{fig:general_spatial_conf}, this analysis would be biased for the local effect. This complication arises because of the statistical dependence in the exposure variable, therefore this notion of confounding pertains solely to the setting with dependent data. Instead, one needs to adjust for the local {\it and} the neighbor's confounder value in order to identify local and interference effects.






\cref{fig:interference} represents a setting with interference and no spatial confounding. 
If the exposure $\bm Z$ is not spatial, interpretable local and interference effects can be estimated without any adjustment (see Section 
B
of the Supplementary Material).
However, when $\bm Z$ is spatial, 
this analysis would be invalid, even in the complete absence of confounding.
%
In turn, in the graph of \cref{fig:predictor_interference}, $\bm U$ is a spatial predictor of the exposure but it is not a confounder. 
To estimate the local effect without accounting for interference, one might adjust for the local variable $U$, or not. The two analyses would return different values for the local effect estimate, none of which is causally interpretable. Therefore, in this scenario, spatial interference could be mis-interpreted as spatial confounding.


\cref{fig:direct_interference} shows a setting with {direct spatial confounding and interference}, where once more the exposure's inherent spatial structure can lead to misleading conclusions if not properly accommodated (see Section
B
of the Supplementary Material).
Lastly, the graph in \cref{fig:general_interference} represents the situation also shown in \cref{fig:dag_expanded_considered} with direct spatial confounding, indirect spatial confounding, and interference, of which the other graphs are special cases. \cng{In all cases, unit-level analyses should investigate local and interference effects simultaneously while controlling for the local covariate value and the covariate values of interfering units.}

In the presence of interference, it has been previously advocated that neighbors' covariate values have to be adjusted \citep{ogburn2014causal, forastiere2021identification}.
However, this prior work does not address non-causal dependencies among observations that might naturally occur in spatial settings. \cite{vansteelandt2007confounding} considers the setting where treatments within a cluster can be correlated due to unmeasured cluster-level variables, but they do not investigate the role of statistical dependencies in confounders and exposures.

We addressed both of these points. We illustrated that adjusting for neighbors' covariate information is particularly important for spatial data analysis.
Moreover, 
we established that the variables' inherent dependence complicates separating statistical and causal dependencies, and accurately attributing causal effects. 
Our investigations illustrated that spatial settings are intrinsically different from settings with independent observations, and that analyses that are valid for the latter can be invalid for the former. Therefore, for drawing causal inferences, we must comprehensively account for the data's spatial structure.

These points are illustrated with a motivating simulation study in Section
C.1 
of the Supplementary Material. We see in practice how researchers might mis-attribute spatial confounding to interference, or mis-identify spatial confounding if interference is not accounted for. We also see that the estimators' biases increase with the strength of the variables' spatial dependence.
Across all scenarios, simultaneously estimating local and interference effects while adjusting for local and neighborhood confounder values eliminates biases, 
an important practical guidance for researchers aiming to draw causal inferences from spatial data.

\section{Causal inference with spatial dependencies on a spatial network}
\label{sec:one_network}

In the case of a spatial network of interconnected units, the population of interest cannot be partitioned in non-interacting groups. 
Estimation techniques that consider blocks of interconnected units (such as the pairs of \cref{sec:pairs}) as the unit of analysis cannot be applied, and unit-level analyses are the only option. In \cref{subsec:po_network} we introduce potential outcomes for units across space based on an exposure mapping and define local and interference effects, and in \cref{subsec:network_ignorability} we introduce ignorability for spatial data with unmeasured confounding.

\cng{In our study, many counties do not have operating power plants, and for those the local exposure is undefined. In this section, we discuss the setting for which all units have both local and neighborhood exposure. The formalization of potential outcomes and estimands for the case with undefined exposures is deferred to {Section I
of the Supplementary Materials.}}

\subsection{Local and interference effects under an exposure mapping}
\label{subsec:po_network}

Let $Z_i \in \mathcal{Z}$ denote the treatment value of unit $i = 1, 2, \dots, n$, which can be binary or continuous with realization $z_i$. Set $\bm Z = (Z_1, Z_2, \dots, Z_n)$ with realization $\bm z$. Continuous treatments are often referred to as exposures, so we use the two terms interchangeably. In full generality, a unit's potential outcomes depend on the treatment level of all units on the spatial network, denoted by $Y_i(\bm z)$ for $\bm z \in \mathcal{Z}^n$. We reduce the number of potential outcomes by assuming the presence of a known interference network and exposure mapping \citep{aronow2017estimating, zigler2025bipartite, forastiere2021identification}. Let $A$ denote a known adjacency matrix of dimension $n \times n$, where $A_{ij} = 1$ reflects that the outcome for unit $i$ might depend on unit $j$'s treatment level, $A_{ij} = 0$ otherwise, and all diagonal elements are 0.
We use $\overline Z_i = \sum_j A_{ij} Z_j / \sum_j A_{ij}$ to denote the average exposure of units with which $i$ is connected through $A$, with realization $\overline z_i$. Let also $\overline{\bm Z} = (\overline Z_1, \overline Z_2, \dots, \overline Z_n)$ with realization $\overline{\bm z}$. We make the following assumption.

\begin{assumption}
Let $\bm z, \bm z'$ be two treatment vectors in $\mathcal{Z}^n$ such that $z_i = z_i'$ and $\overline z_i = \overline z_i'$.
It holds that $Y_i(\bm z) = Y_i(\bm z')$, and the potential outcome $Y_i(\bm z)$ can be denoted as $Y_i(z_i, \overline z_i).$
\label{ass:sutva}
\end{assumption}

For areal data such as the ones in \cref{sec:data}, a binary adjacency matrix $A$ often makes sense. However, for the case of point-referenced data the adjacency matrix could be defined based on the units' geographical distance. Then, the average neighborhood exposure would reflect a weighted average of the exposures of other locations, with weights driven by the locations' geographic proximity.
\cng{The case of interconnected blocks is a special case of network data with the adjacency matrix defined as a block diagonal matrix, and, as a result, the work in this section and \cref{sec:method} applies to this simpler case as well.}

In air pollution studies with interference, exposure mappings have been previously defined based on atmospheric processes in \cite{zigler2025bipartite}, and on population mobility in \cite{shin2023spatial}.
Our discussion below would straightforwardly accommodate an asymmetric, weighted adjacency matrix $A$, or a definition of a more intricate exposure mapping. 

\cng{To define local and interference causal effects, consider for unit $i$ values $z_i, z_i'$ for their individual treatment and values $\overline z_i, \overline{z}_i'$ for their neighborhood treatment, such that the pairs $(z_i, \overline z_i), (z_i, \overline z_i'),$ and $(z_i', \overline z_i)$ 
are possible to occur. Let $\bm z = (z_1, z_2, \dots, z_n)$ denote the vector of hypothesized individual exposures, and $\bm z', \overline{\bm z}, \overline{\bm z}'$ defined similarly. We denote the average potential outcome over the $n$ units in the network under exposures specified in $(\bm z, \overline{\bm z})$ as $\overline Y(\bm z, \overline{\bm z}) = n^{-1} \sum_{i = 1}^n Y_i(z_i, \overline z_i)$.
Then, we define the local causal effect as
\begin{equation}
    \lambda^*(\bm z, \bm z'; \overline{\bm z}) = \E \left[ \overline Y(\bm z', \overline{\bm z}) - \overline Y(\bm z, \overline{\bm z}) \right],
\label{eq:local_effect_network}
\end{equation}
which represents the average change in the potential outcomes for a change in the individual treatment from its value in $\bm z$ to its value in $\bm z'$ when the neighborhood treatment is fixed at $\overline{\bm z}$.
Similarly, we define the interference effect as
\begin{equation}
\iota^*(\overline{\bm z}, \overline{\bm z}'; \bm z) = 
\E \left[ \overline Y(\bm z, \overline{\bm z}') - \overline Y(\bm z, \overline{\bm z}) \right],
\label{eq:interference_effect_network}
\end{equation}
which represents the average causal effect for a change in the neighborhood treatment from $\overline{\bm z}$ to $\overline{\bm z}'$ when the individual treatment is set to $\bm z$.
}

\cng{Common choices for hypothesized exposures include the case where all units have the same individual and neighborhood exposure value. This can be achieved by setting $\bm z = z\bm 1_n, \bm z' = z' \bm 1_n, \overline{\bm z} = \overline z \bm 1_n$ and $\overline{\bm z}' = \overline z' \bm 1_n$, where $\bm 1_n$ is the a vector of 1s of length $n$, and $z, z'$ and $\overline z, \overline z'$ are the hypothesized individual and neighborhood exposure value, respectively. Alternative estimands include the shift-interventions that consider what would occur if the observed exposures where shifted by some constant \citep{haneuse2013estimation}, which have been previously advocated for spatial analyses \citep{gilbert2021causal}.
This can be achieved by setting $\bm z = \bm Z, \bm z' = \bm Z + \delta \bm 1_n, \overline{\bm z} = \overline{\bm Z}$ and $\overline{\bm z}' = \overline{\bm Z} + \overline \delta \bm 1_n$ for constants $\delta, \overline \delta$.
}

\cng{With network data, it is not possible to conceive that units are independently drawn from a superpopulation of interest. Therefore, we interpret the expectation in the definitions in \cref{eq:local_effect_network} and \cref{eq:interference_effect_network} from a model-based perspective, where potential outcomes are viewed as random variables \citep{zigler2025bipartite}. Alternatively, these estimands can be interpreted as local and interference effects over similar networks \citep{ogburn2024causal}. For network data, sample average local and interference effects can be defined eliminating the expectation operator. Our Bayesian formulation in \cref{sec:method} allows for estimation of such estimands, where the model is used to impute missing potential outcomes \citep{ding2018causal, li2023bayesian}. We discuss this in Section J of the Supplementary Materials.}

\subsection{Ignorability in the presence of unmeasured spatial covariates}
\label{subsec:network_ignorability}

For identifying and estimating causal effects from observational data, a set of covariates that satisfies ignorability has to be conditioned on. Let $\widetilde C_i = (C_{i1}, C_{i2}, \dots, C_{ip})^T$ denote unit $i$'s $p$ measured covariates, which include individual and neighborhood characteristics, and $\bm C = (\widetilde C_1, \widetilde C_2, \dots, \widetilde C_n)$ the $n \times p$ matrix of measured covariates. However, these covariates are not a sufficient conditioning set for unconfoundedness of the treatment assignment, 
\begin{equation}
(Z_i, \overline Z_i) \not\!\indep Y_i(z, \overline z) \mid \widetilde C_i \quad \text{for some } z, \overline z,
\label{eq:not_ignorable}
\end{equation}
and biases for estimating causal effects persist. Instead, we assume that ignorability holds conditional on measured covariates, and the local and neighborhood value of an unmeasured covariate. Specifically, let $\bm U = (U_1, U_2, \dots, U_n)$ denote an unmeasured covariate, and $\overline U_i = \sum_j A_{ij}U_j / \sum_j A_{ij}$. We make the following assumption.
\begin{assumption}
\cng{It holds that
\( \displaystyle
(Z_i, \overline Z_i) \indep \{ Y_i(z, \overline z), \text{ for all } z, \overline z\} \mid \widetilde C_i, U_i, \overline U_i,
\)}
and also that
$f(Z_i = z, \overline Z_i = \overline z \mid \widetilde C_i, U_i, \overline U_i) > 0$.
\label{ass:network_ignorability}
\end{assumption}
\cref{ass:network_ignorability} resembles the unconfoundedness assumption on the individual and neighborhood treatment for network data in \cite{forastiere2021identification} and \cite{ zigler2025bipartite}, though here we allow for confounding from 
an unmeasured covariate. In principle, the potential outcomes considered in \cref{eq:not_ignorable} and \cref{ass:network_ignorability} need only correspond to exposures $(z, \overline z)$ that appear in the estimand of interest, though we refrain from writing this explicitly for notational simplicity.

In Section C of the Supplementary Materials, we provide an example of how the variables' inherent spatial structure might occur in a spatial setting.
There,
we also investigate the influence of spatial dependencies in learning causal effects from a network of spatial data. The conclusions are the same as the ones for paired data: \begin{enumerate*}[label=(\alph*)]
\item spatial confounding and interference can manifest as each other, 
\item inherent spatial dependencies can render standard estimation techniques invalid, 
\item controlling for local and neighborhood covariates is necessary, and 
\item local and interference effects should be investigated simultaneously.
\end{enumerate*}

\section{Bayesian inference of local and interference effects with spatial dependencies and unmeasured spatial confounding}
\label{sec:method}

We develop a Bayesian approach for estimating causal effects with spatially dependent data that addresses complications due to spatial causal and statistical dependence and mitigates bias due to local and neighborhood unmeasured spatial confounding. 
In \cref{subsec:bayesian_treatment}, we show that, when missing confounders are present, it is necessary to incorporate the treatment assignment mechanism in a Bayesian causal inference procedure.
These derivations inform us of the assumptions required for the unmeasured confounder and the exposure described in \cref{subsec:UZ_assumptions}. 
In \cref{subsec:identifiability}, we show theoretically that, under a linear model formulation, all model parameters are identifiable from data, even in the presence of unmeasured spatial confounders.
In \cref{subsec:priors}, we design sensible prior distributions for the hyperparameters of the spatial confounder by studying what these choices imply for our prior beliefs on the unmeasured covariate's confounding strength.

\subsection{The role of the treatment assignment mechanism in spatial settings}
\label{subsec:bayesian_treatment}

Bayesian causal inference views unobserved potential outcomes as missing data, and inference on causal effects is acquired from their posterior distribution \citep{rubin1978bayesian, imbens1997bayesian, ding2018causal, li2022bayesian}.
Let 
$Y_i (\cdot)$ denote the set of all potential outcomes for unit $i$, and
$\bm Y(\cdot)  = \{Y_i (\cdot), \text{ for all } i \} $ their collection across units.
We also write $\bm Y(\cdot) = \{ \bm Y, \bm Y^{\text{miss}} \}$ where $\bm Y = (Y_1, Y_2, \dots, Y_n)$ are the observed outcomes and $\bm Y^{\text{miss}}$ is the collection of {\it un}observed potential outcomes.
Bayesian inference specifies
\(
p(\bm Y(\cdot), \bm Z, \overline{\bm Z}, \bm C \mid \theta)
\) 
and a prior distribution \( p(\theta) \),
and imputes missing potential outcomes from
\begin{equation}
\begin{aligned}
& p(\bm Y^{\text{miss}} \mid \bm Y, \bm Z, \overline{\bm Z}, \bm C, \theta)
\propto
\ P(\bm Z, \overline{\bm Z} \mid \bm Y(\cdot), \bm C, \theta) \  P(\bm Y(\cdot) \mid \bm C, \theta) \ P(\bm C \mid \theta).
\end{aligned}
\label{eq:impute_Ymiss}
\end{equation}
If ignorability does not hold based only on measured covariates as in \cref{eq:not_ignorable},
\(
P(\bm Z, \overline{\bm Z} \mid \bm Y(\cdot), \bm C, \theta)
\neq
P(\bm Z, \overline{\bm Z} \mid \bm C, \theta)
\). 
As a result, the treatment assignment mechanism will not ``drop out'' from \cref{eq:impute_Ymiss}, and it will be informative for the imputation of missing potential outcomes \citep{mccandless2007bayesian, ricciardi2020bayesian}.
Instead, we write $p(\bm Y(\cdot), \bm Z, \overline{\bm Z}, \bm C)$ as
\begin{align*}
\int
& p(\bm Y(\cdot) \mid \bm Z, \bm C, \bm U, \theta^*) \ 
p(\bm Z \mid \bm C, \bm U, \theta^*) \ 
p(\bm U \mid \bm C, \theta^*) p(\bm C \mid \theta^*)
\ \mathrm{d} \bm U \ p(\theta^*) \ \mathrm{d}\theta^* 
\end{align*}
where $\overline{\bm Z}$ is excluded since it is uniquely defined based on $\bm Z$, and $\theta^*$ extends $\theta$ to include parameters governing $\bm U$. 

\cng{We make the following assumption which codifies that potential outcomes are conditionally independent across units and only dependent on local and neighborhood information.
\begin{assumption}[Independence and individuality in potential outcomes]
    It holds that
    \[
    p(\bm Y(\cdot) \mid \bm Z, \bm C, \bm U, \theta^*)
= \prod_{i = 1}^n p(Y_i(\cdot) \mid Z_i, \overline Z_i, \widetilde C_i, U_i, \overline U_i, \theta^*).
    \]
\label{ass:pot_out_independence_units}
\end{assumption}
\vspace{-8pt}
If \cref{ass:network_ignorability} also holds, the joint distribution in \cref{ass:pot_out_independence_units} is equal to \( \prod_{i = 1}^n p(Y_i(\cdot) \mid \widetilde C_i, U_i, \overline U_i, \theta^*) \).} Therefore, having access to $\bm U$ (in addition to $\bm C$) renders the treatment assignment ignorable within the Bayesian framework, since $p(\bm Y(\cdot), \bm Z, \overline{\bm Z}, \bm C)$ is now equal to
\begin{equation}
\begin{aligned}
\int \Big[ \prod_{i = 1}^n p(Y_i(\cdot) \mid \widetilde C_i, U_i, \overline U_i, \theta^*) \Big] \ 
p(\bm Z \mid \bm C, \bm U, \theta^*) 
p(\bm U \mid \bm C, \theta^*) \ p(\bm C \mid \theta^*)
\ \mathrm{d} \bm U \ p(\theta^*) \ \mathrm{d}\theta^*.
\end{aligned}
\label{eq:bayesian_framework}
\end{equation}

Since $\bm U$ is unknown, and it plays a role in the distribution of the treatment in \cref{eq:bayesian_framework}, the treatment assignment has to be incorporated in a valid Bayesian procedure for imputing the missing potential outcomes. 
This establishes from a new perspective that, within the Bayesian paradigm, simply including a spatial random effect in the outcome model does \textit{not} account for unmeasured spatial confounding and incorporating an exposure model is necessary.

\subsection{Exposure-confounder assumptions}
\label{subsec:UZ_assumptions}

The derivations in \cref{eq:bayesian_framework} also illustrate that it is necessary to specify joint distributions on the unmeasured and measured variables in order to proceed. As is often done in Bayesian causal inference, we do not model the joint distribution of the covariates $\bm C$, and instead condition on their values \citep{li2022bayesian}.
For the unmeasured spatial variable and for the (continuous) exposure, we specify a class of joint multivariate normal distributions as discussed in the following assumption.
\begin{assumption}
The unmeasured spatial confounder and the spatial exposure have a joint normal distribution conditional on the measured covariates. Specifically,
\begin{equation}
\begin{pmatrix} \bm U \\ \bm Z \end{pmatrix} \Big| \  \bm C \sim N_{2n} \left(
\begin{pmatrix} \bm 0_n \\ \gamma_0 \bm 1_n + \bm C^T \bm \gamma_C \end{pmatrix} ,
\begin{pmatrix} G & Q \\ Q^\top & H \end{pmatrix}^{-1}
\right),
\label{eq:UZ_normal}
\end{equation}
for $\bm \gamma_C$ vector of length $p$, and $G, H$ positive definite matrices. The matrix $Q$ is diagonal with elements $q_i = - \rho \sqrt{g_{ii}h_{ii}}$, where $g_{ii}, h_{ii}$ are the diagonal elements of $G, H$, respectively.
\label{ass:UZ_normal}
\end{assumption}
\noindent

The assumption of multivariate normality on $(\bm U, \bm Z)$ in \cref{ass:UZ_normal} allows us to interpret the parameters of the precision matrix.
Zero elements of the precision matrix specify conditional independence of the corresponding variables. A diagonal $Q$ encodes that $Z_i \indep \bm U_{-i} \mid U_i, \bm Z_{-i}, \bm C_i$, where 
$\bm U_{-i} = (U_1, \dots, U_{i-1}, U_{i + 1}, \dots, U_n)$ includes all the entries in $\bm U$ except the one for unit $i$, and $\bm Z_{-i}$ is defined similarly.
Therefore, the assumption that $Q$ is diagonal is a statistical representation of the absence of an arrow from $U_i$ to $Z_j$ in the graphs of \cref{fig:graphs}, describing that the unmeasured variable is a driver of only the local exposure level.
It is reasonable to specify that $\bm U$ does not depend on $\bm C$ in \cref{ass:UZ_normal} since the part of the unmeasured variable that is correlated with measured covariates is already adjusted for.

The joint distribution in \cref{ass:UZ_normal} has been previously used in \cite{schnell2020mitigating}. Our work provides two new insights on the role and interpretation of this assumption through the lens of Bayesian causal inference. We have shown that adopting a joint distribution on $(\bm U, \bm Z)$ is necessary within the Bayesian framework 
(\cref{subsec:bayesian_treatment} and equation \cref{eq:bayesian_framework}). Moreover, we have linked the distributional \cref{ass:UZ_normal} to the causal relationship of variables viewed through the causal graph representation. Therefore, we illustrate how this statistical assumption relates to the assumptions made on the complex causal dependence of the two variables, and how \cref{ass:UZ_normal} can be adapted under different assumptions.

Since we only have one realization of the spatial exposure $\bm Z$, the conditional precision matrices $G$ and $H$ cannot be estimated without imposing some structure on their elements.
To ease prior elicitation in \cref{subsec:priors} that is consistent for both areal and point-referenced data, we specify $G = \tau_U^2 (\widetilde D - \phi_U \widetilde A)$ in either case, where $\widetilde D$ is the diagonal matrix with entries $\widetilde D_i = \sum_j \widetilde A_{ij}$, and $\tau_U, \phi_U$ are unknown parameters. Similarly, we specify $H = \tau_Z^2 (\widetilde D - \phi_Z \widetilde A)$. The matrix $\widetilde A$ can be the same as or different from the adjacency matrix in the definition of the neighborhood exposure and covariate in Assumptions \ref{ass:sutva} and \ref{ass:network_ignorability}, or in the definition of $G$ and $H$.

Despite the interpretability of the spatial parameters in the precision matrix in the multivariate normal distribution, alternative specifications for the joint distribution of $(\bm U, \bm Z)$ can be accommodated.
In the case of binary exposures, one can use a probit or logistic model for the spatial treatment variable and impose \cref{ass:UZ_normal} on the underlying linear predictor.

\subsection{A model on potential outcomes and identifiability in the presence of unmeasured spatial confounding}
\label{subsec:identifiability}

It is reasonable to ponder whether causal effects are identifiable from the observed data based on our assumptions in the presence of unmeasured local and neighborhood confounding, and spatial structure in the exposure variable. 

\cng{We study this question under a specific formulation of potential outcomes. We assume that potential outcomes satisfy
\begin{equation}
\begin{aligned}
Y_i(z, \overline z) &= \beta_0 + \beta_Z z + \beta_{\bar Z} \overline z + \widetilde C_i^T \bm \beta_C + \beta_U U_i + \beta_{\bar U} \overline U_i + \epsilon_i(z, \overline z).
\end{aligned}
\label{eq:linear_sem}
\end{equation}
where $\epsilon_i(z, \overline z)$ are mean zero random variables with variance $\sigma^2_Y$ which are independent across units, and across exposure levels for the same unit.
Under this model, the parameters $\beta_Z$ and $\beta_{\bar Z}$ describe the local and interference effects for a unit increase in the local and neighborhood exposure, respectively. Specifically, the local effect is $\beta_Z = \lambda^*(\bm z, \bm z + \bm 1_n; \overline{\bm z})$, and the interference effect is $\beta_{\bar Z} = \iota^*(\overline{\bm z} + \bm 1_n, \overline{\bm z}; \bm z)$, for any $\bm z, \overline{\bm z}$.
}

The coefficients $\beta_U, \beta_{\bar U}$, and the parameter $\tau_U$ of the precision matrix $G$ are not identifiable up to scaling of the unmeasured confounder $\bm U$. To see this, note that setting $\bm U' = c \bm U$, $\beta_U' = \beta_U / c$, $\beta_{\bar U}' = \beta_{\bar U} / c$, and ${\tau_U^{2}}' = \tau_U^2 / c^2$ for some $c \neq 0$ leads to the same value of the likelihood for $(\bm Y, \bm Z, \bm U) \mid \bm C$. Therefore, without loss of generality, we set $\beta_U = 1$. Even though at first sight it might appear that we ``force'' $U$ in the outcome model by setting $\beta_U = 1$, we show in \cref{subsec:priors} that our prior for $\tau_U$ ensures that this is not the case.

The next theorem shows that, for a large enough spatial ring network, all model parameters, including the causal effects, are identifiable even in the presence of unmeasured spatial confounding. The definition of a ring graph and the proof are in Section 
D
of the Supplementary Material. Without loss of generality, we consider the case without measured covariates.

\begin{theorem}
Consider the setting of \cref{eq:linear_sem} for units organized on a ring graph with $n$ nodes and without measured covariates. Using data $(\bm Z, \bm Y)$ we can identify whether or not $\rho\phi_U = 0$. If $\rho \phi_U \neq 0$, and for $\beta_U = 1$, we have that all model parameters $(\beta_Z, \beta_{\bar Z}, \beta_{\bar U}, \rho, \tau_U, \phi_U, \tau_Z, \phi_Z, \sigma^2_Y)$
are identifiable from $(\bm Z, \bm Y)$ as $n \rightarrow \infty$.
\label{theorem:identifiability}
\end{theorem}

If $\rho \phi_U = 0$, there is no spatial predictor of the exposure and therefore no unmeasured spatial confounding. Our proof draws from \cite{schnell2020mitigating}. Here, we address complications due to the causal dependence across units which impose the inclusion of the neighborhood confounder and exposure values in the outcome model, and additional model parameters. Crucially, \cref{theorem:identifiability} establishes that we can identify complex confounder structures that are not limited to local restrictions.

The proof provides interesting insights for where the information in the data comes from to identify non-local confounding structures. All spatial parameters, $(\tau_U, \phi_U, \tau_Z, \phi_Z)$, and the neighborhood confounding effects, $\beta_{\bar U}$, are identified based on the covariance structure in the exposure and the outcome residuals, and how it attenuates as a function of distance. Interestingly, neighborhood confounding effects are identified by studying the spatial structure in the outcome model residuals that is not explained by similar structure in the exposure model residuals. 

\cng{We make a few remarks: 
\begin{enumerate*}[label=(\alph*)]
    \item The exposure can be included in the model in \cref{eq:linear_sem} in a different form from the one in the joint distribution of \cref{ass:UZ_normal}.
\item Non-linear functions and interaction terms between the local exposure, the neighborhood exposure, and the covariates can be straightforwardly incorporated in the model.
Since incorrect parametric specifications can bias causal effect estimation, model diagnostics should be used to evaluate deviations from the specifications in (a) and (b), as we illustrate in our study in \cref{sec:application}.
\item The outcome model in \cref{eq:linear_sem} specifies independence of the error terms across units, which corresponds to a structural representation of the absence of outcome dependence in the graphs of \cref{fig:graphs} (missing $Y_1 \leftrightarrow Y_2$). 
\item The independence of the error terms across exposure levels for the same unit implies independence of corresponding potential outcomes \citep{silva2022multiple}. For the estimands in \cref{eq:local_effect_network} and \cref{eq:interference_effect_network}, correlation of potential outcomes does not alter estimation \citep{ding2018causal, li2022bayesian}. For alternative estimands corresponding to sample average local and interference effects, the correlations of potential outcomes can be incorporated as sensitivity parameters (see section J of the Supplementary Material).
\item  Our proof of the identifiability results in \cref{theorem:identifiability} uses the ring graph assumption to acquire a closed form for the data's covariance matrix. 
Our conjecture is that identifiability holds for other spatial dependence structures that allow for a sufficient number of location pairs at different distances. A comprehensive examination of general identifiability under arbitrary spatial dependence is reserved for future research.
\end{enumerate*}
}

\subsection{Prior distributions for confounding adjustment}
\label{subsec:priors}

In Bayesian settings, model performance often depends on the choice of hyperparameters, and non-informative priors can lead to poor performance in certain settings \citep{gelman2008weakly}. We adopt weakly informative prior distributions for intercepts, coefficients of the measured covariates, and the local and neighborhood exposure and confounder values, the variance of the residual error in \cref{eq:linear_sem}, and the parameters of the covariance matrix in \cref{eq:UZ_normal}.

We consider measured covariates, exposure and outcome that are standardized to have mean 0 and variance 1.
We adopt independent $N(0, \sigma^2_{\text{prior}})$ prior distributions for the coefficients of all measured covariates, $\bm \beta_C, \bm \gamma_C$ and for the intercepts $\beta_0, \gamma_0$, with $\sigma^2_{\text{prior}} = 2$. 
We adopt a $N(0, \sigma^2_{\text{prior},\overline U})$ prior distribution for $\beta_{\bar U},$ and we set $\sigma^2_{\text{prior},\overline U} = 0.35^2$ to express the prior belief that the importance of the neighborhood value of the confounder is smaller than that of the local value of the confounder (since $\beta_U = 1$).
For the residual variance of the outcome model in \cref{eq:linear_sem}, we specify $\sigma^2_Y \sim IG(\alpha_Y, \beta_Y)$. We follow a data-driven procedure for choosing the hyperparameters $\alpha_Y, \beta_Y$. We regress the outcome on the local exposure and the measured covariates and acquire the estimated residual variance, $\widetilde \sigma^2_Y$. 
We set $\alpha_Y = 3$ and $\beta_Y = \widetilde \sigma^2_Y / 2$, which leads to a prior distribution on the residual variance that puts most of its weight on values smaller than $\widetilde \sigma^2_Y$, and specifically $P(\sigma^2_Y < \widetilde \sigma^2_Y) \approx 0.91$.

Next, we specify prior distributions for the parameters of the joint precision matrix in \cref{eq:UZ_normal}. We specify flat priors for $\phi_Z, \rho$ on the $(-1, 1)$ interval. We assume that $\phi_U > 0$, and specify $\phi_U \sim \text{Beta}(6, 6)$ to encourage values that imply some spatial dependence ($\phi_U$ away from 0) while avoiding degenerate distributions ($\phi_U$ away from 1). We also require that $\phi_Z < \phi_U$ since the exposure should vary within levels of the confounder, in line with conclusions from \cite{Paciorek2010}, \cite{schnell2020mitigating} and \cite{dupont2022spatial}.

Lastly, we design prior distributions for $\tau_U, \tau_Z$. The priors for these parameters can have a large effect on model performance, and their choice requires careful consideration. For simplicity we discuss in detail the situation where all nodes have the same number of neighbors, and $D = dI_n$ for some $d > 0$ and $I_n$ being the $n \times n$ identity matrix.
We can show that $\Var(U_i) \geq (d \tau_U^2)^{-1}$, where equality holds for $\rho = 0$. Since $U_i$ is a priori centered at 0, if the marginal variance of $U_i$ is small for all $i$, the vector of $\bm U$ is almost indistinguishable from the vector of all zeros, and essentially drops out from the outcome model (even though we set $\beta_U = 1$). 
Conversely, if the marginal variance of $U_i$ is big a priori, then this prior distribution would imply a strong importance of $U$ in the outcome model (considering $\beta_U = 1$).
Therefore, our choice for the prior distribution of the hyperparameter $\tau_U$ should be informed by the implied prior distribution on the unmeasured covariate's confounding strength. We specify a prior distribution for $\tau_U$ which avoids the pathological situation that $U$ has an unrealistically high predictive accuracy for the outcome. Specifically,
we specify that $1 / \tau_U$ has a truncated mean-zero normal distribution with variance $d \sigma^2_{\text{prior}} / 2$, and truncated below at 0. 
The induced prior on $(d \tau_U^2)^{-1}$ ensures that the unmeasured variable's strength in the outcome model resembles, a priori, the measured covariates' strength in the outcome model specified by the $N(0, \sigma^2_{\text{prior}})$ prior distribution on their coefficients.

Similarly, the magnitude of $1 / (d \tau_Z^2)$ can be conceived as the variance in the exposure that cannot be explained by covariates. Therefore, $1 / \tau_Z^2$ should not be too small because we expect {\it some} inherent variability in the exposure. At the same time, it should not be too big in comparison to the residual variance of the regression of $Z$ on the measured covariates, denoted by $\widetilde \sigma^2_Z$. 
Let $\widetilde s^2_Z$ be the observed marginal variance in the exposure across locations. We specify that $1 / \tau_Z$ follows a truncated normal distribution centered at $\sqrt{d \ \widetilde \sigma^2_Z / 2}$ with standard deviation 1, truncated below at $\sqrt{d \ 0.01 \ \widetilde s^2_Z}$ and above at $\sqrt{d \ \widetilde \sigma^2_Z / 0.8}$. 

The prior distributions on $\tau_U$ and $\tau_Z$ are illustrated in Section
E
of the Supplementary Material.
We see that the implied prior distribution on the unmeasured variable's predictive strength allows for all reasonable values.
When the degree is not constant across nodes (which is the case in most networks), we set $d$ to be the median network degree.

We sample from the posterior distribution of model parameters using Markov chain Monte Carlo (MCMC). The algorithm is described in Section 
F
of the Supplementary Material.

\section{Simulations}
\label{sec:sims}

We perform simulations to evaluate the extent to which the approach introduced in \cref{sec:method} mitigates the bias in estimating local and interference causal effects that is caused by direct and indirect unmeasured spatial confounding and the inherent spatial dependence in the exposure.
We simulate data under the data generative mechanisms in \cref{fig:graphs}. We consider observations on a network of interconnected units represented by a line graph, where each unit is connected with two others, except for the first and last units which only have one neighbor each. 
Simulations with pairs of interacting observations are deferred to Section
G
of the Supplementary Material.

For number of units $n \in \{200, 350, 500\}$, we generated four measured covariates from independent $N(0, 1)$ distributions, the unmeasured confounder and the exposure of interest from \cref{eq:UZ_normal}, and calculated $\overline U, \overline Z$ for each observation. The outcome was generated according to the linear model in \cref{eq:linear_sem} with $\gamma_0 = \beta_0 = 0$. The coefficients of the measured covariates were generated randomly once and were fixed to $\bm \gamma_C = (-0.35, -0.64,  0.49,  0.06)$ and $\bm \beta_C = (0.06, 0.85, 0.02, 0.33)$ throughout our simulations. We specified spatial parameters $\phi_U = 0.6, \phi_Z = 0.4,$ and $\rho = 0.35$.
For the network data, for which median node degree is equal to 2, we set $\tau^2_U = \tau^2_Z = 1$. In all cases, we set the outcome residual error variance to one. The default outcome model coefficients of the local and neighborhood exposure and unmeasured confounder were set to $\beta_Z = 1,$ $\beta_{\bar Z} = 0.8,$ $\beta_U = 1$, and $\beta_{\bar U} = 0.5$, except for when the relationship does not exist, in which case the corresponding coefficient is set to 0. These specifications match exactly the motivating simulations for the network data discussed in \cref{sec:one_network} which are presented in detail in Section
C.2
of the Supplementary Material.

We generate 500 data sets under each of the 36 different scenarios which are combinations of the six scenarios in \cref{fig:graphs}, the three sample sizes, and for network and paired data.
We compare the proposed approach to OLS conditional on the measured covariates for estimating local and interference effects.
In the absence of unmeasured spatial confounding, the OLS estimator is most efficient for estimating causal effects since it is based on the correctly specified outcome model. Therefore, this comparison informs us of potential efficiency loss when spatial confounding is considered but it is not truly present.
In the presence of unmeasured confounding, the OLS estimator incurs confounding bias. Therefore, comparing the two approaches in this setting illuminates the extent to which our approach can alleviate this bias.
For our method, we considered two chains with 7,000 iterations as a burn-in period and thinning by 60. We report results for data sets for which the MCMC did not show signs of lack of convergence, evaluated based on the $\widehat R$ statistic \citep{vehtari2021rank} for $\beta_Z$ and $\beta_{\bar Z}$. 

\begin{table*}[!t]
    \centering
    \caption{Simulation results with One Interconnected Network. Bias, root mean squared error (rMSE), and coverage of 95\% intervals based on OLS and the method of \cref{sec:method}, for the local and the interference effect. We show simulation results for the 6 settings in \cref{fig:graphs}, and for sample size equal to 200, 350 and 500. Coverage rates are reported as percentages.
    }
    \resizebox{\linewidth}{!}{%
    \begin{tabular}{*{15}{c}}
       & & \multicolumn{6}{c}{Local effect} & &  \multicolumn{6}{c}{Interference effect} \\
       \cmidrule(lr){3-8} \cmidrule(lr){10-15}
        \multicolumn{2}{c}{True model \&} & \multicolumn{3}{c}{OLS} &  \multicolumn{3}{c}{Our approach} &
        & \multicolumn{3}{c}{OLS} &  \multicolumn{3}{c}{Our approach} \\
        \cmidrule(lr){3-5} \cmidrule(lr){6-8} \cmidrule(lr){10-12} \cmidrule(lr){13-15}  
        \multicolumn{2}{c}{sample size} & Bias & rMSE & Cover & Bias & rMSE & Cover &
        & Bias & rMSE & Cover & Bias & rMSE & Cover \\[5pt]
    
        \hline
        \\[-10pt]
        \multirow{4}{*}{\ref{fig:direct}} & & \multicolumn{13}{c}{$\beta_{\bar Z} = 0$ and $\beta_{\bar U}=0$} \\[0pt]
        \cmidrule{7-11}
        & 200 & 0.492 & 0.505 & 0.3
        & -0.013 & 0.306 & 91.6 &
        & 0.154 & 0.186 & 72.3
        & 0.030 & 0.137 & 93.4  \\ 
        & 350 & 0.499 & 0.507 & 0
        & -0.060 & 0.247 & 94.9 & 
        & 0.146 & 0.167 & 55\phantom{.0} 
        & 0.002 & 0.094 & 96.6 \\
        & 500 & 0.510 & 0.516 & 0
        & -0.073 & 0.222 & 96.5 & 
        & 0.157 & 0.172 & 36\phantom{.0} 
        & 0.009 & 0.086 & 93.8 \\[3pt]
        \hline
        \\[-10pt]

        \multirow{4}{*}{\ref{fig:general_spatial_conf}} & &
        \multicolumn{13}{c}{$\beta_{\bar Z} = 0$}
        \\[0pt]
        \cmidrule{7-11}
        & 200 & 0.616 & 0.630 & 0 
        & -0.169 & 0.302 & 96.9 & 
        & 0.277 & 0.302 & 34.3 
        & 0.027 & 0.146 & 93.3 \\
        & 350 & 0.624 & 0.632 & 0 
        & -0.188 & 0.283 & 93.5 & 
        & 0.273 & 0.288 & 15\phantom{.0} 
        & 0.004 & 0.107 & 95.7  \\
        & 500 & 0.639 & 0.644 & 0 
        & -0.174 & 0.253 & 94.6 & 
        & 0.289 & 0.299 & \phantom{0}3.7 
        & 0.011 & 0.092 & 94.8 \\[3pt]
        \hline
        \\[-10pt]

        \multirow{4}{*}{\ref{fig:interference}} & &
        \multicolumn{13}{c}{$\beta_{UZ} = 0$ and $\beta_U = \beta_{\bar U} = 0$}
        \\[0pt]
        \cmidrule{7-11}
        & 200 & \phantom{-}0.003 & 0.097 & 95.3 
        & -0.027 & 0.120 & 98.8 & 
        & \phantom{-}0.004 & 0.086 & 95.7 
        & \phantom{-}0.002 & 0.087 & 96.2 \\
        & 350 & -0.002 & 0.072 & 96\phantom{.0} 
        & -0.023 & 0.100 & 99.7 & 
        & -0.007 & 0.066 & 94.3 
        & -0.016 & 0.067 & 95.5 \\
        & 500 & \phantom{-}0.001 & 0.067 & 92.7 
        & -0.019 & 0.094 & 98.7 & 
        & -0.001 & 0.056 & 94\phantom{.0} 
        & -0.004 & 0.059 & 94.1 \\[3pt]
        \hline
        \\[-10pt]

        \multirow{4}{*}{\ref{fig:predictor_interference}} & &
        \multicolumn{13}{c}{$\beta_U = 0$ and $\beta_{\bar U} = 0$} \\[0pt]
        \cmidrule{7-11}
        & 200 & 0.002 & 0.085 & 96\phantom{.0} 
        & -0.020 & 0.112 & 99.2 & 
        & \phantom{-}0.003 & 0.081 & 96.3 
        & \phantom{-}0.004 & 0.085 & 95.4 \\
        & 350 & 0.000 & 0.064 & 95.7 
        & -0.027 & 0.125 & 96.7 & 
        & -0.006 & 0.063 & 94\phantom{.0} 
        & -0.017 & 0.068 & 93.9 \\
        & 500 & 0.001 & 0.060 & 92.7 &
        -0.012 & 0.113 & 97.7 & 
        & -0.001 & 0.054 & 94\phantom{.0} 
        & -0.003 & 0.056 & 94.5 \\[3pt]
        \hline
        \\[-10pt]

        \multirow{4}{*}{\ref{fig:direct_interference}} & &
        \multicolumn{13}{c}{$\beta_{\bar U} = 0$} \\[0pt]
        \cmidrule{7-11}
        & 200 & 0.492 & 0.505 & 0.3 
        & 0.037 & 0.296 & 89.6 & 
        & 0.154 & 0.186 & 72.3 
        & \phantom{-}0.030 & 0.141 & 92\phantom{.0} \\
        & 350 & 0.499 & 0.507 & 0 
        & -0.031 & 0.234 & 95.3 & 
        & 0.146 & 0.167 & 55\phantom{.0} 
        & -0.004 & 0.101 & 95\phantom{.0} \\
        & 500 & 0.510 & 0.516 & 0 
        & -0.017 & 0.202 & 98.6 & & 0.157 & 0.172 & 36\phantom{.0} 
        & \phantom{-}0.011 & 0.088 & 92.9 \\[3pt]
        \hline
        \\[-10pt]
       
        \multirow{3}{*}{\ref{fig:general_interference}} 
        & 200 & 0.616 & 0.630 & 0 
        & -0.139 & 0.279 & 97.1 & 
        & 0.277 & 0.302 & 34.3 
        & 0.019 & 0.145 & 94.2 \\ 
        & 350 & 0.624 & 0.632 & 0 
        & -0.155 & 0.254 & 93.9 & 
        & 0.273 & 0.288 & 15\phantom{.0} 
        & 0.000 & 0.106 & 95.7 \\
        & 500 & 0.639 & 0.644 & 0 
        & -0.144 & 0.229 & 95\phantom{.0} & 
        & 0.289 & 0.299 & \phantom{0}3.7 
        & 0.008 & 0.091 & 95\phantom{.0} \\[1pt]
        \hline
        \end{tabular}
     }%
     \label{tab:sims_network}
\end{table*}

Results for the network data simulations are shown in \cref{tab:sims_network}. We show bias, root mean squared error (rMSE), and coverage of 95\% intervals (confidence intervals for OLS, credible intervals for the Bayesian method).
When the unmeasured variable does not confound the relationship of interest (scenarios \ref{fig:interference} and \ref{fig:predictor_interference}) and OLS performs well, our method remains essentially unbiased for both the local and the interference effects. In these settings where our approach is not necessary for controlling for the unmeasured variable, it has slightly larger rMSE than OLS for the local effect, but the two approaches have similar rMSE for estimating the interference effects. These results indicate that the proposed approach might avoid efficiency loss in estimating interference effects when accounting for unmeasured confounding.
In all other cases where unmeasured confounding exists (scenarios \ref{fig:direct}, \ref{fig:general_spatial_conf}, \ref{fig:direct_interference}, and \ref{fig:general_interference}), our approach returns significantly lower bias and achieves substantially lower rMSE in comparison to OLS for all sample sizes, and for both local and interference effects. Furthermore, the proposed approach achieves close to nominal coverage in all scenarios and for both local and interference effects. Our simulations illustrate that our method can protect from biases arising due to unmeasured spatial variables, while ensuring proper inference.
Simulations for paired data are shown in Section
G
of the Supplemntary Material,
and the conclusions are identical.

\section{Estimating the impact of power plant emissions allowing for interference and unmeasured spatial confounding}
\label{sec:application}

We return to our study in \cref{sec:data} to estimate the effects of SO$_2$ emissions from power plants on cardiovascular mortality among the elderly, allowing for the presence of interference and unmeasured confounding by spatial variables. 

\cng{In our data, we have two sets of counties, those with both local and neighborhood exposure, and those with only neighborhood exposure. We incorporate both sets of counties in our analysis by adopting separate models of the form in Equation \cref{eq:linear_sem} and linking their parameters. Specifically, we allow for separate intercepts for the two models, while coefficients of shared variables are assumed to be equal to one another. We provide more information on the exact model formulation in Section I of the Supplementary Material.}

\cng{Since the local emissions are highly skewed, we adopt the joint normality in \cref{ass:UZ_normal} on the log-transformed exposure whose distribution is bell-shaped. To reduce the sensitivity of our estimated effects to parametric specifications of the exposure-response curve, we include the local and neighborhood exposures in the outcome model using penalized B-splines (see Section H of the Supplementary Material for the model formulation using splines). We specify binary adjacency matrices in the CAR structure $G$ and $H$ in \cref{ass:UZ_normal} that are equal to 1 if counties are within 50 and 20 kilometers, respectively.} We also analyzed the data using OLS which ignores potential unmeasured confounding. 

For all analyses, we acquired 1,500 samples from the posterior distribution from each of three chains, using a burn-in period of 6,000 iterations and thinning by 50. We evaluated convergence of our MCMC by investigating traceplots of all parameters. \cng{We also investigated the presence of violations of our model assumptions by studying Q-Q plots of residuals and plots of residuals against covariates (see Section H of the Supplementary Material). We find no obvious violation of the normality assumption for the exposure model residuals in \cref{ass:UZ_normal} or the linearity assumption in either model. The residuals of the outcome model appear to be less dispersed than expected, which we discuss in \cref{sec:discussion}.}

\begin{figure}[!t]
\includegraphics[width=\linewidth]{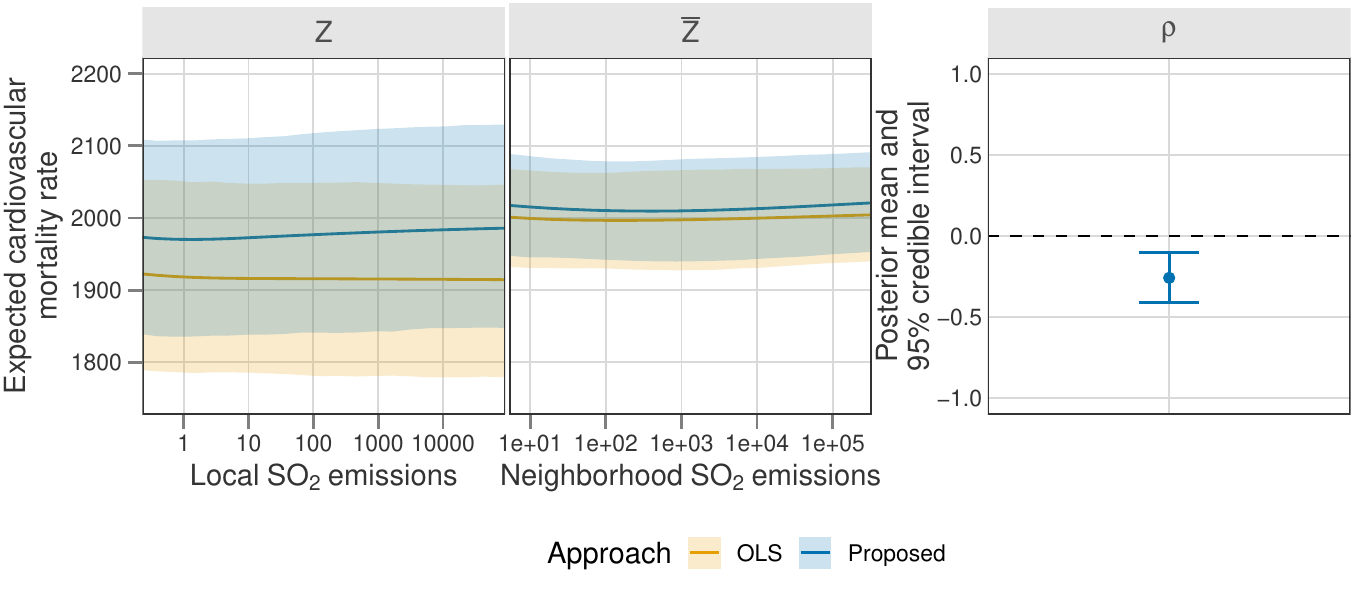}
\caption{Estimated exposure-response (ER) curve for SO$_2$ emissions and cardiovascular mortality rate. The line corresponds to the posterior mean, and the bands to the pointwise 95\% credible intervals. The color reflects whether the proposed approach (blue) or OLS that ignores unmeasured spatial confounding is considered (orange). Left: Estimated ER curve as a function of local emissions when neighborhood emissions are set to the 40$^{th}$ quantile of their observed distribution. Center: Estimated ER curve as a function of neighborhood emissions when local emissions are set to the 40$^{th}$ quantile of their observed distribution. Right: Posterior mean and 95\% credible interval for the correlation parameter between the unmeasured spatial variable $U$ and the exposure $Z$ in \cref{ass:UZ_normal}.}
\label{fig:app_results}
\end{figure}

\cng{The expected cardiovascular mortality rate (per 100,000 persons) as a function of local (left) and neighborhood (center) exposure is shown in \cref{fig:app_results}. (Since the exposures are highly skewed, the x-axis is log-transformed in both figures). Only the counties with local exposure contribute to the results for the local exposure, whereas all counties contribute to the results for neighborhood exposure (see Section I of the Supplementary material), which explains the difference in the width of the credible intervals. Lastly, we show the posterior mean and 95\% credible interval (CI) for $\rho$, which reflects the correlation between the local unmeasured spatial variable and the exposure in \cref{ass:UZ_normal}.}

\cng{The estimates from OLS and the proposed approach are comparable, and the estimated exposure-response (ER) curve based on one approach is within the credible band of the results from the other approach. Therefore, if unmeasured spatial confounding is present of the form that satisfies our assumptions, it does not appear to be strong enough to bias the OLS analysis. Despite that, the value of $\rho$ is significantly away from 0 (posterior mean: $- 0.26$, 95\% CI: $- 0.41$ to $-0.1$), reflecting the presence of an unmeasured variable that is locally correlated with the exposure. However, this unmeasured variable is relatively ``flat'' across space based on the value of $\tau_U$ (posterior mean: $4.5$, 95\% CI: $1.2$ to $11.8$) and, as a result, does not have a strong confounding effect on the relationship under study (see \cref{subsec:priors} for the relationship between $\tau_U$ and the strength of confounding).}

\cng{We performed a number of sensitivity analyses, which we summarize here and present in detail in Section H of the Supplementary Material. First, we evaluated the sensitivity of our results to the choice of the neighborhood structure in
\begin{enumerate*}[label=(\alph*)]
\item the definition of neighborhood exposure,
\item the spatial structure of the unmeasured variable, and
\item the spatial structure of the exposure
\end{enumerate*}
in Assumptions \ref{ass:sutva} and \ref{ass:UZ_normal}.
We find that the results are almost identical across all values when the neighborhood exposure is defined based on distances up to 100 kilometers.
Second, we considered including the exposure variable (in logarithm) linearly in the outcome model which might be more interpretable over the B-spline formulation. We find some evidence of unmeasured spatial confounding with effect estimates of the proposed approach being shifted towards more positive values than based on OLS.
Third, we performed the analysis while excluding the set of variables corresponding to weather characteristics. Since weather-related variables are expected to be spatial, our methodology should capture confounding bias resulting from their exclusion to some extent. We find that including or excluding measured weather covariates does not alter results.
Lastly, we evaluated the sensitivity of our results to the choice of hyperparameters in the prior distributions in \cref{subsec:priors}. Results remained essentially unchanged, reflecting robustness to prior choices.}

\section{Discussion}
\label{sec:discussion}

In this manuscript we discussed the inherent challenges and opportunities that arise in causal inference with spatial data. We illustrated the complications that arise from the data's inherent dependence structure, and discussed the interplay of spatial confounding and interference.  Unmeasured spatial confounding does not necessarily have to exist due to a completely missed covariate. Instead, it is possible that a confounder is mis-measured, or its functional form not correctly included in the outcome model. In these settings, employing a procedure that mitigates bias from unmeasured spatial covariates can improve estimation and inference.

\cng{In our analysis of the effect of local and neighborhood SO$_2$ emissions from power plants, we find limited evidence of unmeasured spatial confounding once demographic, power plant and weather variables are included. If we were to interpret our results causally, our estimates would reflect that emissions do not significantly impact cardiovascular mortality. This can be explained in at least two ways.}

\cng{First, we find that there is some evidence of misspecification in the residuals of the outcome model (see Section H of the Supplementary Material). To address this, it is important that future research investigates the complications that arise in defining and estimating causal effects with inherent spatial dependence in the outcome variable.} In all scenarios we considered here, we assumed that the outcome is not inherently spatial, though it might exhibit spatial dependence when its spatial predictors are not conditioned on. We believe that our advancements in drawing causal graphs in spatial settings in \cref{sec:pairs} and interpreting the spatial dependencies as a common underlying spatial trend in \cref{fig:dag_underlying} can provide a way forward in this setting. An inherently dependent outcome variable is entirely different from outcome dependencies that occur through contagion, and the interplay of the two in estimating causal effects is an interesting question for future research in order to fully understand causality in dependent settings. \cng{Furthermore, future research should also extend this work to non-linear models to address potential violations of the modeling assumptions on the outcome model residuals.}

\cng{Second, we used a neighborhood exposure that is based on geographic proximity. This is likely not accurate in studies of air pollution where emissions from different sources can travel in complicated patterns \citep{henneman2019accountability, zigler2025bipartite}. Future research can focus on how data and atmospheric models can be used simultaneously to identify the correct interference structure in the presence of unmeasured spatial confounding.}

\bibliographystyle{plainnat}
\bibliography{Spatial-Interference}       


\clearpage

\doparttoc 
\faketableofcontents 
\part{} 

\begin{center}
\Large

{\large Supplementary Materials for: \\[5pt]
\Large
Spatial causal inference in the presence of unmeasured confounding and interference}

{\large Georgia Papadogeorgou, 
  Srijata Samanta}

\end{center}

\appendix
\setcounter{page}{1}
\setcounter{section}{0}    
\renewcommand{\thesection}{\Alph{section}}
\setcounter{equation}{0}    
\renewcommand{\theequation}{S.\arabic{equation}}
\setcounter{table}{0}    
\renewcommand{\thetable}{S.\arabic{table}}
\setcounter{figure}{0}    
\renewcommand{\thefigure}{S.\arabic{figure}}
\setcounter{proposition}{0}    
\renewcommand{\theproposition}{S.\arabic{proposition}}
\setcounter{remark}{0}    
\renewcommand{\theremark}{S.\arabic{remark}}
\titleformat{\section}
{\normalfont\bfseries\large\centering}{Supplement \thesection.}{1em}{}
\titleformat{\subsection}
{\normalfont\normalsize\bfseries}{\Alph{section}.\arabic{subsection}}{1em}{}

\vspace{-30pt}
\addcontentsline{toc}{section}{Supplement} 
\part{ } 
\parttoc
\clearpage

\section{Causal estimands for a population of blocks}

\subsection{Alternative definitions of local and interference effects with paired data}
\label{subsec:alternative_definitions_block}

Alternate definitions of local effects draw the treatments for other units from a pre-specified distribution. For $\pi \in [0, 1]$, we use $\lambda_i(\pi)$ to denote the local effect for unit $i$ when the treatment of unit $j$ is drawn from a Bernoulli distribution with probability $\pi$, and $\lambda_i(\pi) = \pi \lambda_i(1) + (1 - \pi) \lambda_i(0)$. Similarly, the interference effect of unit $i$ when their own treatment is drawn from a Bernoulli$(\pi)$ distribution is denoted by $\iota_i(\pi)$ and $\iota_i(\pi) = \pi \iota_i(1) + (1 - \pi) \iota_i(0)$. For $\pi \in \{0, 1\}$ these definitions revert back those in \cref{eq:local_effect} and \cref{eq:interference_effect}.
If there does not exist a natural ordering of the units within a pair, then the local and interference causal effects could be defined as
\( \displaystyle
\lambda(z) = \left( \lambda_1(z) + \lambda_2(z) \right) / 2 \), and
$\iota(z) = \left(\iota_1(z) + \iota_2(z) \right) / 2$, respectively, for $z \in \{0, 1\}$.

\subsection{Blocked interference with more than two units}
\label{subsec:blocks_larger}

For interference blocks that are larger than two units, estimands for local and interference effects can be defined to average over hypothetical distributions of the neighbours' treatments, in agreement to literature on partial interference \citep{Hudgens2008, Tchetgen2012}.
We define the average outcome for unit $i$ when its treatment is set to a fixed value $z \in \{0, 1\}$, and the treatment of the other units in the block, $\bm z_{-i}$, are independent draws from a Bernoulli distribution with probability of success $\pi \in [0, 1]$ as
\(
\overline Y_i(z, \pi) 
= \sum_{\bm z_{-i} } Y_i(z_i = z, \bm z_{-i}) p(\bm z_{-i} ; \pi),
\)
where $p(\cdot ; \pi)$ is the joint probability mass function for independent Bernoulli trials with probability of success $\pi$. Then, the local effect for unit $i$ can be defined as
\(
\lambda_i(\pi) = \E \left[ \overline Y_i(1, \pi) - \overline Y_i(0, \pi) \right],
\)
and the interference effect on unit $i$ as
\(
\iota_i(\pi, \pi'; z) = \E \left[ \overline Y_i(z, \pi') - \overline Y_i(z, \pi) \right].
\)
For $\pi, \pi' \in \{0, 1\}$ and for blocks with two units, these estimands revert back to the estimands in \cref{eq:local_effect} and \cref{eq:interference_effect}, respectively.

Spatial dependence in confounders and exposure values will lead to the same complications in identifying local and interference effects discussed for paired data. To avoid distraction, we do not delve into blocked data with more than two units further. Instead, in \cref{sec:one_network} we focus on data on a single spatial network.

\section{Identifiability of quantities under different graphs}
\label{supp_sec:identifiability}

\cng{We use $i,j$ to denote the two units. We generalize the definition of local effects and interference effects to both units (we only discuss Unit 1 in the main manuscript). Specifically, we define the local effect for unit $i$ when fixing the treatment of unit $j$ to $z$ as
\begin{equation}
\begin{aligned}
\lambda_i(z) 
&= \E[Y_i(z_i = 1, z_j = z) - Y_i(z_i = 0, z_j= z)] \\
&\equiv \E[ Y_i(1, z) - Y_i(0, z)],
\end{aligned}
\end{equation}
and the interference effects for unit $i$ when setting its own treatment level at $z$ as
\begin{equation}
\begin{aligned}
\iota_i(z) &= \E[Y_i(z_i = z, z_j = 1) - Y_i(z_i = z, z_j= 0)] \\
& \equiv \E[Y_i(z, 1) - Y_i(z, 0)].
\end{aligned}
\end{equation}
}


Here, we state and prove all identifiability (and lack of) results stated in \cref{sec:pairs}.
For identifiability, we view the dependencies in \cref{fig:graphs} in terms of the underlying spatial structure shown in \cref{fig:dag_underlying}. 
Causal graphs have been used to establish nonparametric identifiability of causal estimands in a variety of settings \citep[e.g.][]{pearl1995causal, pearl2000models, 
avin2005identifiability, vansteelandt2007confounding}, and in the presence of interference explicitly \citep{ogburn2014causal}. 
Many of the statements we prove here have proofs that are straightforward based on the existing theory on graphical models \citep{spirtes1993causation, pearl1995causal, pearl2000models}, but we include them here for completeness.
Furthermore, we prove any statements about identifiability of local and interference effects that are less obvious because they pertain specifically to the spatial structure of the observations or the presence of spatial interference.

\subsection{Graph \ref{fig:direct}: Direct spatial confounding}

\cng{
Under scenario \ref{fig:direct}, to identify the local causal effects it suffices to control for the local value of the confounder.
However, there are four backdoor paths for $\iota_1(z)$, the interference effect of Unit 2's treatment on Unit 1's outcome:
\begin{enumerate}[label=(\roman*)]
\item $Z_2 \leftarrow U_2 \leftrightarrow U_1 \rightarrow Y_1$,
\item $Z_2 \leftarrow U_2 \leftrightarrow U_1 \rightarrow Z_1 \rightarrow Y_1$,
\item $Z_2 \leftrightarrow Z_1 \leftarrow U_1 \rightarrow Y_1$, and
\item $Z_2 \leftrightarrow Z_1 \rightarrow Y_1$.
\end{enumerate}
Paths (i) and (ii), and paths (iii) and (iv) exist because $\bm U$ and $\bm Z$ are inherently spatial, respectively. If they were {\it not} spatial and the two units were independent, the backdoor paths would not exist, and one could identify interference effects without any adjustments.
With spatial data, these same analyses would mis-attribute spatial statistical dependence to interference, and mis-identify the presence of spatial interference.
Instead, local and interference effects should be investigated simultaneously adjusting for the local value of the confounder. 
These are summarized in the following proposition.
}

\begin{proposition}[Identifiability of local and interference effects under direct spatial confounding] For variables with causal relationships depicted in \cref{fig:direct}, the following statements hold:
\begin{enumerate}[leftmargin=*,topsep=3pt,itemsep=0pt]
\item Local effects can be identified by controlling for local confounder.
\item If $\bm U, \bm Z$ are not spatial, interference effects can be identified without any adjustment.
\item If $\bm U, \bm Z$ are spatial, interference effects can be identified by controlling for the local value of the exposure and the confounder.
\end{enumerate}
\label{supp_prop:direct_identify}
\end{proposition}

\begin{proof}[Proof of \cref{supp_prop:direct_identify}]
The dependencies in this scenario are depicted in the following graph, where we include the underlying $U^u, Z^u$ that drive the covariate's and the exposure's spatial dependence:
\begin{figure}[H]
\centering
\begin{tikzpicture}
        \node (1) {$U_1$};
        \node [below left =0.25 and 0.25 of 1] (10) {$U^u$};
		\node[ right= of 1] (2) {$Z_1$};
        \node [below left =0.25 and 0.25 of 2] (20) {$Z^u$};
		\node[ right= of 2] (3) {$Y_1$};
		\node[below = of 1] (12) {$U_2$};
		\node[ right= of 12] (22) {$Z_2$};
		\node[ right= of 22] (32) {$Y_2$};
		\draw[->] (1) to (2); 
		\draw[->] (12) to (22); 
		\draw[->] (2) to (3); 
		\draw[->] (12) to [out=330,in=210] (32);
		\draw[->] (1) to [out=30,in=150] (3);
		\draw[->] (22) to (32);
		\draw[->] (10) to (1);
		\draw[->] (10) to (12); 
		\draw[->] (20) to (2);
		\draw[->] (20) to (22);
\end{tikzpicture}
\end{figure}
\noindent
We use the theory on identifiability based on graphical models. We investigate the presence of backdoor paths from the local or neighbourhood exposure on the outcome for local and interference effects, respectively. In this setting, interference effects are non-existent and $\iota_i(z) = 0$.

\begin{enumerate}[leftmargin=*]
\item 
Since this is a scenario without interference, potential outcomes can be denoted as $Y_i(z_i, z_j) = Y_i(z_i)$. Therefore, it suffices to study backdoor paths from $Z_i$ to $Y_i$. The only such backdoor path is $Z_i \leftarrow U_i \rightarrow Y_i$. Therefore, controlling for the local value of the confounder, $U_i$, blocks this path, and local effects for Unit $i$ are identified.

\item If $\bm U, \bm Z$ are not spatial, the causal graph is of the form:
\begin{figure}[H]
\centering
\begin{tikzpicture}
        \node (1) {$U_1$};
		\node[ right= of 1] (2) {$Z_1$};
		\node[ right= of 2] (3) {$Y_1$};
		\node[below =0.2 of 1] (12) {$U_2$};
		\node[ right= of 12] (22) {$Z_2$};
		\node[ right= of 22] (32) {$Y_2$};
		\draw[->] (1) to (2); 
		\draw[->] (12) to (22); 
		\draw[->] (2) to (3); 
		\draw[->] (12) to [out=330,in=210] (32);
		\draw[->] (1) to [out=30,in=150] (3);
		\draw[->] (22) to (32);
\end{tikzpicture}
\end{figure}
and clearly $Z_2$ and $Y_1$ are (unconditonally) independent. Therefore the interference effect on Unit 1 is identified as a contrast of Unit 1 outcomes between pairs that have $Z_2 = 1$ and pairs that have $Z_2 = 0$, since
\[
\iota_1(z) = 0 = \E[Y_1 \mid Z_2 = 1] - \E[Y_1 \mid Z_2 = 0].
\]

\item If $\bm U, \bm Z$ are spatial, $Z_2$ and $Y_1$ are not (unconditionally) independent, and therefore we expect that $\E[Y_1 \mid Z_2 = 1] - \E[Y_1 \mid Z_2 = 0] \neq 0$, which means that it does not identify $\iota_1(z)$.

Here, all backdoor paths from $\bm Z$ to $Y_1$ are blocked by $U_1$. Therefore, interference effects can be identified by controlling for the local value of the exposure and the confounder as
\[
\iota_1(z) = \E \left[ \E(Y_1 \mid Z_1 = z, U_1, Z_2 = 1) - \E(Y_1 \mid Z_1 = z, U_1, Z_2 = 0) \right],
\]
where the outer expectation is with respect to $U_1$.

\end{enumerate}

\end{proof}

Therefore, in the case of direct spatial confounding only in Scenario \ref{fig:direct}, an analysis that investigates local and interference effects simultaneously controlling for the local value of the confounder would be valid.

\subsection{Graph \ref{fig:general_spatial_conf}: Direct and indirect spatial confounding}

\cng{In the presence of indirect spatial confounding in Scenario \ref{fig:general_spatial_conf}, adjusting for the neighbor's exposure opens the path $Z_i \leftrightarrow Z_j \leftarrow U_j \rightarrow Y_i$ on which $Z_j$ is a collider, and breaks identifiability of the local effect. In this scenario, one needs to adjust for the local {\it and} the neighbor's confounder value in order to identify local and interference effects.
Since this path is only present when $\bm Z$ is spatial, this notion of confounding pertains solely to the setting with dependent data, and it is not met in settings with independent observations. These results are summarized in the following proposition.}

\begin{proposition}[Identifiability of local and interference effects under direct and indirect spatial confounding] For variables with causal relationships depicted in \cref{fig:general_spatial_conf}, the following statements hold:
\begin{enumerate}[leftmargin=*,topsep=3pt,itemsep=0pt]
\item Local and interference effects can be identified by controlling for the local and neighbourhood covariate value.
\item When the exposure is inherently spatial, controlling for the neighbourhood exposure and the local covariate does not suffice to identify local effects.
\end{enumerate}
\label{supp_prop:general_spatial_conf_identify}
\end{proposition}

\begin{proof}[Proof of \cref{supp_prop:general_spatial_conf_identify}]
We again focus on local and interference effects for Unit 1. Results for the effects on Unit 2 are identical.

\begin{enumerate}[leftmargin=*]
\item
The graph describing this setting is depicted below: 
\begin{figure}[H]
\centering
\begin{tikzpicture}
        \node (1) {$U_1$};
        \node [below left =0.25 and 0.25 of 1] (10) {$U^u$};
        \node [below left =0.25 and 0.25 of 2] (20) {$Z^u$};
		\node[ right= of 1] (2) {$Z_1$};
		\node[ right= of 2] (3) {$Y_1$};
		\node[below = of 1] (12) {$U_2$};
		\node[ right= of 12] (22) {$Z_2$};
		\node[ right= of 22] (32) {$Y_2$};
		\draw[->] (1) to (2); 
		\draw[->] (12) to (22); 
		\draw[->] (2) to (3); 
		\draw[->] (12) to [out=330,in=210] (32);
		\draw[->] (1) to [out=45,in=45,looseness=1.35] (32);
		\draw[->] (12) to [out=315,in=315,looseness=1.35] (3);
		\draw[->] (1) to [out=30,in=150] (3);
		\draw[->] (22) to (32);
		\draw[->] (10) to (1);
		\draw[->] (10) to (12); 
		\draw[->] (20) to (2);
		\draw[->] (20) to (22);
\end{tikzpicture}
\end{figure}

The vector $\bm U$ blocks all backdoor paths from the vector $\bm Z$ to the outcome unit 1, $Y_1$, and therefore it holds that $\bm Z \indep Y_1(z_1, z_2) \mid \bm U$. This implies that
\[
\E[Y_1(z_1, z_2)] = \E \left[ \E(Y_1 \mid Z_1 = z_1, Z_2 = z_2, \bm U) \right],
\]
where the outer expectation is with respect to $\bm U$. Therefore, local and interference effects are identified while controlling for the local and neighbourhood covariate.

\item This is a scenario where interference is absent, and we can denote potential outcomes for Unit 1 based only on the treatment of the unit itself, $Y_1(z_1, z_2) = Y_1(z_1)$.
The neighbourhood exposure, $Z_2$ is a collider on the backdoor path $Z_1 \leftarrow Z^u \rightarrow Z_2 \leftarrow U_2 \rightarrow Y_1$. Therefore, When $Z_2$ is conditioned on, this backdoor path is open. This implies that $Z_1 \not\indep Y_1(z_1) \mid Z_2, U_1$ and local effects are not identified.

\end{enumerate}

\end{proof}

\subsection{Graph \ref{fig:interference}: Spatial interference}

The definitions of the local and interference effects $\lambda_i(\pi)$ and $\iota_i(\pi)$ that appear in this subsection are given in Supplement \ref{subsec:alternative_definitions_block}.
These local effects are different than the ones in \cref{sec:pairs}, though they are still interpretable.

Propositions \ref{supp_prop:interf_nospat_identif_local} and \ref{supp_prop:interf_nospat_identif_interf} refer to the case of \cref{fig:interference} with $\bm Z$ not spatial. This case corresponds to the graph:
\begin{figure}[H]
\centering
\begin{tikzpicture}
        \node (1) {$U_1$};
        \node [below left =0.25 and 0.25 of 1] (10) {$U^u$};
		\node[ right= of 1] (2) {$Z_1$};
		\node[ right= of 2] (3) {$Y_1$};
		\node[below = of 1] (12) {$U_2$};
		\node[ right= of 12] (22) {$Z_2$};
		\node[ right= of 22] (32) {$Y_2$};
		\draw[->] (2) to (3); 
		\draw[->] (2) to (32);
		\draw[->] (22) to (3);
		\draw[->] (22) to (32);
		\draw[->] (10) to (1);
		\draw[->] (10) to (12); 
\end{tikzpicture}
\end{figure}
\cng{In this case, the local effect for Unit $i$ can be identified by contrasting the average outcome for this unit among pairs with $Z_i = 1$ and pairs with $Z_i = 0$, without any adjustment. This is summarized in the following proposition.}

\begin{proposition}
For variables with causal relationships depicted in \cref{fig:interference}, if $\bm Z$ is not spatial, then 
$\lambda_1(\pi)$ for $\pi = P(Z_2 = 1)$ is identifiable based on the difference of averages of Unit 1 outcomes for pairs with $Z_1 = 1$ and pairs with $Z_1 = 0$, i.e.,
\[
\lambda_1(\pi) = \E(Y_1 \mid Z_1 = 1) - \E(Y_1 \mid Z_1 = 0)
\]
The case for $\lambda_2(\pi')$, for $\pi' = P(Z_1 = 1)$ is symmetric.
\label{supp_prop:interf_nospat_identif_local}
\end{proposition}

\begin{proof}[Proof of \cref{supp_prop:interf_nospat_identif_local}]
Note that 
\begin{align*}
\lambda_1(\pi)
&= \E [ \pi Y_1(1, 1) - \pi Y_1(0, 1) + (1 - \pi) Y_1(1, 0) - (1 - \pi) Y_1(0, 0)] \\
&= \E [ \pi Y_1(1, 1) + (1 - \pi) Y_1(1, 0)] - \E [ \pi Y_1(0, 1) + (1 - \pi) Y_1(0, 0)],
\end{align*}
so it suffices to identify $\E[ \pi Y_1(z, 1) + (1 - \pi) Y_1(z, 0)]$, for $z = 0, 1$.
The proof is similar to that on Page 566 of \cite{ogburn2014causal}.
\begin{align*}
\E & \left[ \pi Y_1(z, 1) + (1 - \pi) Y_1(z, 0) \right] \\
&= \pi  \E \left[ Y_1(z, 1) \right] + (1 - \pi) \E \left[ Y_1(z, 0) \right] \\
&= \pi  \E \left[ Y_1(z, 1) \mid Z_1 = z, Z_2 = 1 \right] +
(1 - \pi) \E \left[ Y_1(z, 0) \mid Z_1 = z, Z_2 = 0 \right]
\tag{Ignorability} \\
&= \pi  \E \left[ Y_1 \mid Z_1 = z, Z_2 = 1 \right] +
(1 - \pi) \E \left[ Y_1 \mid Z_1 = z, Z_2 = 0 \right]
\tag{Consistency of potential outcomes} \\
&= P(Z_2 = 1)  \E \left[ Y_1 \mid Z_1 = z, Z_2 = 1 \right] +
P(Z_2 = 0) \E \left[ Y_1 \mid Z_1 = z, Z_2 = 0 \right] \\
&= P(Z_2 = 1 \mid Z_1 = z)  \E \left[ Y_1 \mid Z_1 = z, Z_2 = 1 \right] +
P(Z_2 = 0 \mid Z_1 = z) \E \left[ Y_1 \mid Z_1 = z, Z_2 = 0 \right]
\tag{Treatment values are independent} \\
&= \E \left[ Y_1 \mid Z_1 = z \right].
\end{align*}
\end{proof}

\begin{proposition}
For the variables' with causal dependence depicted in \cref{fig:interference}, when the exposure $\bm Z$ is spatial, the difference of averages of Unit 1 outcomes for pairs with $Z_1 = 1$ and pairs with $Z_1 = 0$ does {\it not} identify an interpretable local effect for Unit 1. Interpretable local causal effects for Unit 1 can be identified by adjusting for the neighbour's exposure.
\label{supp_prop:interf_identif_local}
\end{proposition}

\begin{proof}[Proof of \cref{supp_prop:interf_identif_local}]
Following the steps of the proof of \cref{supp_prop:interf_nospat_identif_local} backwards, we have that the average outcome of Unit 1 among pairs with $Z_1 = z$ estimates the quantity
\begin{align*}
\E(Y_1 \mid Z_1 = z)
&= 
P(Z_2 = 1 \mid Z_1 = z)  \E \left[ Y_1(z, 1) \right] +
P(Z_2 = 0 \mid Z_1 = z) \E \left[ Y_1(z, 0) \right]
\end{align*}
where we used consistency of potential outcomes, and that $\bm Z \indep Y_1(z_1, z_2)$ under the causal dependence depicted in the graph \ref{fig:interference}.
This quantity is the average outcome when Unit 1's treatment is set to $z$ and Unit 2's treatment is equal to 1 with probability $P(Z_2 = 1 \mid Z_1 = z)$ and 0 otherwise.

Then, the contrast of average Unit 1 outcomes among pairs with $Z_1 = 1$ and $Z_1 = 0$ estimates the peculiar contrast
\begin{align*}
\E(Y_1 \mid & Z_1 = 1) - \E(Y_1 \mid Z_1 = 0)  = \\ 
&= \left\{ P(Z_2 = 1 \mid Z_1 = 1)  \E \left[ Y_1(1, 1) \right] +
P(Z_2 = 0 \mid Z_1 = 1) \E \left[ Y_1(1, 0) \right] \right\} \\
& \hspace{20pt} - \left\{ P(Z_2 = 1 \mid Z_1 = 0)  \E \left[ Y_1(0, 1) \right] +
P(Z_2 = 0 \mid Z_1 = 0) \E \left[ Y_1(0, 0) \right] \right\},
\end{align*}
where not only Unit 1's treatment changes from 1 to 0, but also the probability of treatment for Unit 2 changes. Therefore, it is unclear whether this estimated quantity has a causal interpretation. At the least, it does not isolate the effect of a change in $Z_1$ and cannot be interpreted as a local causal effect.

Since the ignorability of the (vector) exposure holds unconditionally, $\bm Z \indep Y_1(z_1, z_2)$, we have that Unit 1's local effects defined in \cref{sec:pairs} can be identified when $Z_2$ is adjusted.

\end{proof}

\begin{remark}
When $\bm Z$ is not spatial, $P(Z_2 = 1 \mid Z_1 = z) = P(Z_2 = 1)$, and \cref{supp_prop:interf_identif_local} reverts back to \cref{supp_prop:interf_nospat_identif_local}.
The results in Propositions \ref{supp_prop:interf_nospat_identif_local} and \ref{supp_prop:interf_identif_local} combined establish explicitly how, in this very simple case without any confounding, an identification strategy for local effects can be invalidated by statistical dependence in the exposure variable.
\end{remark}

\begin{proposition}
For variables with causal relationships depicted in \cref{fig:interference}, if $\bm Z$ is not spatial, then $\iota_1(\pi)$ for $\pi = P(Z_1 = 1)$ is identifiable based on the difference of averages of Unit 1 outcomes for pairs with $Z_2 = 1$ and pairs with $Z_2 = 0$. The case for $\iota_2(\pi')$, for $\pi' = P(Z_2 = 1)$ is symmetric.
\label{supp_prop:interf_nospat_identif_interf}
\end{proposition}
\noindent
The proof of \cref{supp_prop:interf_nospat_identif_interf} is identical to the proof of \cref{supp_prop:interf_nospat_identif_local}, hence it is omitted.

\subsection{Graph \ref{fig:predictor_interference}: Interference with spatial predictor of the exposure}

\cng{For the local effect, there is one additional open backdoor path, $Z_1 \leftarrow U_1 \leftrightarrow U_2 \rightarrow Z_2 \rightarrow Y_1$, compared to the setting in \cref{fig:interference}.
A researcher interested in estimating local effects might ignore the potential for interference. Their analysis might adjust for the local variable $U$, or not. The two analyses will return different values for the local effect estimate, none of which is causally interpretable. This is summarized in the following result.}

\begin{proposition}
For the variables with causal relationships depicted in \cref{fig:predictor_interference}, the difference of averages of Unit 1 outcomes for pairs with $Z_1 = 1$ and pairs with $Z_1 = 0$ conditionally on $U_1$ and unconditionally identify different quantities, none of which is an interpretable causal effect.
\label{supp_prop:predictor_interf_identif}
\end{proposition}

\begin{proof}[Proof of \cref{supp_prop:predictor_interf_identif}]
We consider again two estimators, one of which is unconditional and the other is conditional on $U_1$. These estimators are of the form
\[
\widehat \tau = \E(Y_1 \mid Z_1 = 1) - \E(Y_1 \mid Z_1 = 0),
\]
and
\[
\widehat \tau^{\mid U} = \E[ \E(Y_1 \mid Z_1 = 1, U_1) - \E(Y_1 \mid Z_1 = 0, U_1)],
\]
respectively, where for the second estimator the outer expectation is with respect to the distribution of $U_1$ across pairs.

For the unconditional estimator, $\widehat \tau$, since $\bm Z \indep Y_1(z_1, z_2)$ under the causal dependencies represented in \ref{fig:predictor_interference}, we can again derive that
\begin{align*}
\widehat \tau &= \E(Y_1 \mid Z_1 = 1) - \E(Y_1 \mid Z_1 = 0)  \\ 
&= \left\{ P(Z_2 = 1 \mid Z_1 = 1)  \E \left[ Y_1(1, 1) \right] +
P(Z_2 = 0 \mid Z_1 = 1) \E \left[ Y_1(1, 0) \right] \right\} \\
& \hspace{20pt} - \left\{ P(Z_2 = 1 \mid Z_1 = 0)  \E \left[ Y_1(0, 1) \right] +
P(Z_2 = 0 \mid Z_1 = 0) \E \left[ Y_1(0, 0) \right] \right\},
\end{align*}
identically to the derivations in the proof of \cref{supp_prop:interf_identif_local}.

Furthermore, since $\bm Z \indep Y_1(z_1, z_2) \mid U_1$ also holds, we can similarly derive that
\begin{align*}
\E[ \E(Y_1 \mid & Z_1 = z, U_1) ] = \\
&= \E\{ P(Z_2 = 1 \mid Z_1 = z, U_1) \ \E[ Y_1(z, 1) \mid U_1] + \\
& \hspace{40pt} + P(Z_2 = 0 \mid Z_1 = z, U_1) \ \E[Y_1(z, 0) \mid U_1] \} \\
&= \E[ P(Z_2 = 1 \mid Z_1 = z, U_1)] \ \E[ Y_1(z, 1) ] + \\
& \hspace{40pt} + \E [P(Z_2 = 0 \mid Z_1 = z, U_1)] \ \E[Y_1(z, 0)],
\end{align*}
where, in the last equation, we have used the fact that $U_1$ does not predict $Y_1$ except through $Z$. Therefore the (conditional) contrast of average outcomes, $\widehat \tau^{\mid U}$ estimates
\begin{align*}
\widehat \tau^{\mid U} 
&= \E[ \E( Y_1 \mid Z_1 = 1, U_1) ] - \E[ \E(Y_1 \mid Z_1 = 0, U_1) ] \\
&= \left\{ \E[ P(Z_2 = 1 \mid Z_1 = 1, U_1)] \ \E[ Y_1(1, 1) ] + \E [P(Z_2 = 0 \mid Z_1 = 1, U_1)] \ \E[Y_1(1, 0)] \right\} - \\
& \hspace{20pt} - 
\left\{
\E[ P(Z_2 = 1 \mid Z_1 = 0, U_1)] \ \E[ Y_1(0, 1) ] + \E [P(Z_2 = 0 \mid Z_1 = 0, U_1)] \ \E[Y_1(0, 0)]
\right\}.
\end{align*}

In general, it holds that
\[
P(Z_2 = 1 \mid Z_1 = z) \neq \E[ P(Z_2 = 1 \mid Z_1 = z, U_1)],
\]
since the outer expectation on the right-hand side is with respect to the (marginal) distribution of $U_1$, rather than the distribution of $U_1$ given $Z_1 = z$. Since these two quantities are different, in general it holds that $\widehat \tau \neq \widehat \tau^{\mid U}$. Therefore, the conditional and unconditional estimators estimate different quantities.

The proof that none of these estimate an interpretable causal effect is identical to the one in the proof of \cref{supp_prop:interf_identif_local}, and it relates to the fact that these contrast consider a distribution for $Z_2$ that changes based on the value of $Z_1$.
\end{proof}

\begin{remark}
From the proof of \cref{supp_prop:predictor_interf_identif}, we can identify some interesting cases where the conditional and unconditional estimators estimate the same quantity, or they return interpretable local causal effects:
\begin{itemize}
\item When $\bm Z$ is not spatial, we have that
\[ \E[ P(Z_2 = 1 \mid Z_1 = z, U_1)] = \E[ P(Z_2 = 1 \mid U_1)] = P(Z_2 = 1), \]
and therefore the conditional estimator, $\widehat \tau^{\mid U}$, estimates the interpretable local causal effect $\lambda_1(P(Z_2 = 1))$.
\item When $\bm Z$ is not spatial, but the spatial predictor $\bm U$ is present,
\( P(Z_2 = 1 \mid Z_1 = z) \neq P(Z_2 = 1), \)
and the unconditional estimator, $\widehat \tau$, still fails to estimate an interpretable causal effect.
\item When $\bm U$ is not spatial, $ \E[ P(Z_2 = 1 \mid Z_1 = z, U_1)] = P(Z_2 = 1 \mid Z_1 = z$, and the two estimators estimate the same quantity.
\end{itemize}
\end{remark}

\subsection{Graph \ref{fig:direct_interference}: Direct spatial confounding and interference}

\cng{\cref{fig:direct_interference} shows a setting with {direct spatial confounding and interference}, which combines the scenarios in Figures \ref{fig:direct} and \ref{fig:interference}. When $\bm Z$ is inherently spatial, it is necessary to condition on the local value of the confounder {\it and} the neighbor's exposure to identify local effects. In contrast, if $\bm Z$ is not inherently spatial, interpretable local effects can be identified conditioning only on the local value of the confounder. Similar conclusions can be drawn about the identifiability of interference effects. We formalize these below. These results illustrate once more that the exposure's inherent spatial structure can lead to misleading conclusions if not properly accommodated.}

\begin{proposition}
When the variables' causal relationships are depicted in \cref{fig:direct_interference}, 
\begin{enumerate}[leftmargin=*,topsep=3pt,itemsep=0pt]
\item Controlling for the local confounder and neighbourhood exposure suffices to identify local effects.
\item Failing to adjust for the local confounder or the neighbourhood exposure returns estimates that cannot be interpreted as causal effects.
\end{enumerate}
\label{supp_prop:direct_interf_ident_local}
\end{proposition}

\begin{proof}[Proof of \cref{supp_prop:direct_interf_ident_local}]
$ $
\begin{enumerate}[leftmargin=*]
\item The local confounder blocks all backdoor paths from $\bm Z$ into $Y_1$. As a result, all potential outcomes of the form $\E[Y_1(z_1, z_2)]$, and hence local (and interference) effects, can be identified.
\item Without conditioning on $U_1$, there is a backdoor path from the vector $\bm Z$ to the outcome $Y_1$, and we have that $\bm Z \not\indep Y_1(z_1, z_2)$. Therefore, local (or interference) effects cannot be identified.

Without conditioning on $Z_2$, this setting is almost identical to the one discussed in \cref{supp_prop:predictor_interf_identif} where $U_1$ is conditioned on or not, and estimated quantities cannot be interpreted as causal effects.
\end{enumerate}
\end{proof}

Next, we focus on identification of interpretable causal effects when the exposure is not spatial. \cref{supp_prop:direct_interf_nospat_identif} shows that, when $\bm Z$ is not spatial, one would need to adjust only for the local confounder in order to acquire interpretable local causal effects, even if interference is present. First, we define these {\it new} type of interpretable effects.

We define conditional average local effects. First, let
\[
\lambda_i(z; u_i) = \E[ Y_i(z_i = 1, z_j = z) - Y_i(z_i = 0, z_j = 0) \mid U_i = u_i]
\]
denote the expected change in unit $i$'s outcome for changes in its own treatment when the neighbour's treatment is set to $z$, among clusters with $U_i = u_i$. This is the equivalent to the local effects defined in \cref{eq:local_effect}, where we now also condition on the unit's covariate values.

We also consider expected conditional average local effects, where we average over a distribution for the neighbour's treatment. Specifically,
let $\pi(u_i) = P(Z_j = 1 \mid U_i = u_i)$. We define
\[
\lambda_i(\pi(u_i) ; u_i) = \pi(u_i) \lambda_i(1; u_i) + (1 - \pi(u_i)) \lambda_i(0; u_i),
\]
representing the average change in unit $i$'s outcome among clusters with $U_i = u_i$ for changes in unit $i$'s own treatment, and when the treatment of its neighbour is distributed according to $\pi(\cdot)$. These effects are the conditional equivalent to effects $\lambda_i(\pi)$ in Supplement \ref{subsec:alternative_definitions_block}.

\begin{proposition}
When the variables' causal relationships can be described in the graph of \cref{fig:direct_interference}, if $\bm Z$ is not spatial, it holds that
\[
\E_{U_1} [\lambda_1(\pi(U_1); U_1)] = \E_{U_1} [ \E \left( Y_1 \mid Z_1 = 1, U_1 \right) - \E \left( Y_1 \mid Z_1 = 0, U_1 \right) ]
\]
The case for $\E_{U_2}[\lambda_2(\pi'(U_2); U_2)]$, for $\pi'(u_2) = P(Z_1 = 1 \mid U_2 = u_2)$ is symmetric.
\label{supp_prop:direct_interf_nospat_identif}
\end{proposition}

\begin{proof}[Proof of \cref{supp_prop:direct_interf_nospat_identif}]
We follow steps that are similar to those in the proof of  \cref{supp_prop:interf_nospat_identif_local}. However, here, we have to account for the fact that we average over a distribution of $\pi(U_1)$.
\begin{align*}
\E_{U_1} & \left[ \E \left( Y_1 \mid Z_1 = 1, U_1 \right) - \E \left( Y_1 \mid Z_1 = 0, U_1 \right) \right] = \\
&= \E_{U_1} \big\{ \big[
\E \left( Y_1 \mid Z_1 = 1, Z_2 = 1, U_1 \right) P(Z_2 = 1 \mid Z_1 = 1, U_1) + \\
& \hspace{100pt}
\E \left( Y_1 \mid Z_1 = 1, Z_2 = 0, U_1 \right) P(Z_2 = 0 \mid Z_1 = 1, U_1)
\big]  -  \\
& \hspace{60pt}
\big[
\E \left( Y_1 \mid Z_1 = 0, Z_2 = 1, U_1 \right) P(Z_2 = 1 \mid Z_1 = 0, U_1) + \\
& \hspace{160pt}
\E \left( Y_1 \mid Z_1 = 0, Z_2 = 0, U_1 \right) P(Z_2 = 0 \mid Z_1 = 0, U_1)
\big] \big\} \\
&= \E_{U_1} \big\{ \big[
\E \left( Y_1(1, 1) \mid Z_1 = 1, Z_2 = 1, U_1 \right) P(Z_2 = 1 \mid Z_1 = 1, U_1) + \\
& \hspace{100pt}
\E \left( Y_1(1, 0) \mid Z_1 = 1, Z_2 = 0, U_1 \right) P(Z_2 = 0 \mid Z_1 = 1, U_1)
\big]  -  \\
& \hspace{60pt}
\big[
\E \left( Y_1(0, 1) \mid Z_1 = 0, Z_2 = 1, U_1 \right) P(Z_2 = 1 \mid Z_1 = 0, U_1) + \\
& \hspace{160pt}
\E \left( Y_1(0, 0) \mid Z_1 = 0, Z_2 = 0, U_1 \right) P(Z_2 = 0 \mid Z_1 = 0, U_1)
\big] \big\}
\tag{Consistency of potential outcomes} \\
&= \E_{U_1} \big\{ \big[
\E \left( Y_1(1, 1) \mid U_1 \right) P(Z_2 = 1 \mid Z_1 = 1, U_1) + 
\E \left( Y_1(1, 0) \mid U_1 \right) P(Z_2 = 0 \mid Z_1 = 1, U_1)
\big]  -  \\
& \hspace{60pt}
\big[
\E \left( Y_1(0, 1) \mid U_1 \right) P(Z_2 = 1 \mid Z_1 = 0, U_1) +
\E \left( Y_1(0, 0) \mid U_1 \right) P(Z_2 = 0 \mid Z_1 = 0, U_1)
\big] \big\}
\tag{Ignorability $Z_1, Z_2 \indep Y_1(z_1, z_2) \mid U_1$ implied by the graph \ref{fig:direct_interference}} \\
&= \E_{U_1} \big\{ \big[
\E \left( Y_1(1, 1) \mid U_1 \right) P(Z_2 = 1 \mid U_1) + 
\E \left( Y_1(1, 0) \mid U_1 \right) P(Z_2 = 0 \mid U_1)
\big]  -  \\
& \hspace{60pt}
\big[
\E \left( Y_1(0, 1) \mid U_1 \right) P(Z_2 = 1 \mid U_1) +
\E \left( Y_1(0, 0) \mid U_1 \right) P(Z_2 = 0 \mid U_1)
\big] \big\}
\tag{$Z_1 \indep Z_2 \mid U_1$ according to the graph \ref{fig:direct_interference}} \\
&= \E_{U_1} \big\{ \big[
\E \left( Y_1(1, 1) \mid U_1 \right) \pi(U_1) + 
\E \left( Y_1(1, 0) \mid U_1 \right) (1 - \pi(U_1))
\big]  -  \\
& \hspace{60pt}
\big[
\E \left( Y_1(0, 1) \mid U_1 \right) \pi(U_1) +
\E \left( Y_1(0, 0) \mid U_1 \right) (1 - \pi(U_1))
\big] \big\} \\
&= \E_{U_1} \big[ \lambda_1(\pi(U_1); U_1) \big]
\end{align*}
\end{proof}

We can define and identify interference effects similarly, without adjusting for the local exposure value.

\section{Illustrating the bias due to statistical and causal dependence in paired and network data}
\label{supp_sec:motivating_sims}

\subsection{Motivating simulation study with paired data}
\label{subsec:illustrate_bias_pairs}

To illustrate the points made in \cref{subsec:pairs_graphs} and show how interference and spatial confounding can manifest as each other and affect estimation of local and interference effects, we perform a small simulation study. We simulate pairs of $\bm U$ from a bivariate Normal distribution with mean 0, and covariance matrix $\Sigma_U = \left( \begin{smallmatrix} 1 & \phi_U \\ \phi_U & 1 \end{smallmatrix} \right)$. We also simulate a bivariate normal error term $\bm \epsilon_Z = (\epsilon_{Z,1}, \epsilon_{Z, 2})$ with marginal variances equal to 1 and correlation parameter $\phi_Z$. The binary exposure is generated from a Bernoulli distribution with a logistic link function and linear predictor $\beta_{UZ} U_i + \epsilon_{Z,i}$. Higher values of $\phi_U, \phi_Z$ correspond to stronger inherent spatial dependence for $\bm U$ and $\bm Z$.
The outcome is generated independently across locations from a normal distribution with mean $\beta_Z Z_i + \beta_{\bar Z} \overline Z_i + \beta_U U_i + \beta_{\bar U} \overline U_i$ and variance 1, where $\overline Z_i$ and $\overline U_i$ represent the value of the exposure and the covariate for the neighbour of location $i$, respectively.
Under this model, $\beta_Z$ and $\beta_{\bar{Z}}$ correspond to the local and interference effects, respectively.
We consider the six different scenarios presented in \cref{fig:graphs} by setting different parameters to zero. 
We simulate 300 data sets of 200 pairs each, and fit ordinary least squares (OLS) using different sets of predictor variables.
The data generating model and hyperparameters for each of these scenarios are listed in Table \ref{tab:pairdata}, along with the bias for the OLS estimators of $\beta_Z$ and $\beta_{\bar{Z}}$.

\begin{table}[p]
\spacingset{1}
\small
\centering
\caption{Motivating Simulation Study with Paired Data. For the graphs of \cref{fig:graphs}, we illustrate the induced biases in estimating local and interference causal effects due to spatial dependencies.
In these simulations, the parameters that drive the data generative mechanism are $\phi_U, \phi_Z, \beta_{UZ}, \beta_Z, \beta_{\bar Z}, \beta_U, \beta_{\bar U}.$ The different scenarios of \cref{fig:graphs} correspond to different set of parameters fixed at 0, shown below. Unless otherwise noted, the parameters are fixed at $\phi_U = 0.7$, $\phi_Z = 0.5, \beta_{UZ} = 1, \beta_Z = 1, \beta_{\bar Z} = 0.8, \beta_U = 1, \beta_{\bar U} = 0.5$. We generate 300 data sets of 200 pairs each. We regress the outcome on a different set of variables (columns), and report the bias of the OLS estimator for the local effect estimator, $\beta_Z$, and the interference effect estimator, $\beta_{\bar Z}$, when $\overline Z$ is included in the conditioning set. Values are rounded to the third decimal point, and those in {\bf bold} are discussed in the main text.
}
\spacingset{1}
    \resizebox{0.88\textwidth}{!}{%
    \begin{tabular}{*{10}{c}}
        \hline
        & & & & &  & \\[-12pt]
        & &  \multicolumn{8}{c}{Conditioning set} \\
        \multirow{2}{*}{\shortstack[c]{True \\[5pt] Model}} & \multirow{3}{*}{\shortstack[c]{Alternative \\[5pt] spatial \\[5pt] parameters}} & \multicolumn{8}{c}{\& estimated parameter} \\[2pt]
        \cmidrule{3-10} 
        & & \\[-15pt]
        & &  $(Z) $ & $(Z, U)$  & \multicolumn{2}{c}{  $(Z, \bar{Z})$} & \multicolumn{2}{c}{  $(Z, \bar{Z}, U)$} & \multicolumn{2}{c}{  $(Z, \bar{Z}, U, \bar{U})$}\\
        \cmidrule(lr){3-3} \cmidrule(lr){4-4} \cmidrule(lr){5-6} \cmidrule(lr) {7-8} \cmidrule(lr) {9-10}
        & & $\beta_Z$ & $\beta_Z$ & $\beta_Z$ & $\beta_{\bar{Z}}$ & $\beta_Z$ & $\beta_{\bar{Z}}$ & $\beta_Z$ & $\beta_{\bar{Z}}$\\
        & & & & & & \\[-8pt]
        \hline \\[-8pt]
\multirow{2}{*}{\ref{fig:direct}} & & \multicolumn{8}{c}{$\beta_{\bar Z} = 0$ and $\beta_{\bar U}=0$} \\[2pt]
        \cmidrule{5-8}
        & & 0.726 & -0.003 & 0.660 & {\bf 0.406} & -0.003 & {\bf -0.002} & -0.002 & 0.000  \\
        \midrule \\[-8pt]
\multirow{2}{*}{\ref{fig:general_spatial_conf}} & & 
        \multicolumn{8}{c}{$\beta_{\bar Z} = 0$}
        \\[2pt]
        \cmidrule{5-8}
         & 
         & 0.983 & 0.002 & 0.863 & 0.737 & -0.013 & {\bf 0.198} & -0.002 & {\bf 0.000}  \\ 
         \midrule \\[-8pt]
\multirow{4}{*}{\ref{fig:interference}} & & 
\multicolumn{8}{c}{$\beta_{UZ} = 0$ and $\beta_U = \beta_{\bar U} = 0$}
        \\[2pt]
        \cmidrule{5-8}
        & $\phi_z = 0.7$ & {\bf 0.152} & 0.087 & 0.001 & -0.002 & -0.001 & -0.003 & 0.000 & -0.002  \\
         & $\phi_z = 0.5$ & {\bf 0.129} & 0.060 & -0.001 & -0.001 & -0.003 & -0.002 & -0.002 & 0.000 \\
        & $\phi_z = 0.3$ & {\bf 0.105} & 0.032 & -0.002 & 0.001 & -0.003 & 0.000 & -0.003 & 0.001 \\
        \midrule \\[-8pt]
\multirow{4}{*}{\ref{fig:predictor_interference}} & 
& \multicolumn{8}{c}{$\beta_U = 0$ and $\beta_{\bar U} = 0$} \\[2pt]
\cmidrule{5-8}
        & $\beta_{UZ} = 1.5$ & {\bf 0.173} & {\bf 0.052} & 0.000 & 0.001 & -0.002 & 0.000 & -0.002 & 0.003 \\
        & $\beta_{UZ} = 1\phantom{.a}$ &  0.129 & 0.060 & -0.001 & -0.001 & -0.003 & -0.002 & -0.002 & 0.000  \\
        & $\beta_{UZ} = 0.5$ & 0.083 & 0.061 & -0.004 & -0.003 & -0.006 & -0.004 & -0.005 & -0.003 \\
\midrule \\[-8pt]
\multirow{3}{*}{\ref{fig:direct_interference}} & & 
\multicolumn{8}{c}{$\beta_{\bar U} = 0$} \\[2pt]
        \cmidrule{5-8}
& $\phi_Z = 0.5$ & 0.856 & {\bf 0.060} & 0.660 & 0.406 & -0.003 & -0.002 & -0.002 & 0.000 \\ 
& $\phi_Z = 0\phantom{.5}$ & 0.800 & {\bf -0.001} & 0.684 & 0.445 & 0.000 & -0.003 & 0.000 & -0.001 \\
\midrule \\[-8pt]
\ref{fig:general_interference}
         & & 1.113 & 0.064 & 0.863 & 0.737 & -0.013 & 0.198 & -0.002 & 0.000 \\
        & & & & & &  \\[-8pt]
          \hline
        \end{tabular}
     }%
\label{tab:pairdata}
\end{table}

In the presence of only direct spatial confounding (Scenario \ref{fig:direct}), we see that failing to adjust for the local spatial confounder returns biased interference effect estimates ($\beta_{\bar Z}$ in the model with $Z, \overline Z$ in \cref{tab:pairdata}). Therefore, in the presence of inherently spatial data, adjusting for spatial confounders is crucial for learning interference effects, even if spatial confounding is direct only. When spatial confounding is both direct and indirect (Scenario \ref{fig:general_spatial_conf}), adjusting only for the local spatial confounder and exposure values can still return misleading interference effects ($\beta_{\bar Z}$ in the model with $Z, \overline Z, U$), and it is necessary to also account for the neighbour's covariate value. In the presence of interference (Scenario \ref{fig:interference}) and when the exposure is inherently spatial, the local effect estimator is biased when the neighbour's exposure value is not conditioned on, and the bias is larger for stronger spatial dependence. Instead, local and interference effects can be unbiasedly estimated when they are considered simultaneously ($\beta_Z, \beta_{\bar Z}$ in the model with $Z, \overline Z$).
In Scenario \ref{fig:predictor_interference}, the local effect of the exposure for unit $i$ is biased regardless of whether $U_i$ is adjusted for or not. At the same time, the estimates when $U_i$ is included in the model or not are substantially different, which could be interpreted as $U_i$ confounding the local effect. Therefore, in this scenario, the inherent spatial structure in the confounders and exposure could lead to  interference being mistakenly interpreted as spatial confounding.
In Scenario \ref{fig:direct_interference}, we see that when the exposure is not inherently spatial ($\phi_Z = 0$), we can learn local effects without adjusting for the neighbour's exposure. However, this estimator is biased when the exposure has an inherent spatial structure, illustrating practically that spatial dependencies can hinder some analyses invalid if not properly taken into account.
Of course, when all the possible dependencies are present in Scenario \ref{fig:general_interference}, one would need to condition on local and neighbourhood covariates to properly estimate local and interference effects. The estimator that account for all of local and neighbourhood exposure and confounding values returns unbiased effect estimates across all scenarios.

\subsection{
An example of inherent spatial structure in network data}

We provide an example of how the variables' inherent spatial structure might occur in a spatial setting. This is merely an illustration, and it is not required in our work. We return to viewing the inherent spatial structure in $\bm U$ as driven from an underlying covariate $U^u$ as in \cref{fig:dag_underlying}. For $U^u = (U_1^u, U_2^u, \dots, U_n^u)$ vector of independent random variables, set $U_i = \sum_{j = 1}^n w_{ij} U_j^u + \epsilon_{i}$, for $w_{ij}$ not all zero and $\epsilon_i$ independent errors. Then the elements of $\bm U$ that share elements of $U^u$ are statistically dependent. If the weights $w_{ij}$ are based on the spatial proximity of $i$ and $j$, this dependence structure will be {\it spatially} driven. We can similarly conceive $Z^u$ and $\bm Z$.

\subsection{Motivating simulation study in a setting with one spatial network of observations}
\label{supp_sec:motivating_one_network}

Under the different scenarios of \cref{fig:graphs}, the ignorability \cref{ass:network_ignorability} might also hold conditional on $\widetilde C$ only, conditional on $\widetilde C$ and $U$, or might only hold conditional on all of $\widetilde C, U$ and $\overline U$. Here, we investigate the influence of spatial dependencies in learning local and interference effects from a single interconnected network of spatial data.

We consider a graph with $n$ nodes. We assume that this graph is a line graph, in that the first and last nodes are connected only to the second and second to last, respectively, and node $i$ is connected to nodes $i - 1$ and $i + 1$ for $i = 2, 3, \dots, n-1$. This implies the following adjacency and degree matrices:
\[
A = \begin{pmatrix}
0 & 1 & 0 & 0 & \dots & 0 & 0 & 0 \\
1 & 0 & 1 & 0 & \dots & 0 & 0 & 0 \\
0 & 1 & 0 & 1 & \dots & 0 & 0 & 0 \\
& & \vdots & & \dots & & \vdots &  \\
0 & 0 & 0 & 0 & \dots & 1 & 0 & 1 \\
0 & 0 & 0 & 0 & \dots & 0 & 1 & 0 
\end{pmatrix} \quad \text{and} \quad
D = \begin{pmatrix}
1 & 0 & 0 & \dots & 0 & 0 \\
0 & 2 & 0 & \dots & 0 & 0 \\
0 & 0 & 2 & \dots & 0 & 0 \\
& & \vdots & \dots & \vdots & \\
0 & 0 & 0 & \dots & 2 & 0 \\
0 & 0 & 0 & \dots & 0 & 1
\end{pmatrix}.
\]
We generate $\bm U = (U_1, U_2, \dots, U_n)$ and $\bm Z = (Z_1, Z_2, \dots, Z_n)$ simultaneously from a multivariate normal distribution as follows
\[
\begin{pmatrix} \bm U \\ \bm Z \end{pmatrix} \sim N_{2n} \left( \bm 0_{2n},
\begin{pmatrix} G & Q \\ Q & H \end{pmatrix}^{-1}
\right),
\]
where $\bm 0_{2n}$ is a vector of length $2n$ of all $0$s. We specify $G$ and $H$ according to a conditional autoregressive distribution as $G = \tau_U^2 (D - \phi_U A)$ and $H = \tau_Z ^2 (D - \phi_Z A)$. Then, $Q$ is specified to be diagonal with elements $Q_{ii} = - \rho \sqrt{G_{ii} H_{ii}}.$ Note that different values of $\phi$ for the same value of $\tau$ lead to different marginal variances for the entries of $\bm U$ and $\bm Z$.

We exclude measured covariates for simplicity. Once $\bm U$ and $\bm Z$ are generated, the outcome is generated according to the model \cref{eq:linear_sem}, where $\epsilon \sim N(0, 1)$ independent.
Unless otherwise specified, the hyperparameters for these simulations are set to the values reported in \cref{tab:motivating_network}.

\begin{table}[p]
\centering
\spacingset{1}
\caption{
Motivating Simulation Study with One Interconnected Network. Unless otherwise noted, the parameters are fixed at $\phi_U = 0.6$, $\phi_Z = 0.4$, $\tau_U = \tau_Z = 1$, $\rho = 0.35$, $\beta_Z = 1, \beta_{\bar Z} = 0.8, \beta_U = 1, \beta_{\bar U} = 0.5$. We generate 200 data sets with $n = 100$. We regress the outcome on a different set of variables (columns), and report the bias of the OLS estimator for the local effect estimator, $\beta_Z$, and the interference effect estimator, $\beta_{\bar Z}$, when $\overline Z$ is included in the conditioning set. Values are rounded to the third decimal point. We bold the entries corresponding to the same cells as in \cref{tab:pairdata}. The qualitative conclusions remain unchanged.}
    \resizebox{0.82\textwidth}{!}{%
    \begin{tabular}{*{10}{c}}
        \hline
        & & & & &  & \\[-5pt]
        & &  \multicolumn{8}{c}{Conditioning set} \\
        \multirow{2}{*}{\shortstack[c]{True \\[5pt] Model}} & \multirow{3}{*}{\shortstack[c]{Alternative \\[5pt] spatial \\[5pt] parameters}} & \multicolumn{8}{c}{\& estimated parameter} \\[5pt]
        \cmidrule{3-10} 
        & & \\[-5pt]
        & &  $(Z) $ & $(Z, U)$  & \multicolumn{2}{c}{  $(Z, \bar{Z})$} & \multicolumn{2}{c}{  $(Z, \bar{Z}, U)$} & \multicolumn{2}{c}{  $(Z, \bar{Z}, U, \bar{U})$}\\
        \cmidrule(lr){3-3} \cmidrule(lr){4-4} \cmidrule(lr){5-6} \cmidrule(lr) {7-8} \cmidrule(lr) {9-10}
        & & $\beta_Z$ & $\beta_Z$ & $\beta_Z$ & $\beta_{\bar{Z}}$ & $\beta_Z$ & $\beta_{\bar{Z}}$ & $\beta_Z$ & $\beta_{\bar{Z}}$\\
        & & & & & & \\[-5pt]
        \hline \\[-5pt]
\multirow{2}{*}{\ref{fig:direct}} & & \multicolumn{8}{c}{$\beta_{\bar Z} = 0$ and $\beta_{\bar U}=0$} \\[5pt]
        \cmidrule{5-8}
        & & 0.550 & -0.005 & 0.428 & {\bf 0.370} & -0.007 & {\bf 0.006} & -0.007 & -0.004  \\
        \midrule \\[-5pt]
\multirow{2}{*}{\ref{fig:general_spatial_conf}} & & 
        \multicolumn{8}{c}{$\beta_{\bar Z} = 0$}
        \\[5pt]
        \cmidrule{5-8}
         & 
         & 0.683 & 0.045 & 0.489 & 0.595 & -0.008 & {\bf 0.237} & -0.013 & {\bf 0.030} \\ 
         \midrule \\[-5pt]
\multirow{4}{*}{\ref{fig:interference}} & & 
\multicolumn{8}{c}{$\rho = 0$ and $\beta_U = \beta_{\bar U} = 0$}
        \\[5pt]
        \cmidrule{5-8}
        & $\phi_z = 0.6$ & {\bf 0.437} & 0.334 & -0.012 & 0.000 & -0.007 & 0.002 & -0.007 & 0.011 \\
         & $\phi_z = 0.4$ & {\bf 0.261} & 0.193 & -0.002 & -0.008 & 0.003 & -0.004 & 0.002 & -0.007 \\
        & $\phi_z = 0.2$ & {\bf 0.151} & 0.088 & 0.006 & -0.002 & 0.002 & -0.005 & 0.003 & 0.000 \\
        \midrule \\[-5pt]
\multirow{4}{*}{\ref{fig:predictor_interference}} & 
& \multicolumn{8}{c}{$\beta_U = 0$ and $\beta_{\bar U} = 0$} \\[5pt]
\cmidrule{5-8}
        & $\rho = 0.15$ & 0.183 & 0.170 & 0.003 & 0.003 & 0.004 & 0.005 & 0.004 & 0.002 \\
        & $\rho = 0.35$ & 0.275 & 0.186 & 0.011 & -0.012 & 0.004 & -0.018 & 0.004 & -0.018  \\
        & $\rho = 0.45$ & {\bf 0.431} & {\bf 0.246} & 0.014 & -0.009 & 0.013 & -0.011 & 0.012 & -0.011 \\
        \midrule \\[-5pt]
\multirow{3}{*}{\ref{fig:direct_interference}} & & 
\multicolumn{8}{c}{$\beta_{\bar U} = 0$} \\[5pt]
        \cmidrule{5-8}
& $\phi_Z = 0.4$ & 0.838 & {\bf 0.195} & 0.456 & 0.345 & 0.001 & 0.021 & 0.001 & 0.019 \\ 
& $\phi_Z = 0\phantom{.5}$ & 0.501 & {\bf -0.010} & 0.444 & 0.283 & 0.001 & -0.017 & 0.001 & -0.019 \\
\midrule \\[-5pt]
\ref{fig:general_interference}
         & & 0.973 & 0.222 & 0.507 & 0.611 & -0.008 & 0.209 & -0.014 & -0.002 \\
        & & & & & &  \\
          \hline
          
        \end{tabular}
     }%
\label{tab:motivating_network}
\end{table}

\cref{tab:motivating_network} shows the results.  The conclusions are the same as the ones for paired data: \begin{enumerate*}[label=(\alph*)]
\item spatial confounding and interference can manifest as each other, 
\item inherent spatial dependencies complicate standard estimation strategies and can render them invalid even in simple settings, 
\item controlling for local and neighborhood covariates is crucial for adjusting for confounding and estimating causal effects unbiasedly, and 
\item local and interference effects should be investigated simultaneoulsy in the presence of spatial dependencies.
\end{enumerate*}

\section{Identifiability of model parameters}
\label{supp_sec:identifiability_ours}

\begin{proof}[Proof of \cref{theorem:identifiability}]

Consider a spatial network of observations that are organized on a ring graph, with symmetric adjacency matrix $A_{ij} = 1$ if $|i-j| = 1$, $(i = 1, j=n)$ and $(i =n, j=1)$, and 0 otherwise. Intuitively, under this structure, each unit has two neighbors, the ones with adjacent indices, and units 1 and $n$ are connected. 

Without loss of generality, we consider the case without measured covariates and where the exposure has mean 0.
We show that all coefficients, parameters in the precision matrix, and residual variance are identifiable, hence the causal effects of interest are also identifiable.

\paragraph{A few useful derivations}

\begin{itemize}[leftmargin=*]
\item We can write $\overline {\bm U} = D^{-1} A \bm U$ and $\overline {\bm Z} = D^{-1}A \bm Z$, where $D$ is the degree and $A$ is the adjacency matrix. 

\item We have that
\[
\begin{pmatrix}
G & Q \\ Q^\top & H
\end{pmatrix}^{-1} = 
\begin{pmatrix}
G^{-1} + G^{-1} Q (H - Q^\top G^{-1} Q)^{-1} Q^\top G^{-1} &
- G^{-1} Q (H - Q^\top G^{-1} Q)^{-1} \\
- (H - Q^\top G^{-1} Q)^{-1} Q^\top G^{-1} & (H - Q^\top G^{-1} Q)^{-1} 
\end{pmatrix},
\]
and using the known formulas for the multivariate normal distribution, we have that
\[
\bm U \mid \bm Z \sim N(-G^{-1}Q\bm Z, G^{-1}).
\]
\item The diagonal elements of the diagonal matrix $Q$ are $Q_{ii} = - \rho \sqrt{g_{ii} h_{ii}}$. Under the CAR form of $G = \tau_U^2 (D - \phi_U A)$ and $H = \tau_Z^2 (D - \phi_Z A)$, we have that $Q_{ii} = -2 \rho \tau_U \tau_Z$. These imply that
\[
G^{-1}Q = -2 \rho \tau_U \tau_Z G^{-1}.
\]

\item 
Putting these together, we have that
\begin{align*}
\E( \bm Y \mid \bm Z) 
&= \beta_Z \bm Z + \beta_{\bar Z} \overline {\bm Z} + \E[\bm U \mid \bm Z] + \beta_{\bar U} \E[\overline {\bm U} \mid \bm Z] \\
&= \beta_Z \bm Z + \beta_{\bar Z} D^{-1} A \bm Z
- G^{-1} Q \bm Z  + \beta_{\bar U} D^{-1} A (- G^{-1} Q \bm Z) \\
&= \beta_Z \bm Z + \beta_{\bar Z} (D^{-1} A \bm Z) + 2\rho\tau_U \tau_Z ( G^{-1} \bm Z)  + 2 \rho \tau_U \tau_Z \beta_{\bar U} (D^{-1} A G^{-1} \bm Z).
\numberthis
\label{supp_eq:EYmidZ}
\end{align*}
and that
\begin{align*}
\Var(\bm Y \mid \bm Z) &= \Var [ (I_n + \beta_{\bar U} D^{-1}A ) U \mid Z] + \sigma^2_Y I_n \\
&= (I_n + \beta_{\bar U} D^{-1}A ) \Var(\bm U \mid \bm Z) (I_n + \beta_{\bar U} D^{-1}A )^\top + \sigma^2_Y I_n \\
&= (I_n + \beta_{\bar U} D^{-1}A ) G^{-1} (I_n + \beta_{\bar U} D^{-1}A )^\top + \sigma^2_Y I_n,
\numberthis
\label{supp_eq:VarYmidZ}
\end{align*}
where
\[
I_n + \beta_{\bar U} D^{-1}A = \begin{pmatrix}
    1 & \frac{\beta_{\bar U}}2 &  &  & \dots &  & & \frac{\beta_{\bar U}}2 \\
    \frac{\beta_{\bar U}}2 & 1 & \frac{\beta_{\bar U}}2 &  & \dots &  &  \\
     & \frac{\beta_{\bar U}}2 & 1 & \frac{\beta_{\bar U}}2 & \dots \\
    & & & &  \vdots \\
     & & & &  \dots & \frac{\beta_{\bar U}}2 & 1 &     \frac{\beta_{\bar U}}2 \\
    \frac{\beta_{\bar U}}2 & & & &  \dots & & \frac{\beta_{\bar U}}2 & 1
\end{pmatrix}
\]

\end{itemize}

\paragraph{}
Theorem 1 of \cite{schnell2020mitigating} shows that we can identify whether $\rho\phi_U = 0$ or not based on $\bm Z$. If $\rho\phi_U \neq 0$, they show that $(\tau_Z, \phi_Z, \phi_U, |\rho|)$ are identifiable. Their theorem applies here directly since their results are based on the same specification of the joint distribution $(\bm Z, \bm U)$ on the ring graph.

Our proof deviates from theirs on the specification of the outcome structure and, as a result, the identifiability results for the remaining parameters. These differences stem from allowing for potential interference effects and for including additional parameters on the unmeasured spatial confounder due to non-local confounding, which leads to the inclusion of additional terms and additional unknown parameters in the distribution of $\bm Y$ given $\bm Z$ in equations \cref{supp_eq:EYmidZ} and \cref{supp_eq:VarYmidZ}.

In linear models, expectations and variances are separately identifiable. So we can identify $\Var(\bm Y \mid \bm Z)$ and $\E(\bm Y \mid \bm Z)$.

\paragraph{Identifiability of parameters from $\Var(\bm Y \mid \bm Z)$}

We acquire the form of the entries in $G^{-1}$ as $n \rightarrow \infty$ based on Theorem 4 in the Supplementary Materials of \cite{schnell2020mitigating}. Let entry $(i, j)$ of matrix $G^{-1}$ be denoted by $g_{ij}$. Then, we have that
\begin{align}
\lim_{n \rightarrow \infty} g_{ij}
= 
\frac{\tau_U^2}{2 \sqrt{1 - \phi_U^2}} \left( \frac{\phi_U}{1 + \sqrt{1 - \phi_U^2}} \right)^{|i-j|}
\label{supp_eq:Ginv}
\end{align}
After some mundane matrix multiplications, we have that the entries of the variance in 
\cref{supp_eq:VarYmidZ} can be written as
\begin{align*}
[\Var(\bm Y \mid \bm Z)]_{ij} &= g_{ij} + \frac{\beta_{\bar U}}2 (g_{(i-1)j} + g_{(i+1)j} + g_{i(j-1)} + g_{i(j+1)}) + \\
& \hspace{20pt} + \frac{\beta_{\bar U}^2}4
(g_{(i-1)(j-1)} + g_{(i-1)(j+1)} + g_{(i+1)(j-1)} + g_{(i+1)(j+1)})
+ \sigma^2_Y I(i = j).
\end{align*}
Therefore, we can write the limit of the $(i,j)$ entry for the outcome conditional variance $\lim_{n \rightarrow \infty} \Var(\bm Y \mid \bm Z)_{ij}$ as a function of $k = |i-j|$:

\begin{itemize}[leftmargin=*]
\item
For $k = 0$, and $i = j$:
\end{itemize}
\begin{align*}
\lim_{n \rightarrow \infty} \Var(\bm Y \mid \bm Z)_{ij} 
&= \frac{\tau_U^2}{2 \sqrt{1 - \phi_U^2}} + 4 \frac{\beta_{\bar U}}2 \frac{\tau_U^2}{2 \sqrt{1 - \phi_U^2}} \frac{\phi_U}{1 + \sqrt{1 - \phi_U^2}}
\\
& \quad
+ 4 \frac{\beta_{\bar U}^2}4 \frac{\tau_U^2}{2 \sqrt{1 - \phi_U^2}} \left( \frac{\phi_U}{1 + \sqrt{1 - \phi_U^2}} \right)^2 + \sigma^2_Y \\
&= \frac{\tau_U^2}{2 \sqrt{1 - \phi_U^2}} \left[ 1 + 2 \beta_{\bar U}  \frac{\phi_U}{1 + \sqrt{1 - \phi_U^2}} + \beta_{\bar U}^2 \left(\frac{\phi_U}{1 + \sqrt{1 - \phi_U^2}} \right)^2 \right] + \sigma^2_Y \\
&= \frac{\tau_U^2}{2 \sqrt{1 - \phi_U^2}} \left[ 1 + \beta_{\bar U}  \frac{\phi_U}{1 + \sqrt{1 - \phi_U^2}} \right]^2 + \sigma^2_Y
\end{align*}

\begin{itemize}[leftmargin=*]
\item For $k = 1$:
\end{itemize}
\begin{align*}
\lim_{n \rightarrow \infty} \Var(\bm Y \mid \bm Z)_{ij} &=
\frac{\tau_U^2}{2 \sqrt{1 - \phi_U^2}} \frac{\phi_U}{1 + \sqrt{1 - \phi_U^2}} + \\
& \quad 
+\frac{\beta_{\bar U}}2 \left[ 2 \frac{\tau_U^2}{2 \sqrt{1 - \phi_U^2}} +
2 \frac{\tau_U^2}{2 \sqrt{1 - \phi_U^2}} \left( \frac{\phi_U}{1 + \sqrt{1 - \phi_U^2}} \right)^2\right] + \\
& \quad +\frac{\beta_{\bar U}^2}4
\left[ 3 \frac{\tau_U^2}{2 \sqrt{1 - \phi_U^2}} \frac{\phi_U}{1 + \sqrt{1 - \phi_U^2}}  + \frac{\tau_U^2}{2 \sqrt{1 - \phi_U^2}} \left( \frac{\phi_U}{1 + \sqrt{1 - \phi_U^2}} \right)^3 \right] \\
&=
\frac{\tau_U^2}{2 \sqrt{1 - \phi_U^2}} \left\{ \frac{\phi_U}{1 + \sqrt{1 - \phi_U^2}}+ \beta_{\bar U} \left[ 1 + \left( \frac{\phi_U}{1 + \sqrt{1 - \phi_U^2}} \right)^2 \right] \right. + \\
& \left. \hspace{100pt} +\frac{\beta_{\bar U}^2}4
\left[ 3 \frac{\phi_U}{1 + \sqrt{1 - \phi_U^2}}  + \left( \frac{\phi_U}{1 + \sqrt{1 - \phi_U^2}} \right)^3 \right] \right\}
\end{align*}

\begin{itemize}[leftmargin=*]
\item For $k \geq 2$:
\end{itemize}
\begin{align*}
& \lim_{n \rightarrow \infty} \Var(\bm Y \mid \bm Z)_{ij} = \\
&=
\frac{\tau_U^2}{2 \sqrt{1 - \phi_U^2}} \left( \frac{\phi_U}{1 + \sqrt{1 - \phi_U^2}} \right)^k + \\
&\quad 
+\frac{\beta_{\bar U}}2 \left[ 2 \frac{\tau_U^2}{2 \sqrt{1 - \phi_U^2}}\left( \frac{\phi_U}{1 + \sqrt{1 - \phi_U^2}} \right)^{k -1} +
2 \frac{\tau_U^2}{2 \sqrt{1 - \phi_U^2}} \left( \frac{\phi_U}{1 + \sqrt{1 - \phi_U^2}} \right)^{k+1} \right] + & \\[10pt]
& \quad +\frac{\beta_{\bar U}^2}4
\left[ 2 \frac{\tau_U^2}{2 \sqrt{1 - \phi_U^2}} \left(\frac{\phi_U}{1 + \sqrt{1 - \phi_U^2}}\right)^k  + \frac{\tau_U^2}{2 \sqrt{1 - \phi_U^2}} \left( \frac{\phi_U}{1 + \sqrt{1 - \phi_U^2}} \right)^{k-2} +
\right. \\[10pt]
& \hspace{200pt} \left.
+ \frac{\tau_U^2}{2 \sqrt{1 - \phi_U^2}} \left( \frac{\phi_U}{1 + \sqrt{1 - \phi_U^2}} \right)^{k + 2} \right] \\
&=
\frac{\tau_U^2}{2 \sqrt{1 - \phi_U^2}} \left\{ \left( \frac{\phi_U}{1 + \sqrt{1 - \phi_U^2}} \right)^k + \right. \\
&
\hspace{60pt}
+\beta_{\bar U} \left[ \left( \frac{\phi_U}{1 + \sqrt{1 - \phi_U^2}} \right)^{k -1} +
\left( \frac{\phi_U}{1 + \sqrt{1 - \phi_U^2}} \right)^{k+1} \right] + & \\
& 
\hspace{60pt}
\left. +\frac{\beta_{\bar U}^2}4
\left[ 2 \left(\frac{\phi_U}{1 + \sqrt{1 - \phi_U^2}}\right)^k  + \left( \frac{\phi_U}{1 + \sqrt{1 - \phi_U^2}} \right)^{k-2} +
\left( \frac{\phi_U}{1 + \sqrt{1 - \phi_U^2}} \right)^{k + 2} \right] \right\} \\
&=
\frac{\tau_U^2}{2 \sqrt{1 - \phi_U^2}} \left\{ \left( \frac{\phi_U}{1 + \sqrt{1 - \phi_U^2}} \right)^k + \right. \\
&
\hspace{90pt}
+\beta_{\bar U} \left( \frac{\phi_U}{1 + \sqrt{1 - \phi_U^2}} \right)^{k -1} \left[ 1 +
\left( \frac{\phi_U}{1 + \sqrt{1 - \phi_U^2}} \right)^2 \right] + & \\
& 
\hspace{90pt}
\left. +\frac{\beta_{\bar U}^2}4
\left(\frac{\phi_U}{1 + \sqrt{1 - \phi_U^2}}\right)^{k-2}
\left[ 1 +
\left( \frac{\phi_U}{1 + \sqrt{1 - \phi_U^2}} \right)^2 \right]^2 \right\} \\
&=
\frac{\tau_U^2}{2 \sqrt{1 - \phi_U^2}}
\left(\frac{\phi_U}{1 + \sqrt{1 - \phi_U^2}}\right)^{k-2}
\left\{ \left( \frac{\phi_U}{1 + \sqrt{1 - \phi_U^2}} \right)^2 + \right. \\
&
\hspace{200pt}
+\beta_{\bar U} \frac{\phi_U}{1 + \sqrt{1 - \phi_U^2}} \left[ 1 +
\left( \frac{\phi_U}{1 + \sqrt{1 - \phi_U^2}} \right)^2 \right] + & \\
& 
\hspace{200pt}
\left. +\frac{\beta_{\bar U}^2}4
\left[ 1 +
\left( \frac{\phi_U}{1 + \sqrt{1 - \phi_U^2}} \right)^2 \right]^2 \right\} \\
&=
\frac{\tau_U^2}{2 \sqrt{1 - \phi_U^2}}
\left(\frac{\phi_U}{1 + \sqrt{1 - \phi_U^2}}\right)^{k-2}
\left\{ \frac{\phi_U}{1 + \sqrt{1 - \phi_U^2}} 
+
\frac{\beta_{\bar U}}2
\left[ 1 +
\left( \frac{\phi_U}{1 + \sqrt{1 - \phi_U^2}} \right)^2 \right] \right\}^2
\end{align*}

The parameter $\phi_U$ has been identified based on $\bm Z$. Identifiability of $\phi_U$ can also be achieved by comparing the variances of pairs at distance $k$ and $k'$ (when $k, k' \geq 2$), since for a pair $(i,j)$ with distance $k$ (e.g., $|i-j| = k$), and a pair $(i',j')$ with distance $k'$ (e.g., $|i'-j'| = k'$), we have that
\[
\lim_{n \rightarrow \infty} \frac{\Var(\bm Y \mid \bm Z)_{i'j'}}{\Var(\bm Y \mid \bm Z)_{ij}} = \left( \frac{\phi_U}{1 + \sqrt{1 - \phi_U^2}} \right)^{k'-k}.
\]
Therefore, the spatial parameter $\phi_U$ can be identified by studying how spatial correlation in the outcome attenuates with distance.

Once $\phi_U$ is identified, we can identify $\beta_{\bar U}$ by comparing the variances of pairs at distance 1 (pair $(i,j)$ with $|i-j| = k$) with that of pairs at distance 2 (pair $(i',j')$ with $|i'-j'| = 2$), for which
\[
\lim_{n \rightarrow \infty} \frac{\Var(\bm Y \mid \bm Z)_{i'j'}}{\Var(\bm Y \mid \bm Z)_{ij}} =
\frac
{
\left\{ \frac{\phi_U}{1 + \sqrt{1 - \phi_U^2}} 
+
\frac{\beta_{\bar U}}2
\left[ 1 +
\left( \frac{\phi_U}{1 + \sqrt{1 - \phi_U^2}} \right)^2 \right] \right\}^2}
{ \frac{\phi_U}{1 + \sqrt{1 - \phi_U^2}}+ \beta_{\bar U} \left[ 1 + \left( \frac{\phi_U}{1 + \sqrt{1 - \phi_U^2}} \right)^2 \right]  +\frac{\beta_{\bar U}^2}4
\left[ 3 \frac{\phi_U}{1 + \sqrt{1 - \phi_U^2}}  + \left( \frac{\phi_U}{1 + \sqrt{1 - \phi_U^2}} \right)^3 \right] }
\]
is a bijective function of the parameter $\beta_{\bar U}$. Once $\beta_{\bar U}$ is identified, $\tau_U$ can be trivially identified based on the variance of pairs at distance 1 $(|i-j|=1)$. Subsequently, the residual variance $\sigma^2_Y$ can be identified based on the diagonal elements of the covariance matrix $(i = j)$.

\paragraph{Identifiability of parameters from $\E(\bm Y \mid \bm Z)$}

As long as the linear predictors in $\E(\bm Y \mid \bm Z)$ are not perfectly collinear, their corresponding coefficients are identifiable. These linear predictors are $\bm Z, \overline {\bm Z} = D^{-1}A \bm Z$, $G^{-1} \bm Z$, and $D^{-1}A (G^{-1} \bm Z)$, where all parameters in $G$ have been identified.
We assume that the vector $\bm Z$ is not constant, and there is some variation in the exposure across units. Then, the local and neighbourhood exposures are not perfectly correlated, and the first two predictors are not collinear. Note that a unit's entry in $\overline {\bm Z}$ is the average of the values in $\bm Z$ for its two neighbours. Based on the form of $G^{-1}$, we see that a unit's entry in $G^{-1}\bm Z$ corresponds to a weighted average of entries in $\bm Z$ of all other units, with weights specified according to \cref{supp_eq:Ginv}, which means that $G^{-1}\bm Z$ is not collinear with the previous ones. Lastly, the same argument holds for $D^{-1}A G^{-1}\bm Z$, as long as $G^{-1} \bm Z$ is not constant across units.

From \cref{supp_eq:EYmidZ}, we have that the coefficients $(\beta_Z, \beta_{\bar Z}, \rho\tau_U\tau_Z, \rho\tau_U\tau_Z\beta_{\bar U})$ are identifiable. Since $\tau_U$ and $\tau_Z$ have already been identified, we now have that the parameter $\rho$ is identifiable as well.

\end{proof}

\section{Illustration of prior distributions}
\label{supp_sec:priors}

As discussed in \cref{subsec:priors}, the prior specifications on the spatial parameters $\tau_U, \tau_Z$ have implications on the implied prior on the the confounding strength due to $U$ and the variance of the  exposure $Z$, respectively. Here, we provide a simulation-based illustration for the implied prior properties discussed in the manuscript.

Before we delve into this illustration, we discuss briefly the matrices $G, H$ in the joint precision matrix of \cref{ass:UZ_normal}, which in the absence of measured covariates states that
\begin{equation}
\begin{pmatrix} \bm U \\ \bm Z \end{pmatrix} \Big| \sim N_{2n} \left(
\bm 0_{2n} ,
\begin{pmatrix} G & Q \\ Q & H \end{pmatrix}^{-1}
\right).
\label{supp_eq:UZ_normal_no_covs}
\end{equation}
Since the joint distribution is parameterized through its precision matrix, $G^{-1}$ and $H^{-1}$ are the marginal covariance matrices of $\bm U$ and $\bm Z$, respectively, {\it only} when $\rho = 0$, and
the distribution of $\bm U$ when drawn from $N_n(\bm 0_n, G^{-1})$ is different from the distribution of $\bm U$ when drawn from \cref{supp_eq:UZ_normal_no_covs} for $\rho \neq 0$. In our illustrations below, we will consider vectors $\bm U$ and $\bm Z$ which are drawn from $N_n(\bm 0_n, G^{-1})$ and $N_n(\bm 0_n, H^{-1})$, respectively, or simultaneously from \cref{supp_eq:UZ_normal_no_covs} for $\rho \neq 0$.

\subsection{Prior distribution for $\tau_U$}

\begin{figure}[p]
\centering
\includegraphics[width = 0.8\textwidth,trim=0 57 0 0,clip]{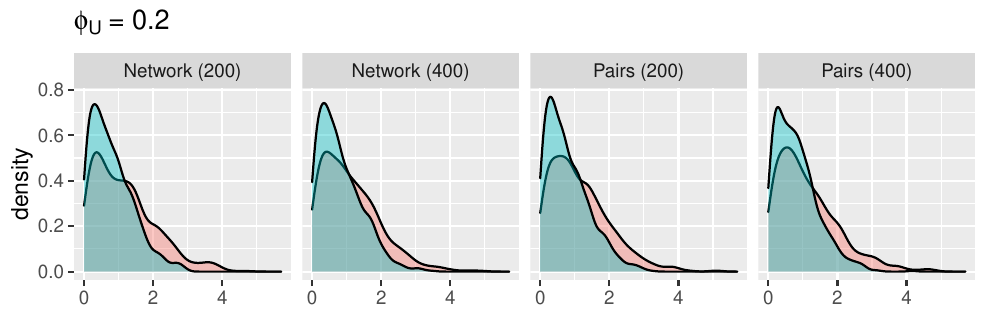} \\
\includegraphics[width = 0.8\textwidth,trim=0 57 0 0,clip]{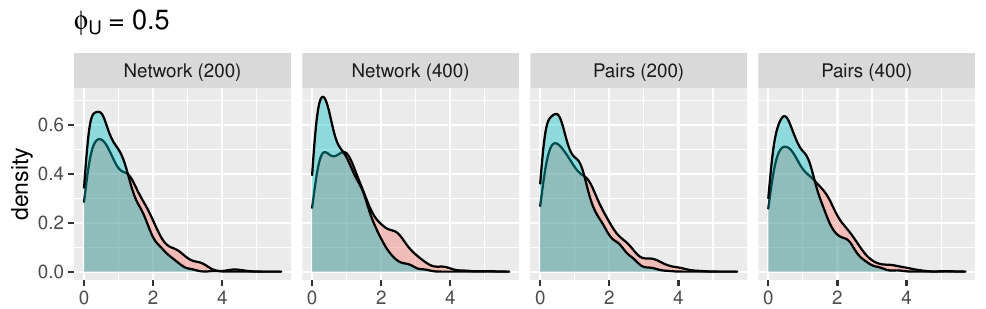} \\
\includegraphics[width = 0.8\textwidth,trim=0 57 0 0,clip]{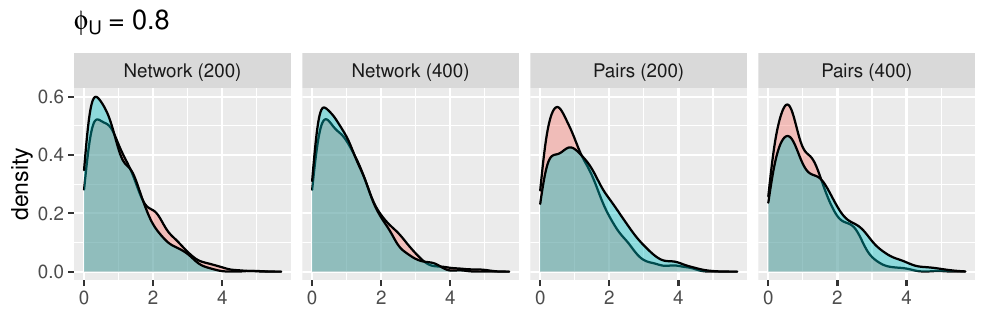}
\includegraphics[width = 0.9\textwidth,trim=0 0 0 160,clip]{figures/tauU_prior_phiU08.pdf}
\caption{Density plot for the implied prior distribution on the amount of outcome variability explained by a measured covariate and the unmeasured covariate $U$ based on the prior distribution for $\tau_U$. We consider network and pair data of sample sizes 200 and 400. $\bm U$ is generated from $N_n(\bm 0_n, G^{-1})$ for $\tau_U$ sampled from its prior distribution and $\phi_U \in \{0.2, 0.5, 0.8\}$.}
\label{supp_fig:prior_tauU}
\end{figure}

The strength of a measured covariate $C$ with variance 1 in the outcome model corresponds to the magnitude of its coefficient $\beta_C$, or (equivalently) the standard deviation of $\beta_C C_i$ across units $i$. Similarly, since the coefficient of $U_i$ is set to $\beta_U = 1$, the strength of the unmeasured $U_i$ in the outcome model can be measured by the standard deviation of the unmeasured confounder. For network and paired data of sizes $n \in \{200, 400\}$, we performed the following procedure 1,000 times:
\begin{enumerate*}[label=(\alph*)]
\item we drew $\beta_C \sim N(0, \sigma^2_{prior})$,
\item we drew $1 / \tau_U$ from the prior distribution described in \cref{subsec:priors},
\item we generated $\bm U = (U_1, U_2, \dots, U_n)$ from $N_n(\bm 0_n, G^{-1})$ where $G$ has a CAR structure with $\tau_U$ the one drawn at the previous step and $\phi_U \in \{0.2, 0.5, 0.8\}$.
\end{enumerate*}
Each time, we calculated the absolute value of $\beta_C$ and the standard deviation of $U_i$ across $i$. Their distributions are shown in \cref{supp_fig:prior_tauU}, using red for the measured covariate $C$ and blue for the unmeasured covariate $U$.
Considering that the outcome is standardized to have variance 1, the prior distribution for the strength of the unmeasured confounder in the outcome model allows for all reasonable values and it is relatively similar to the corresponding prior distribution for a measured covariate, across all configurations.

The two distributions are similar across all choices of $\sigma^2_{prior}$ we explored. Since prior distributions on model coefficients are well-explored and understood in the literature, the prior distribution for $\tau_U$ we designed can be used straightforwardly without requiring additional tuning. Specifically, a researcher can simply specify $\sigma^2_{prior}$ for the prior distribution of a coefficient in the outcome model, and our specification for the prior distribution of $\tau_U$ would automatically translate the choice of $\sigma^2_{prior}$ to an equivalent prior for the confounding strength of the unmeasured covariate.

\subsection{Prior distribution for $\tau_Z$}

\begin{figure}[p]
\centering
\includegraphics[width = 0.8\textwidth,trim=0 19 0 0,clip]{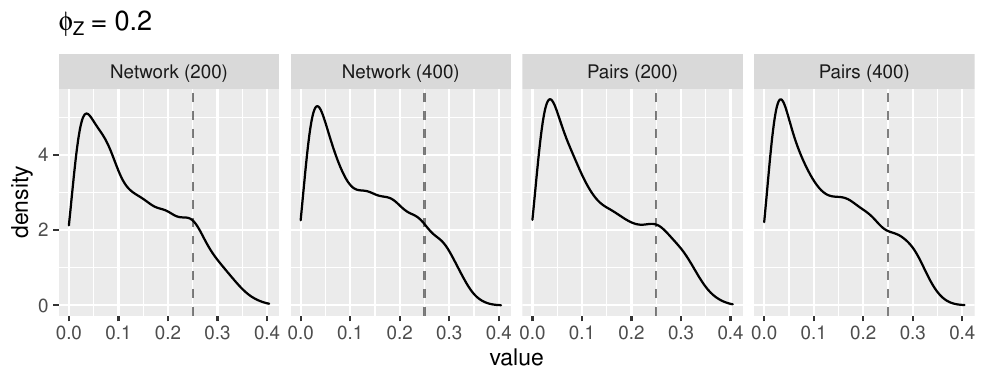} \\[5pt]
\includegraphics[width = 0.8\textwidth,trim=0 19 0 0,clip]{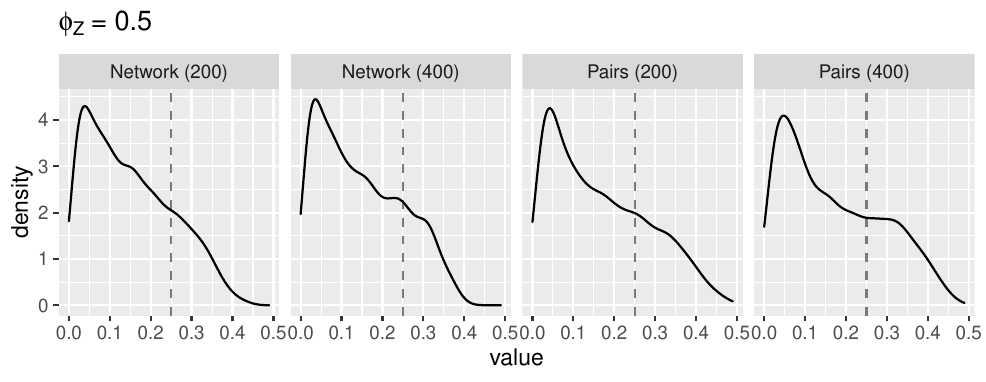} \\[5pt]
\includegraphics[width = 0.8\textwidth,trim=0 19 0 0,clip]{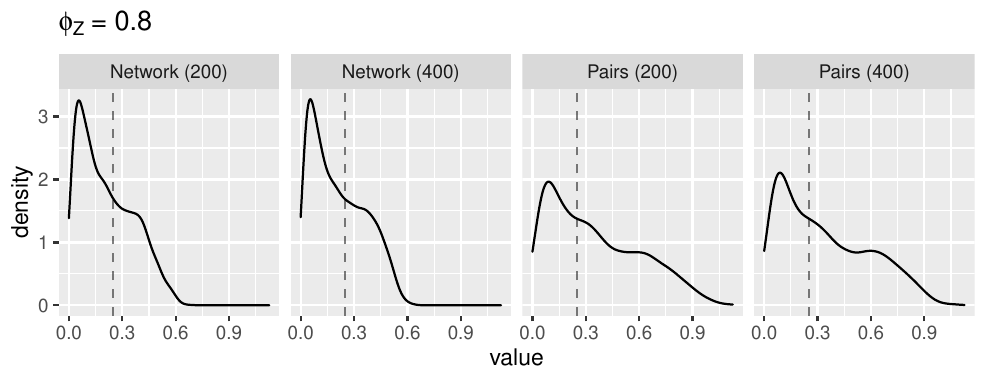} \\[5pt]
\caption{Implied prior distribution on the exposure variability implied by the specified prior distribution for $\tau_Z$. We consider network and pair data of sample sizes 200 and 400, and $\bm Z$ is drawn from $N_n(\bm 0_n, H^{-1})$ where $H$ has a CAR structure with $\tau_Z$ sampled from its prior distribution and $\phi_Z \in \{0.2, 0.5, 0.8\}$.}
\label{supp_fig:prior_tauZ}
\end{figure}

We also investigated the prior on the exposure's variance as implied by the prior on $\tau_Z$ discussed in \cref{subsec:priors}. We set the hypothesized marginal variance of $\bm Z$ to $\widetilde s^2_Z = 1$ and the hypothesized residual variance of $\bm Z$ to $\widetilde \sigma^2_Z = 0.5^2$. We repeated the following procedure 2,000 times:
\begin{enumerate*}[label=(\alph*)]
\item we drew $\tau_Z$ from its prior distribution,
\item we generated $\bm Z$ from $N_n(\bm 0_n, H^{-1})$, where $H$ is specified as CAR with $\tau_Z$ the draw from the previous step and $\phi_Z \in \{0.2, 0.5, 0.8\}$, and
\item we calculated the exposure variance across locations.
\end{enumerate*}
We did so for network and paired data of sample sizes 200 and 400. The distribution of this variance is shown in \cref{supp_fig:prior_tauZ}, where the dashed vertical line represents the hypothesized residual variance of the exposure conditional on measured covariates, $\widetilde \sigma^2_Z$. We see that the implied exposure variability takes values in the neighbourhood of $\widetilde \sigma^2_Z$, as expected.

\subsection{Implied prior distributions when $\rho \neq 0$}

Our prior distributions as described in \cref{subsec:priors} are designed based on approximations of the variability in the unmeasured covariate $\bm U$ and the exposure $\bm Z$ when the two variables are independent. Here, we illustrate using simulation that these prior distributions also imply reasonable prior distributions on the strength of confounding due to $U$ and the inherent exposure variability even when $\rho \neq 0$.

\begin{figure}[!t]
\centering
\includegraphics[width = 0.85\textwidth,trim=0 58 0 0,clip]{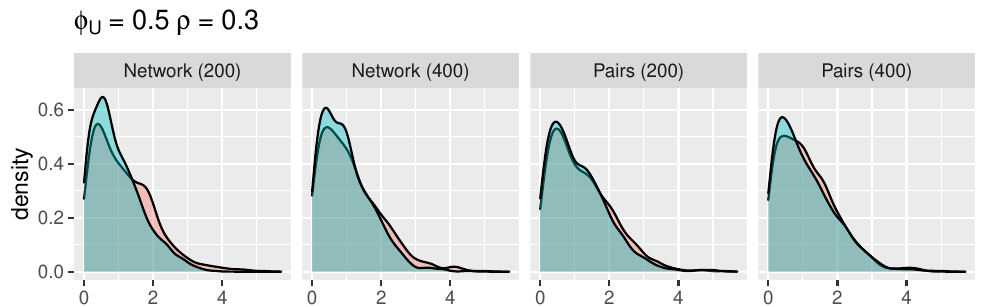} \\
\includegraphics[width = 0.85\textwidth,trim=0 0 0 170,clip]{figures/tauU_prior_phiU05_rho03.pdf} \\
\includegraphics[width = 0.85\textwidth,trim=0 19 0 0,clip]{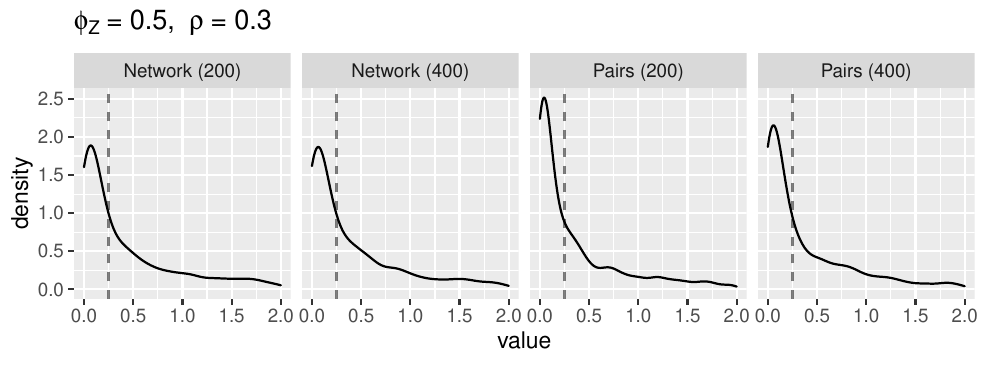}  \\
\caption{Implied priors when the exposure and the unmeasured covariate are correlated according to \cref{ass:UZ_normal} with $\phi_U = \phi_Z = 0.5$ and $\rho = 0.3$. Top: Prior distribution of predictive strength of a measured and the unmeasured covariates (equivalent of \cref{supp_fig:prior_tauU}). Bottom: Prior distribution on the exposure variability (equivalent of \cref{supp_fig:prior_tauZ}).}
\label{supp_fig:prior_tauU_tauZ_rho03}
\end{figure}

We performed the following procedure 1,000 times:
\begin{enumerate*}[label=(\alph*)]
\item we drew $\tau_U$ and $\tau_Z$ from their prior distributions,
\item for these values and for $\phi_U = \phi_Z = 0.5$ and $\rho = 0.3$, we constructed the matrices $G, H$, and $Q$ and the precision matrix \cref{supp_eq:UZ_normal_no_covs},
\item we drew $(\bm U, \bm Z)$ from their joint distribution.
\end{enumerate*}
Based on the 1,000 samples from $(\bm U, \bm Z)$ we calculated the standard deviation of $\bm U$ across locations, and the standard deviation of $\bm Z$ across locations. \cref{supp_fig:prior_tauU_tauZ_rho03} is an equivalent to those in Figures \ref{supp_fig:prior_tauU} and \ref{supp_fig:prior_tauZ} for correlated exposure and unmeasured covariate. Specifically, at the top of \cref{supp_fig:prior_tauU_tauZ_rho03}, we compare the standard deviation of $U$ against the absolute value for draws from the $N(0, \sigma^2_{prior})$ distribution, and we find that the implied confounding strength for a measured and the unmeasured covariate have similar prior distributions. At the bottom of \cref{supp_fig:prior_tauU_tauZ_rho03}, we plotted the distribution of the exposure variance against the hypothesized residual variance $\widetilde \sigma^2_Z$, and we see that the implied prior still allow for a reasonable range of values.

\section{Posterior distribution sampling scheme}
\label{supp_sec:mcmc}

We describe the MCMC updates for approximating the posterior distribution. 
We write $p(\theta \mid \cdot)$ to denote the posterior distribution of $\theta$ conditional on all other parameters. 
We use the following definitions:
\begin{itemize}[leftmargin=*,label=-]
\item \underline{Exposure model residuals:} We use $\bm Z_{res}$ to denote the vector of length $n$ including the exposure residuals based on the current values of the parameters $\gamma_0, \bm \gamma_C$. Specifically, the $i^{th}$ entry of $\bm Z_{res}$ is
$Z_i - \gamma_0 - \widetilde C_i^T \bm \gamma_C$.

\item \underline{Outcome model residuals:} We consider three versions of outcome model residuals, conditional on all covariates, the measured ones only, and the unmeasured covariate only. We denote them by $\bm Y_{res}$, $\bm Y_{res}^C$, and $\bm Y_{res}^U$ with $i^{th}$ entries
\begin{align*}
Y_{res,i} &= Y_i - \beta_0 - \beta_Z Z_i - \beta_{\bar Z}\overline Z_i - \widetilde C_i \bm \beta_C - \beta_U U_i - \beta_{\bar U} \overline U_i \\
Y_{res,i}^C &= Y_i - \beta_0 - \beta_Z Z_i - \beta_{\bar Z}\overline Z_i - \widetilde C_i \bm \beta_C,
\quad \text{and} \\
Y_{res,i}^U &= Y_i - \beta_U U_i - \beta_{\bar U} \overline U_i,
\end{align*}
respectively.

\item \underline{The ``coefficient matrix'' of the unmeasured covariate in the outcome model:} If $A_U$ denotes the adjacency matrix that drives the neighbourhood confounder values $\overline{\bm U}$ in terms of $\bm U$, and $D_U$ is the corresponding degree matrix, then we have that $\overline{\bm U} = D_U^{-1} A_U \bm U$. Therefore, the vector $\bm U$ is included in the outcome model through
\(
\beta_U \bm U + \beta_{\bar U} \overline{\bm U} = (I_n + \beta_{\bar U} D_U^{-1} A_U) \bm U
\),
when $\beta_U = 1$. We definite $M_U = I_n + \beta_{\bar U} D_U^{-1} A_U$ which will play a role for updating the values of the unmeasured covariate $\bm U$. The matrix $M_U$ depends on the current value of $\beta_{\bar U}$ so it is itself updated during the MCMC every time $\beta_{\bar U}$ is updated.

\item \underline{The design matrices:} the $n \times (p + 3)$  design matrix for the outcome model based on measured variables $\bm X = (\bm 1 \ \bm Z \ \overline{\bm Z} \ \bm C)$, and the $n \times (p + 1)$ design matrix for the exposure model  $\bm X_{-z} = (\bm 1 \ \bm C)$.

\end{itemize}

\noindent
The full list of parameters and the corresponding MCMC updates are described below. We use superscripts $\tor$ to denote the $r^{th}$ posterior sample of a given parameter. The updates below describe how the $\tor[r+1]^{th}$ sample is acquired. Most parameters are drawn using Gibbs updates, and Metropolis-Hastings is used for the spatial parameters.
\begin{enumerate}[leftmargin=*,label=(\alph*)]
\item $\bm U^{\tor[r+1]}$ is drawn from its full conditional posterior distribution which is a multivariate normal with mean $\mu_{new, \bm U}$ and variance $\Sigma_{new, \bm U}$ where
\begin{align*}
\Sigma_{new, \bm U} &= \left[ G^{\tor} + \big( M_U^{\tor} \big)^T M_U^{\tor} / \sigma^{2\tor}_Y \right]^{-1}, \quad \text{and}
\\
\mu_{new,\bm U} &= \Sigma_{new, \bm U} \left[ 
\big( M_U^{\tor} \big)^T \bm Y_{res}^{C,\tor} / \sigma^{2\tor}_Y
- Q^{\tor} \bm Z_{res}^{\tor}
\right].
\end{align*}
We update the values of $\overline{\bm U}$ based on $\bm U^{\tor[r+1]}$, and we calculate $\bm Y_{res}^{U\tor[r+1]}$.

\item We draw the intercept and the coefficients of the local exposure, neighbourhood exposure, and the measured covariates in the outcome model, $(\beta_0, \beta_Z, \beta_{\bar Z}, \bm \beta_C)$, from their joint full conditional distribution which is a multivariate normal with mean $\mu_{new, \beta}$ and variance $\Sigma_{new, \beta}$, where
\begin{align*}
\Sigma_{new, \beta} &= \left[ \bm X^T \bm X / \sigma^{2\tor}_Y + I_{p + 3} / \sigma^2_{prior} \right]^{-1},
\quad \text{and} \\
\mu_{new, \beta} &= \Sigma_{new, \beta} \bm X^T \bm Y_{res}^{U,\tor[r+1]} / \sigma^{2\tor}_Y
\end{align*}
We calculate $\bm Y_{res}^{\tor[r+1]}$ and $\bm Y_{res}^{C,\tor[r+1]}$ based on the new $\beta$-values.

\item We draw the intercept and the coefficients of the measured covariates in the exposure model, $(\gamma_0, \bm \gamma_C)$, from their joint full conditional distribution which is a multivariate normal with mean $\mu_{new, \gamma}$ and variance $\Sigma_{new, \gamma}$, where
\begin{align*}
\Sigma_{new, \gamma} &= \left[ \bm X_{-z}^T H^{\tor} \bm X_{-z}  + I_{p + 1} / \sigma^2_{prior} \right]^{-1},
\quad \text{and} \\
\mu_{new, \gamma} &= \Sigma_{new, \gamma} \bm X_{-z}^T \left( H^{\tor} \bm Z + (Q^{\tor})^T \bm U^{\tor[r+1]} \right).
\end{align*}
We update the exposure residuals $\bm Z_{res}$ based on the new $\gamma-$values.

\item We draw the residual outcome model variance from an inverse gamma with shape parameter $\alpha_{new,Y} = \alpha_Y + n / 2$, and rate parameter $\beta_{new,Y} = 
\beta_Y + (\bm Y_{res}^{\tor[r+1]})^T \bm Y_{res}^{\tor[r+1]} / 2. $

\item We draw the coefficient of the neighbourhood unmeasured covariate from a normal distribution with mean $\mu_{new,\bar U}$ and variance $\sigma^2_{new, \bar U}$ where
\begin{align*}
\sigma^2_{new, \bar U} &= \left[ 
\big( \overline{\bm U}^{\tor[r+1]} \big)^T
\overline{\bm U}^{\tor[r+1]} / \sigma^{2\tor[r+1]}_Y + 1 / \sigma^2_{prior, \bar U} \right]^{-1},
\quad \text{and} \\
\mu_{new, \bar U} &=  \sigma^2_{new, \bar U}
\big( \overline{\bm U}^{\tor[r+1]} \big)^T
\big( \bm Y_{res}^{C\tor[r+1]} - \beta_U \bm U^{\tor[r+1]} \big) / \sigma^{2\tor[r+1]}_Y.
\end{align*}
We update $\bm Y_{res}$ and $\bm Y_{res}^U$ based on the new value of $\beta_{\bar U}$.

\item We have specified CAR structure for $G, H$ with two parameters each ($\phi_U, \tau_U, \phi_Z, \tau_Z$) and one parameter ($\rho$) for their correlation. We update all parameters using a Metropolis-Hastings step. Consider the function $\text{dexpit}: \mathbb{R} \rightarrow (-1, 1)$ with $\text{dexpit}(x) = 2 / (1 + \exp(-x)) -1$ and its inverse $\text{dexpit}^{-1}: (-1, 1)  \rightarrow \mathbb{R}$ with $\text{dexpit}^{-1}(x) = \log(1 + x) - \log(1 - x).$ 
If  $\phi_U^{\tor}, \tau_U^{\tor}, \phi_Z^{\tor}, \tau_Z^{\tor}, \rho^{\tor}$ are the current values of the parameters, we propose values $\phi_U^{prop}, \tau_U^{prop},$ $\phi_Z^{prop}, \tau_Z^{prop}, \rho^{prop}$ as follows:
\begin{itemize}[leftmargin=*,label=-]
\item Draw $\epsilon_{\phi_U}$ from $N(0, 0.35^2 s^2)$ and set
$\phi_U^{prop} = \text{dexpit} ( \text{dexpit}^{-1} (\phi_U^{\tor}) + \epsilon_{\phi_U}) $.
\item Draw $\epsilon_{\tau_U}$ from $N(0, 0.2^2s^2)$ and set
$\tau_U^{prop} = \exp( \log( \tau_U^{\tor} ) + \epsilon_{\tau_U} )$.
\item Set $\phi_Z^{prop}$ and $\tau_Z^{prop}$ similarly.
\item  Draw $\epsilon_{\rho}$ from $N(0, 0.5^2 s^2)$ and set
$\rho^{prop} = \text{dexpit} ( \text{dexpit}^{-1} (\rho^{\tor}) + \epsilon_{\rho}) $.
\end{itemize}
Create matrices $G^{prop}, H^{prop}$ and $Q^{prop}$ based on the proposed values.

The acceptance probability for the joint move is given by the ratio of the posterior probabilities of the proposed values versus the current values:
\[
\frac{p( \phi_U^{prop}, \tau_U^{prop}, \phi_Z^{prop}, \tau_Z^{prop}, \rho^{prop} \mid \cdot)}
{p(\phi_U^{\tor}, \tau_U^{\tor}, \phi_Z^{\tor}, \tau_Z^{\tor}, \rho^{\tor} \mid \cdot)},
\]
where 
\(
p( \phi_U, \tau_U, \phi_Z, \tau_Z, \rho \mid \cdot) 
\)
is proportional to the likelihood of \cref{eq:UZ_normal} based on the current values $\gamma_0^{\tor[r+1]}, \bm \gamma_C^{\tor[r+1]}$ and $\bm U^{\tor[r+1]}$ times the prior distribution for these spatial parameters evaluated at the proposed (numerator) or current (denominator) values.
If $\phi_Z^{prop} > \phi_U^{prop}$, these values do not satisfy the prior constraint, and the proposal will be rejected.

\end{enumerate}

\section{Simulation results on pairs of data}
\label{supp_sec:sims_pairs}

For pairs of observations, we specified the adjacency matrix as block diagonal, where each block was the $2\times 2$ matrix
$
\left(
\begin{smallmatrix}
0 & 1 \\ 1 & 0
\end{smallmatrix}
\right).
$
For the simulations on network data in \cref{sec:sims}, the network has median degree 2, and we set $\tau^2_U = \tau^2_Z = 1$. For the pair data, for which median node degree is equal to 1, we set $\tau^2_U = \tau^2_Z = 2$, in order to ensure similar marginal variability in the exposure and the unmeasured confounder in the network and paired data settings

\begin{table}[p]
\spacingset{1.15}
    \centering
    \vspace{20pt}
    \caption{Simulation results for paired data. Results show the bias, root mean squared error and coverage of 95\% intervals for the local and interference effects based on the OLS estimator and our approach.}
    \resizebox{0.98\textwidth}{!}{%
    \begin{tabular}{*{15}{c}}
        \hline
        \\[-5pt]
       & & \multicolumn{6}{c}{Local effect} & &  \multicolumn{6}{c}{Interference effect} \\
       \cmidrule(lr){3-8} \cmidrule(lr){10-15}
        \multicolumn{2}{c}{True model \&} & \multicolumn{3}{c}{OLS} &  \multicolumn{3}{c}{Our approach} &
        & \multicolumn{3}{c}{OLS} &  \multicolumn{3}{c}{Our approach} \\
        \cmidrule(lr){3-5} \cmidrule(lr){6-8} \cmidrule(lr){10-12} \cmidrule(lr){13-15}  
        \multicolumn{2}{c}{sample size} & Bias & RMSE & Cover & Bias & RMSE & Cover &
        & Bias & RMSE & Cover & Bias & RMSE & Cover \\[10pt]
    
        \hline
        \\[-8pt]
        
        \multirow{4}{*}{\ref{fig:direct}} & & \multicolumn{13}{c}{$\beta_{\bar Z} = 0$ and $\beta_{\bar U}=0$} \\[2pt]
        \cmidrule{7-11}
        & 200 & 0.660 & 0.669 & 0 
        & -0.037 & 0.300 & 95.6 & 
        & 0.151 & 0.172 & 55\phantom{.0} 
        & 0.017 & 0.107 & 93.6 \\
        & 350 & 0.660 & 0.664 & 0
        & -0.019 & 0.236 & 98.1 & 
        & 0.144 & 0.155 & 38\phantom{.0} 
        & 0.009 & 0.077 & 95.4 \\
        & 500 & 0.670 & 0.673 & 0 
        & -0.058 & 0.238 & 94\phantom{.0} & 
        & 0.147 & 0.155 & 17.7 
        & 0.009 & 0.064 & 97.9 \\[3pt]
        \hline
        \\[-8pt]

        \multirow{4}{*}{\ref{fig:general_spatial_conf}} & &
        \multicolumn{13}{c}{$\beta_{\bar Z} = 0$}
        \\[2pt]
        \cmidrule{7-11}
        & 200 & 0.923 & 0.932 & 0 
        & -0.154 & 0.296 & 95.2 & 
        & 0.269 & 0.285 & 21.7 
        & 0.021 & 0.124 & 96.1 \\
        & 350 & 0.920 & 0.925 & 0 
        & -0.158 & 0.260 & 94.9 & 
        & 0.265 & 0.273 & \phantom{0}2.3 
        & 0.008 & 0.096 & 96.5 \\
        & 500 & 0.933 & 0.936 & 0 
        & -0.154 & 0.241 & 92.6 & 
        & 0.270 & 0.276 & \phantom{0}0.7 
        & 0.012 & 0.080 & 96\phantom{.0} \\[3pt]
        \hline
        \\[-8pt]

        \multirow{4}{*}{\ref{fig:interference}} & & 
        \multicolumn{13}{c}{$\beta_{UZ} = 0$ and $\beta_U = \beta_{\bar U} = 0$}
        \\[2pt]
        \cmidrule{7-11}
        & 200 & 0.004 & 0.095 & 96\phantom{.0} 
        & -0.027 & 0.125 & 99.1 & 
        & 0.005 & 0.064 & 94.7 
        & 0.003 & 0.064 & 96.5 \\
        & 350 & -0.003 & 0.069 & 95.7 
        & -0.017 & 0.107 & 99.6 & 
        & -0.004 & 0.047 & 95.3 
        & -0.006 & 0.048 & 98.4 \\
        & 500 & 0.000 & 0.065 & 92.3 
        & -0.030 & 0.108 & 99.6 & 
        & 0.001 & 0.041 & 94.3 
        & 0.003 & 0.043 & 95.1 \\[3pt]
        \hline
        \\[-8pt]

        \multirow{4}{*}{\ref{fig:predictor_interference}} & & 
        \multicolumn{13}{c}{$\beta_U = 0$ and $\beta_{\bar U} = 0$} \\[2pt]
        \cmidrule{7-11}
        & 200 & 0.002 & 0.079 & 95.3 
        & -0.026 & 0.122 & 98.3 & 
        & 0.005 & 0.061 & 94.7 
        & -0.001 & 0.063 & 96.6 \\ 
        & 350 & -0.002 & 0.057 & 95.7 
        & -0.032 & 0.112 & 98\phantom{.0} & 
        & -0.003 & 0.044 & 95.7 
        & -0.011 & 0.052 & 95\phantom{.0} \\ 
        & 500 & 0.000 & 0.054 & 91.7 
        & -0.063 & 0.181 & 88.6 & 
        & 0.001 & 0.039 & 93.7 
        & -0.001 & 0.048 & 92.4 \\[3pt]
        \hline
        \\[-8pt]

        \multirow{4}{*}{\ref{fig:direct_interference}} & &
        \multicolumn{13}{c}{$\beta_{\bar U} = 0$} \\[2pt]
        \cmidrule{7-11}
        & 200 & 0.660 & 0.669 & 0 
        & 0.020 & 0.262 & 96.9 & 
        & 0.151 & 0.172 & 55\phantom{.0} 
        & 0.017 & 0.107 & 94.4 \\
        & 350 & 0.660 & 0.664 & 0 
        & 0.019 & 0.212 & 96.6 & 
        & 0.144 & 0.155 & 38\phantom{.0} 
        & 0.009 & 0.078 & 96.6 \\ 
        & 500 & 0.670 & 0.673 & 0 
        & 0.033 & 0.186 & 96.4 & 
        & 0.147 & 0.155 & 17.7 
        & 0.012 & 0.063 & 96.8 \\[3pt]
        \hline
        \\[-8pt]

        \multirow{3}{*}{\ref{fig:general_interference}} 
        & 200 & 0.923 & 0.932 & 0 
        & -0.111 & 0.266 & 96.9 & 
        & 0.269 & 0.285 & 21.7 
        & 0.014 & 0.124 & 95.7 \\
        & 350 & 0.920 & 0.925 & 0 
        & -0.113 & 0.225 & 95.7 & 
        & 0.265 & 0.273 & \phantom{0}2.3 
        & 0.004 & 0.095 & 96\phantom{.0} \\
        & 500 & 0.933 & 0.936 & 0 
        & -0.107 & 0.199 & 95.3 & 
        & 0.270 & 0.276 & \phantom{0}0.7 
        & 0.009 & 0.078 & 96.7 \\[3pt]
        \hline

        \end{tabular}
     }%
     \label{tab:sims_pairs}
\end{table}

\cref{tab:sims_pairs} shows the simulation results for pairs of data with 100, 175, and 250 pairs of observations (total number of observations 200, 350, and 500). We present bias, root mean squared error and coverage of 95\% intervals for the OLS estimator and for our approach, for the local and the interference effects. These results mirror the results for network data shown in \cref{tab:sims_network}, and the conclusions from the two settings are unaltered.

\section{Additional study information}
\label{supp_sec:application}

\subsection{The data set}

We assemble a data set on power plant emissions and characteristics, population demographics, weather, and information on cardiovascular mortality among the elderly, measured at the level of US counties. We describe the data set here.

We acquire power plant emissions and characteristics for 2004 based on the publicly available data from \cite{papadogeorgou2019adjusting} available here: \url{https://dataverse.harvard.edu/dataset.xhtml?persistentId=doi:10.7910/DVN/DKXXSN}. Power plant information includes the number of power plant units in the facility, whether the plant uses mostly natural gas or coal (an important predictor of SO$_2$ emissions), its total emissions, heat input and operating capacity, whether it has a technology installed for oxides of nitrogen control, and whether the plant participated in Phase II of the Acid Rain Program. Our data set includes 906 power plant facilities in 596 counties. We aggregate power plant information at the county level, and define the total SO$_2$ emissions from all power plants in the county as the exposure of interest.

We acquire health information from the United States Centers for Disease Control and Prevention (CDC) WONDER query system. We consider deaths due to the diseases of the circulatory system (I-00 to I-99 codes) among population aged 65 years or older, and define the outcome of interest as the number of deaths per 100,000 residents in 2005.

We considered demographic information as potential confounders. Specifically, we consider population characteristics such as percentages of urbanicity, of white and Hispanic population, of population with at least a high school diploma, of population that lives below the poverty limit, of female population, of population having lived in the area for less than 5 years, of housing units that are occupied, and population per square mile from the 2000 Census, and also county-level smoking rates acquired using the CDC Behavioral Risk Factor Surveillance System data.

\cng{We merge the county-specific data sets on power plant information, cardiovascular mortality, and demographic information, and we drop counties with missing health or demographic information. This results in 2,720 counties, with 586 counties with power plants.}


We downloaded county level weather data for 2004 from the National Oceanic and Atmospheric Administration's (NOAA) data base, available at \url{ftp://ftp.ncdc.noaa.gov//pub/data/cirs/climdiv/}. Specifically, we acquired data for each county describing the maximum, minimum and average temperature, and total precipitation for each month in 2004. We aggregated the data across the twelves months by considering the total yearly precipitation, the second most extreme of the monthly maximum and minimum temperatures, the average maximum and minimum temperatures, and the average, maximum and minimum of the average monthly temperatures. After examining the correlation matrix, we deduced that many covariates were highly correlated, and used only the three mentioned above (total precipitation, second maximum and minimum temperatures).

\cng{To define the neighborhood exposure and the spatial structure in the matrices $G, H$ in \cref{ass:UZ_normal}, we need to specify appropriate adjacency matrices. To do so, we calculate the matrix, $\Delta$, of pairwise county distances. The $(i,j)$ entry of $\Delta$, $\delta_{ij}$, represents the minimum distance between the spatial polygons of counties $i$ and $j$. We consider adjacency matrices $A^*$ whose off-diagonal elements $a_{ij}^*$ are of the form $a_{ij}^* = I(\delta_{ij} \leq \delta^\text{max})$ for some cutoff value $\delta^\text{max}$, and diagonal elements equal to 0.
We use $\delta^{\text{max}, \overline Z}$, $\delta^{\text{max}, G}$ and $\delta^{\text{max}, H}$ to represent the three values, respectively, which we vary.}

\cng{We maintain counties with local or neighborhood exposure (or both). A larger value of $\delta^{\text{max}, \overline Z}$ in the definition of neighborhood exposure leads to a larger number of counties with available exposure. We drop counties with local exposure that have no neighbors based on $\delta^{\text{max}, H}$ with local exposure. We do so to ensure that the CAR matrix $H$ is non-singular, since otherwise the exposure at this location would not be included in the variable's spatial structure. Lastly, we maintain counties that have at least one neighbor based on $\delta^{\text{max}, \overline Z}$ so that their value of $\overline U$ is well-defined\footnote{We note here that these counties do not necessarily have neighborhood exposure since to have a neighborhood exposure one needs to have a neighbor based on $\delta^{\text{max}, \overline Z}$ who also has local exposure.}, and that they have at least one neighbor based on $\delta^{\text{max}, G}$ to avoid singularities of the matrix $G$.}

\begin{figure}[!ht]
\centering
\includegraphics[width = 0.45\textwidth]{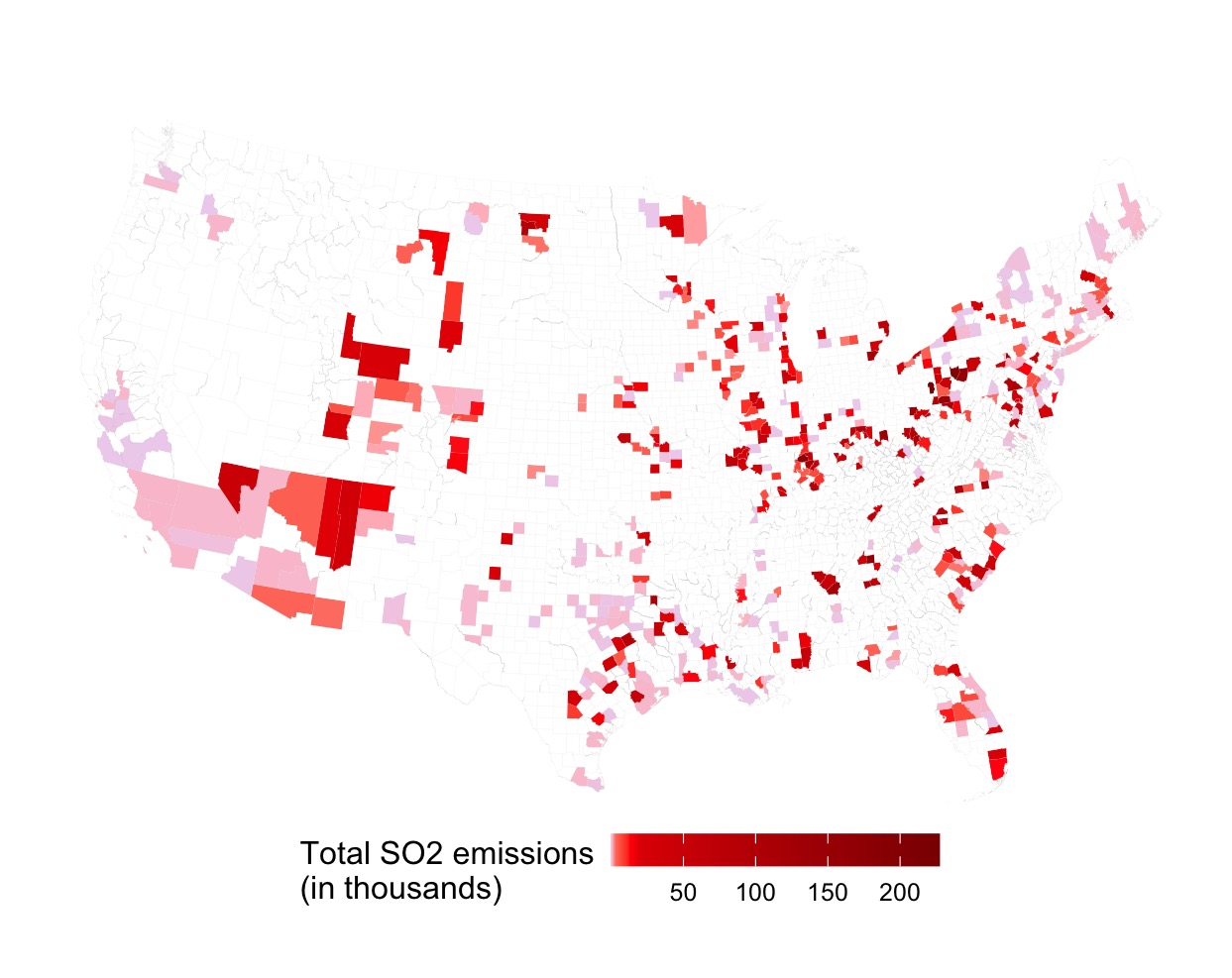}
\vspace{-3pt}
\caption{County-level local exposure.}
\label{app_fig:data}
\end{figure}

\begin{figure}[!ht]
\centering
\includegraphics[width = 0.45\textwidth]{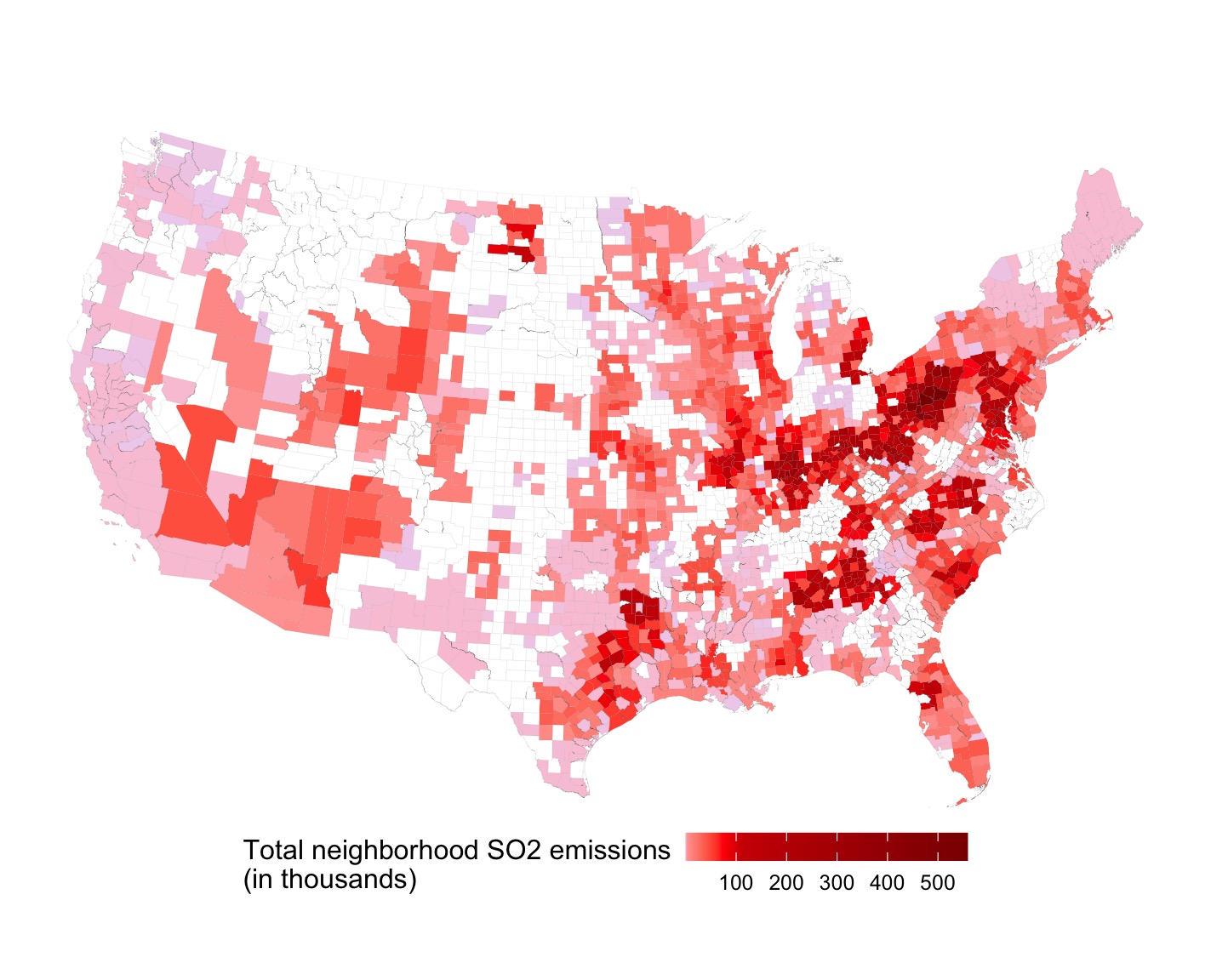}
\includegraphics[width = 0.45\textwidth]{figures/app_neighborhood_exposure.jpeg}\\
\includegraphics[width = 0.45\textwidth] {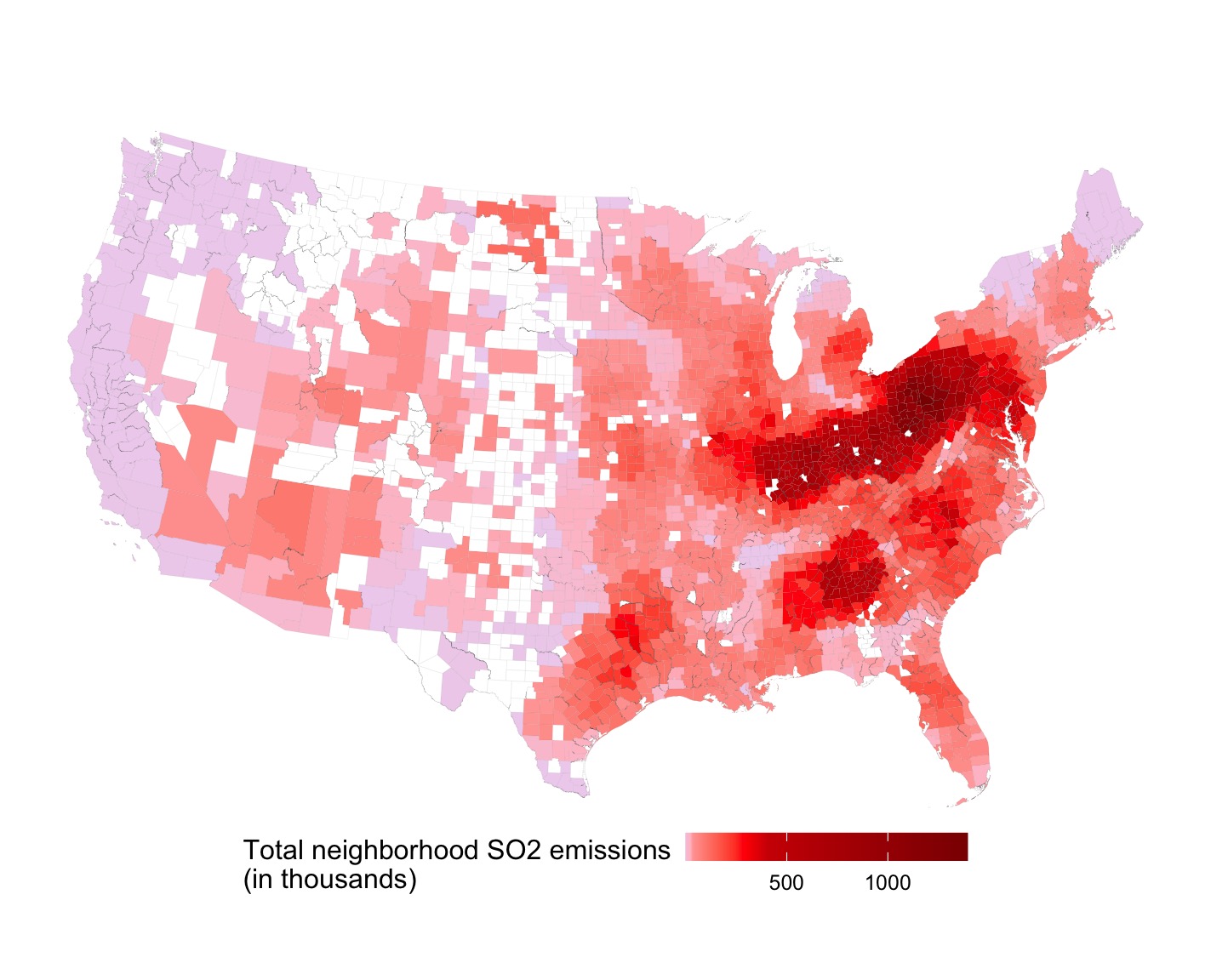}
\includegraphics[width = 0.45\textwidth]{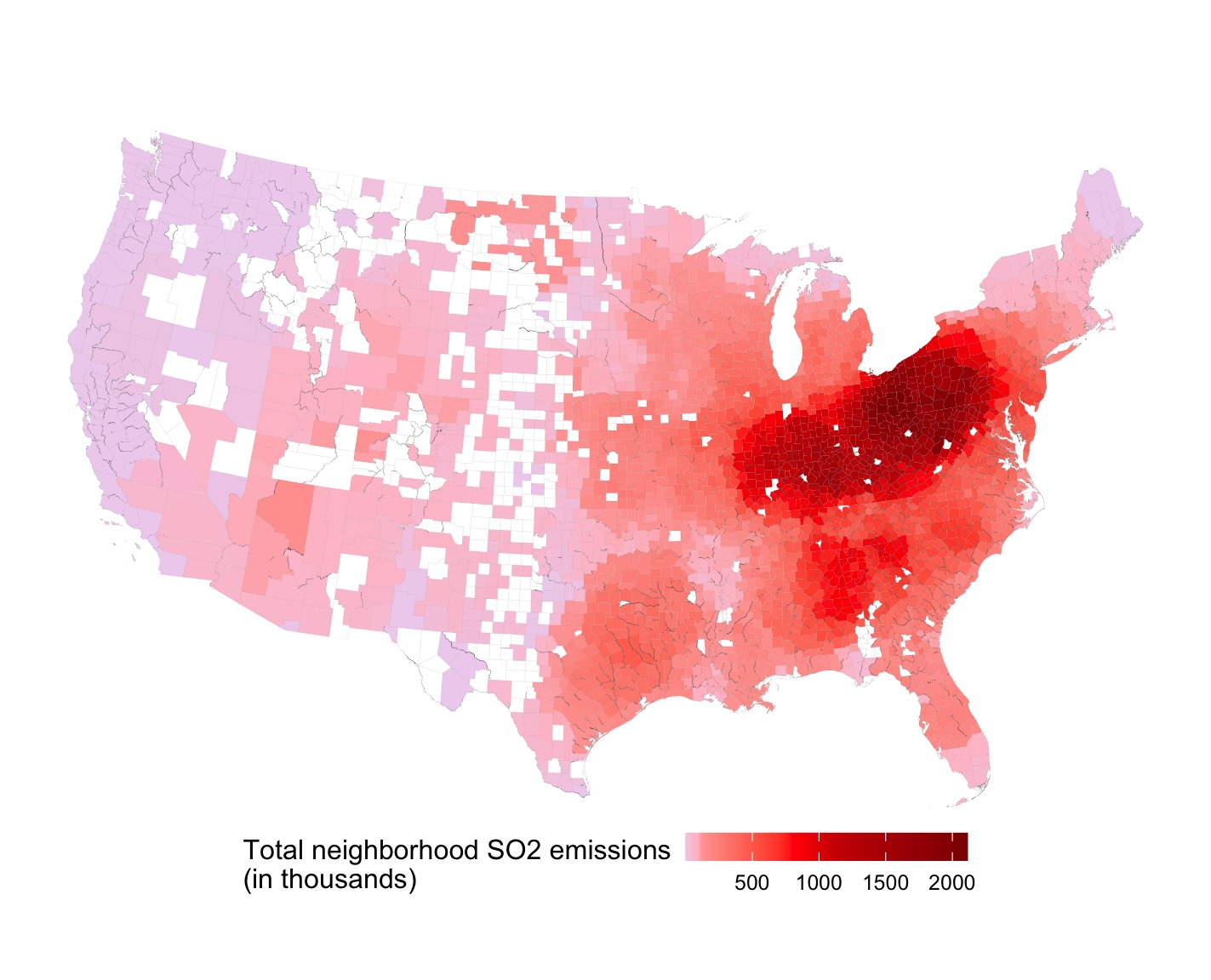}
\caption{Neighborhood exposure based on different values of $\delta^{\text{max}, \overline Z}$. Top Left: 20 kilometers, Top Right: 50 kilometers (used in the main analysis), Bottom Left: 100 kilometers, Bottom Right: 200 kilometers.}
\label{app_fig:app_neighborhood_exposure_all}
\end{figure}

\cng{In \cref{app_tab:app_sample_size} we present the resulting sample sizes based on the different choices of $\delta^\text{max}$ we considered. \cref{app_fig:data} shows the local exposure. \cref{app_fig:app_neighborhood_exposure_all} shows the county-level neighborhood exposure based on different choices of $\delta^{\text{max}, \overline Z} \in \{20, 50, 100, 200\}$ kilometers.}

\begin{table}[!t]
\caption{Sample size after data cleaning and based on different specification of the neighborhood structure in the neighborhood exposure and the spatial structure in the unmeasured spatial variable and the exposure. First 3 columns: maximum distance (in kilometers) that two counties are adjacent. Next 5 columns: starting sample size, sample size after restricting to counties with local or neighborhood exposure, sample size after removing counties without weather information, sample size after removing counties that lead to singularities in the covariance matrix for the exposure, and then the covariance matrix for the spatial variable. Last 3 columns: Number of counties with both exposures, with local exposure only, and with neighborhood exposure only. The row in blue corresponds to the analysis in the main manuscript.}
\label{app_tab:app_sample_size}
\centering
\footnotesize
\begin{tabular}{ccc| c c c c c | ccc}
\toprule
\multirow{3}{*}{$\delta^{\text{max},\overline Z}$} &
\multirow{3}{*}{$\delta^{\text{max},G}$} &
\multirow{3}{*}{$\delta^{\text{max},\overline H}$} &
\multirow{3}{*}{\shortstack[c]{Starting\\Sample\\Size}} &
\multirow{3}{*}{\shortstack[c]{With\\exposure}} &
\multirow{3}{*}{\shortstack[c]{With\\weather}} &
\multirow{3}{*}{\shortstack[c]{Non-\\singular\\H}} &
\multirow{3}{*}{\shortstack[c]{With $\overline U$,\\non-\\singular\\G}} &
\multirow{3}{*}{\shortstack[c]{With\\$Z, \overline Z$}} &
\multirow{3}{*}{\shortstack[c]{With $Z$\\only}} &
\multirow{3}{*}{\shortstack[c]{With $\overline Z$\\only}} \\
& & & & & & & & & & \\ \\[-3pt]
\midrule
20  & 50  & 20  & 2720 & 2154 & 2153 & 2067 & 2065 & 476 & 0  & 1589 \\
\textcolor{blue}{50}  & \textcolor{blue}{50}  &
\textcolor{blue}{20}  & \textcolor{blue}{2720} &
\textcolor{blue}{2527} & \textcolor{blue}{2526} &
\textcolor{blue}{2440} & \textcolor{blue}{2440} &
\textcolor{blue}{476} & \textcolor{blue}{0}  &
\textcolor{blue}{1964} \\
100 & 50  & 20  & 2720 & 2690 & 2689 & 2603 & 2602 & 476 & 0  & 2126 \\
200 & 50  & 20  & 2720 & 2715 & 2714 & 2628 & 2628 & 476 & 0  & 2152 \\
\addlinespace
20  & 100 & 20  & 2720 & 2154 & 2153 & 2067 & 2065 & 476 & 0  & 1589 \\
50  & 100 & 20  & 2720 & 2527 & 2526 & 2440 & 2440 & 476 & 0  & 1964 \\
100 & 100 & 20  & 2720 & 2690 & 2689 & 2603 & 2603 & 476 & 0  & 2127 \\
200 & 100 & 20  & 2720 & 2715 & 2714 & 2628 & 2628 & 476 & 0  & 2152 \\
\addlinespace
20  & 50  & 50  & 2720 & 2154 & 2153 & 2128 & 2127 & 476 & 61 & 1590 \\
50  & 50  & 50  & 2720 & 2527 & 2526 & 2501 & 2501 & 537 & 0  & 1964 \\
100 & 50  & 50  & 2720 & 2690 & 2689 & 2664 & 2663 & 537 & 0  & 2126 \\
200 & 50  & 50  & 2720 & 2715 & 2714 & 2689 & 2689 & 537 & 0  & 2152 \\
\addlinespace
20  & 100 & 50  & 2720 & 2154 & 2153 & 2128 & 2127 & 476 & 61 & 1590 \\
50  & 100 & 50  & 2720 & 2527 & 2526 & 2501 & 2501 & 537 & 0  & 1964 \\
100 & 100 & 50  & 2720 & 2690 & 2689 & 2664 & 2664 & 537 & 0  & 2127 \\
200 & 100 & 50  & 2720 & 2715 & 2714 & 2689 & 2689 & 537 & 0  & 2152 \\
\addlinespace
20  & 100 & 100 & 2720 & 2154 & 2153 & 2148 & 2147 & 476 & 81 & 1590 \\
50  & 100 & 100 & 2720 & 2527 & 2526 & 2521 & 2521 & 537 & 20 & 1964 \\
100 & 100 & 100 & 2720 & 2690 & 2689 & 2684 & 2684 & 557 & 0  & 2127 \\
200 & 100 & 100 & 2720 & 2715 & 2714 & 2709 & 2709 & 557 & 0  & 2152 \\
\bottomrule
\end{tabular}
\end{table}

\subsection{Semiparametric formulation using penalized B-splines}

\cng{To reduce the sensitivity of our estimator to the parametric specification of the outcome model, we replace the linear terms corresponding to the local and neighborhood exposures with basis functions of splines. Specifically, the relationship between the local exposure and the outcome was represented using a B-spline basis of degree 3 with 20 degrees of freedom. We modeled the relationship between the neighborhood exposure and the outcome in the same manner.}

\cng{To control smoothness, we imposed a second-order difference penalty on the spline coefficients. This penalty approximates the integrated squared second derivative and discourages overly wiggly fits. An inverse-Gamma prior with parameters (2, 1) was adopted on the spline coefficients, separately for the local and neighborhood exposures.}

\subsection{Model diagnostics}

\begin{figure}[!ht]
    \centering
    \includegraphics[width=0.8\linewidth]{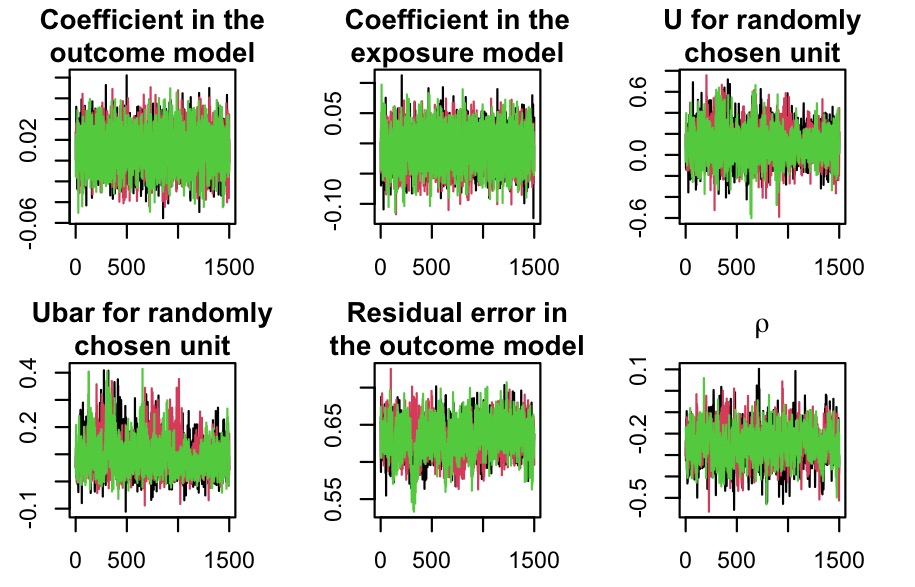}
    \caption{Evaluating MCMC convergence using traceplots. We include a randomly chosen coefficient in the outcome and exposure model, the value of $U$ and $\overline U$ for a randomly chosen unit, and the correlation parameter between the unmeasured spatial variable and the exposure $\rho$.}
    \label{app_fig:app_traceplots}
\end{figure}

\cng{We first investigated convergence of our Bayesian procedure by investigating traceplots of all model parameters across the three chains. A subset of those traceplots for our main analysis is shown in \cref{app_fig:app_traceplots}. We find no obvious signs of lack of convergence. Similar-looking traceplots are acquired as well across our sensitivity analyses (discussed in Section~\ref{app_subsec:sensitivity}). Since we perform a large number of alternative analyses, we refrain from including traceplots for all of them here.}

\begin{figure}[!ht]
    \centering
    \includegraphics[width=0.7\linewidth]{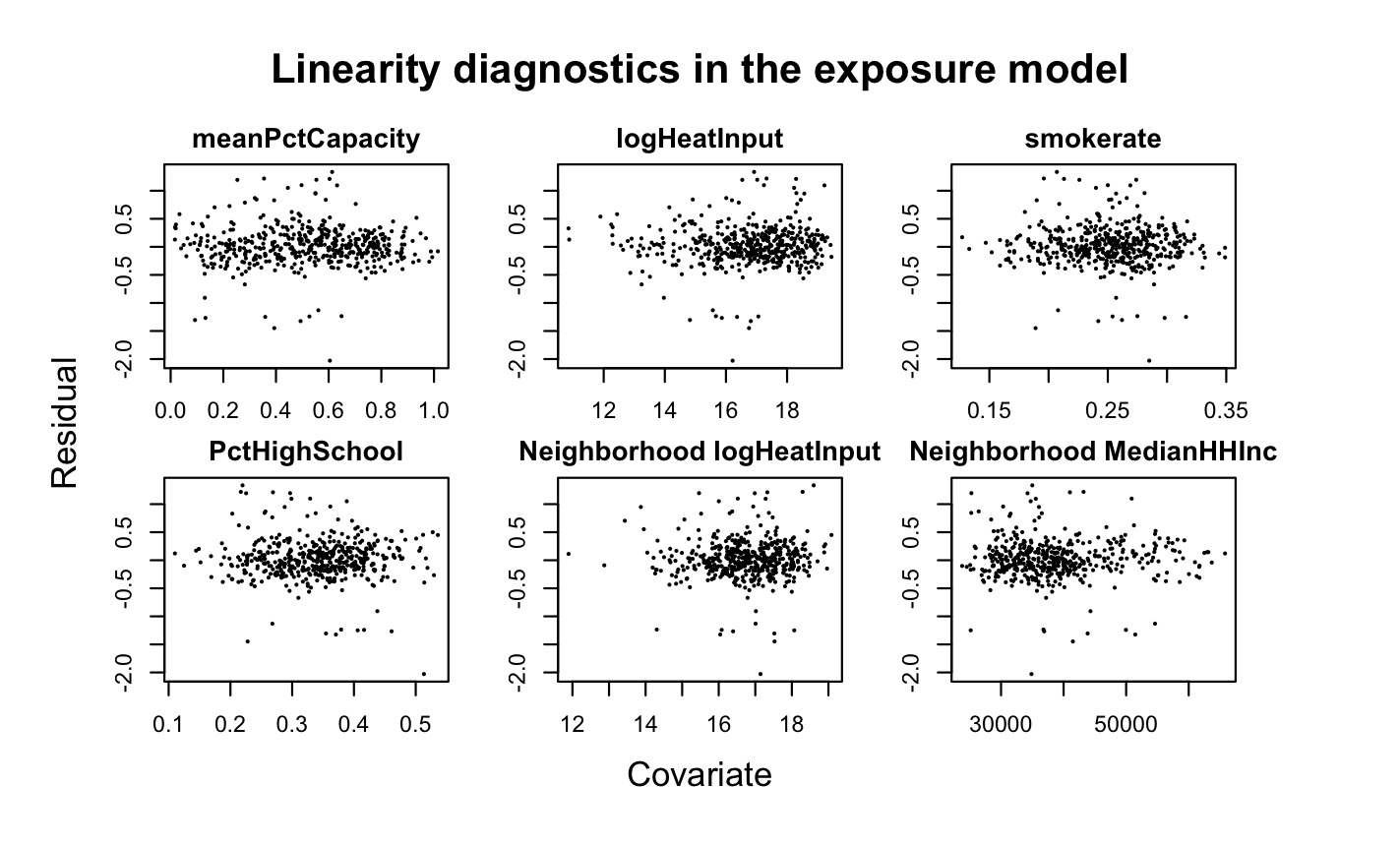}
    \includegraphics[width=0.7\linewidth]{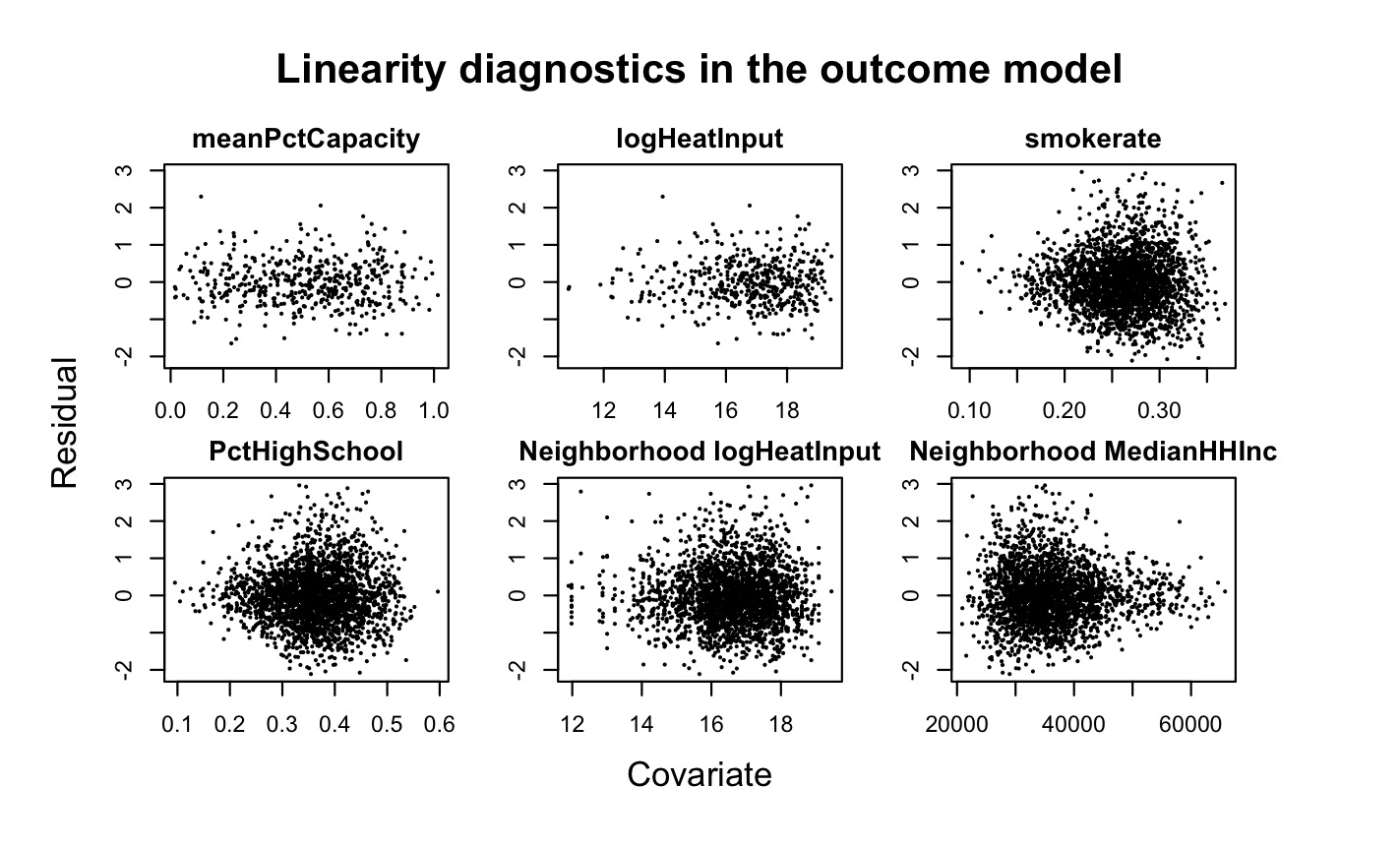}
    \caption{Plotting the residuals of the exposure (top) and outcome (bottom) model against a subset of the measured covariates to investigate potential violations of the linearity assumption for the corresponding model.}
    \label{app_fig:app_linearity}
\end{figure}

\cng{We also evaluated potential violations of the linear relationship specified between the measured covariates and the exposure and the outcome. We did so by plotting the residuals of the corresponding model against all the covariates. A subset of these plots for both the exposure and the outcome model for a randomly selected set of continuous covariates is shown in \cref{app_fig:app_linearity}. We find no obvious violation of linearity. Similarly-looking plots are acquired across our sensitivity analyses as well, which are not shown here.}

\cng{Lastly, we evaluated the normality assumption on the residuals of the exposure and the outcome model using Q-Q plots. First, we discuss the residuals of the exposure model. Since the exposure is allowed to be inherently statistically dependent, the residuals of the exposure model are correlated. To draw the Q-Q plot, we first ``un-correlate'' the residuals in the following manner. Let $\widehat \epsilon_i^{(r)}$ denote the residual of the $i^{th}$ unit in the $r^{th}$ MCMC iteration, and $\widehat{\bm \epsilon}^{(r)}$ the vector of $\widehat \epsilon_i^{(r)}$ across $i$. Let also $\widehat V^{(r)}$ denote the (marginal) covariance matrix of $\bm Z$ at the $r^{th}$ iteration from its distribution in \cref{ass:UZ_normal}. We consider the standardized residuals defined as
\[
\widehat{\bm \epsilon}^{(r)}_\text{stand} = \left( V^{(r)} \right)^{-1/2} \widehat{\bm \epsilon}^{(r)},
\]
and calculate their posterior mean.}

\begin{figure}[!ht]
    \centering
    \includegraphics[width=0.8\linewidth]{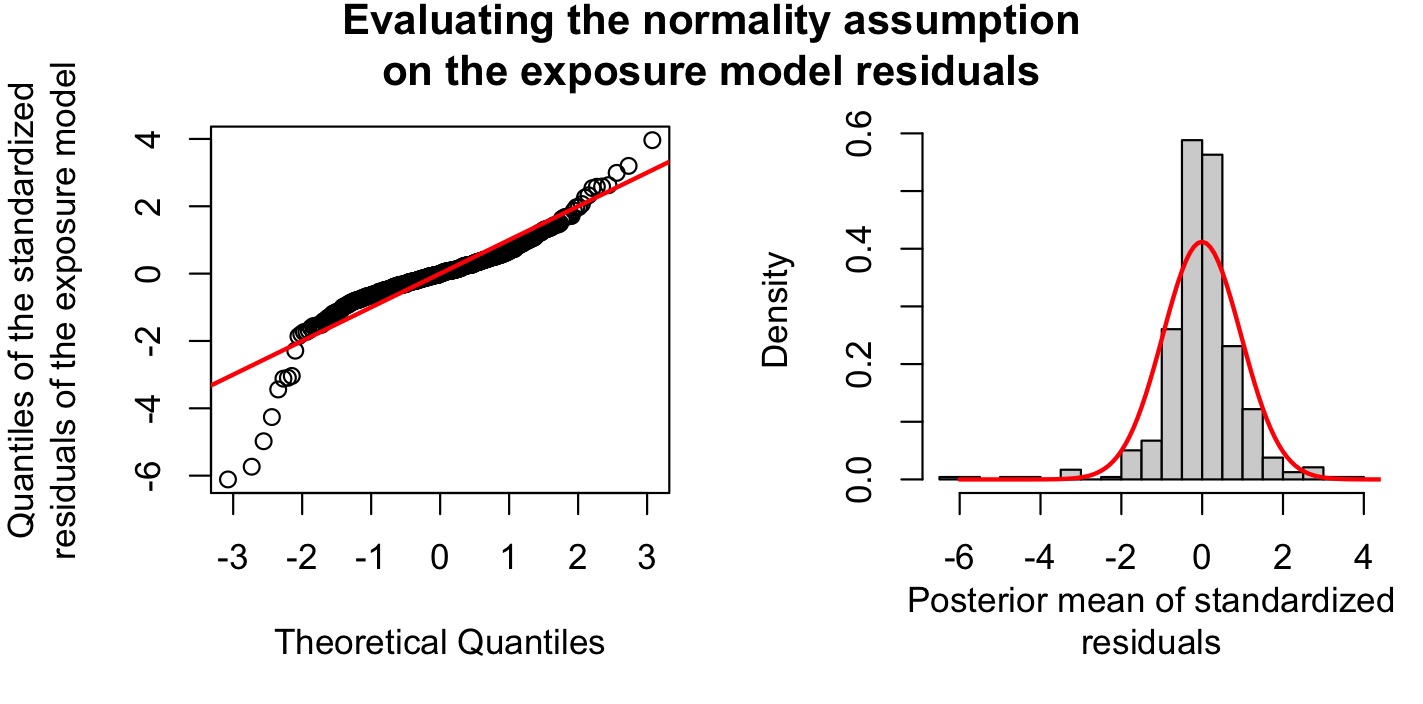}
    \caption{Q-Q plot and histogram of the posterior mean for the standardized exposure model residuals. The red line shows the 45 degree line (left) and the density of the normal distribution with mean and standard deviation that of the standardized residuals (right).}
    \label{app_fig:QQ_exp}
\end{figure}

\cng{\cref{app_fig:QQ_exp} shows the Q-Q plot against the 45 degree line and the histogram of the residuals. There are some extreme values along the tails, but overall the Q-Q plot looks ok. The histogram matches closely the density of the normal distribution. We believe that these results indicate no strong violation of the normality assumption for the exposure residuals.}

\cng{Similar investigations across all sensitivity analyses returned similar conclusions on the normality assumption on the exposure variable. However, we find that the Q-Q plot for the exposure model residuals was slightly closer to the diagonal when we used 20 kilometers as the cutoff in the adjacency matrix for $H$. For that reason, we used this value in the analysis in the main manuscript. We also evaluated posing the normality distribution in \cref{ass:UZ_normal} directly on the local emission variable, instead of log-transforming it. The Q-Q plots for the exposure model residuals from such analyses showed violations of the normality assumption, and for this reason these analyses were not pursued further.}

\begin{figure}[!ht]
    \centering
    \includegraphics[width=0.8\linewidth]{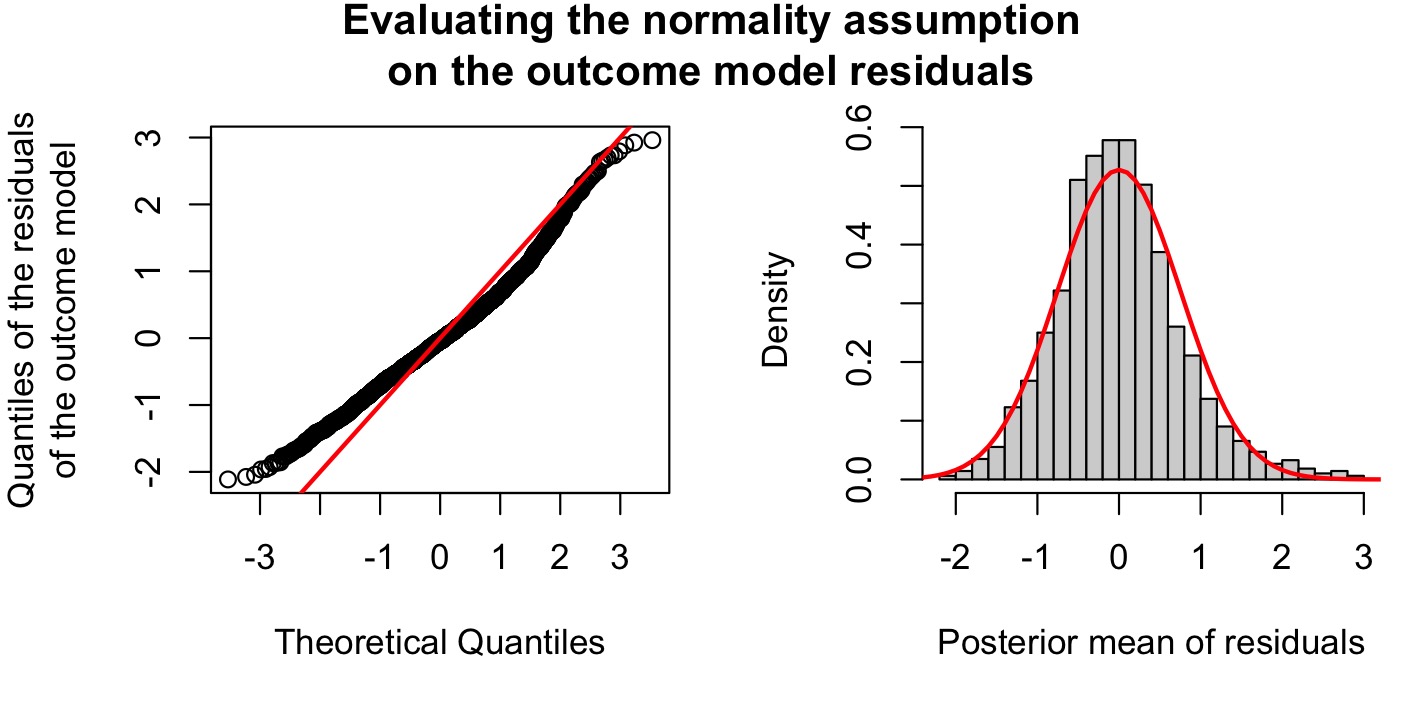}
    \caption{Q-Q plot and histogram of the posterior mean for the outcome model residuals. The red line shows the 45 degree line (left) and the density of the normal distribution with mean the mean and standard deviation that of the standardized residuals (right).}
    \label{app_fig:QQ_out}
\end{figure}

\cng{Lastly, we performed a similar investigation of the normality assumption on the outcome model residuals. \cref{app_fig:QQ_out} shows the results. Even though the histogram seems to match the density of the normal distribution relatively fine, the Q-Q plot indicates violations of the normality assumption. This might be due to spatial correlation in the outcome model residuals due to inherent statistical dependence. The complications in spatial causal inference due to inherent dependence in the outcome model is an interesting topic of future work. The ``flat'' Q-Q plot might also be because we analyze the crude cardiovascular mortality rate as a continuous outcome. An extension to non-linear models would provide further insights.}

\subsection{Sensitivity analysis}
\label{app_subsec:sensitivity}

\cng{We perform a number of sensitivity analyses to evaluate the robustness of estimated quantities to different model specifications.}

\subsubsection{The choice of neighborhood structure}

\cng{First of all, we evaluate the robustness of estimated quantities to the specification of the adjacency matrices in
\begin{enumerate*}[label=(\alph*)]
    \item the definition of neighborhood exposure,
    \item the CAR structure on the unmeasured spatial variable in $G$, and 
    \item the CAR structure on the exposure model residuals in $H$.
\end{enumerate*}}

\begin{figure}[p]
    \centering
    \includegraphics[width=0.8\linewidth, trim = 0 95 0 0 , clip]{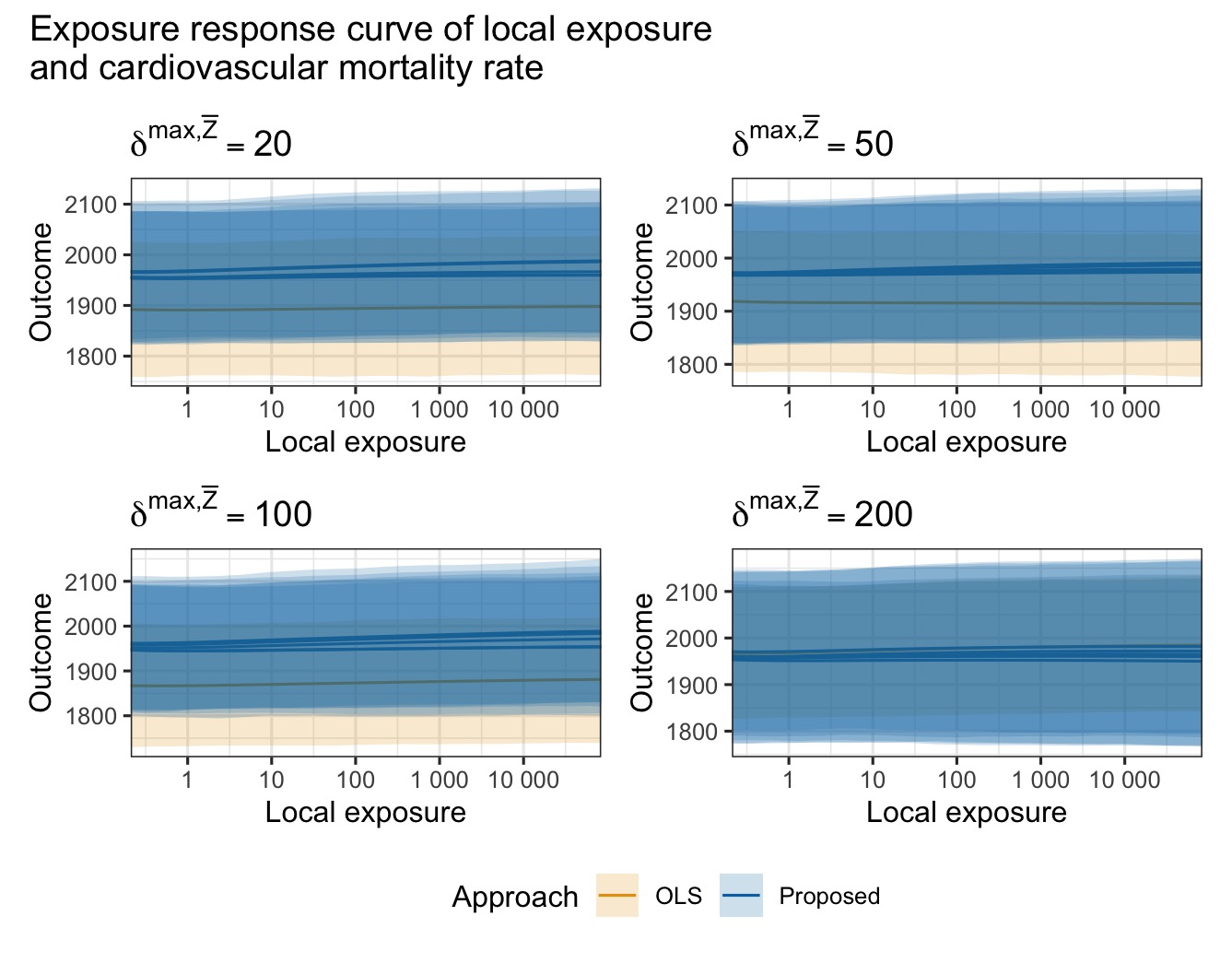} \\
    \includegraphics[width=0.8\linewidth]{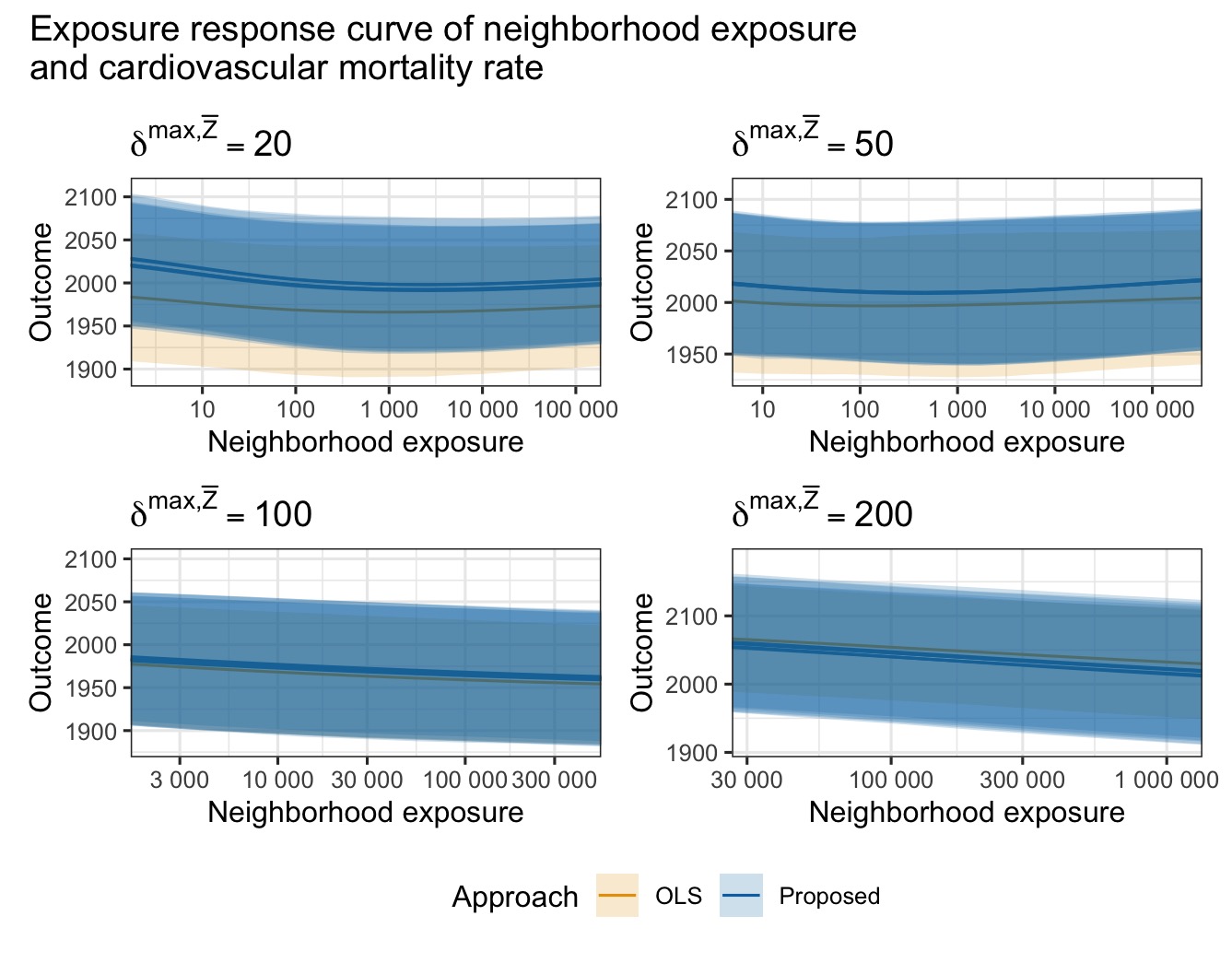}
    \caption{Exposure response curve estimates for local (top) and neighborhood (bottom) emissions on cardiovascular mortality rate. The different panels correspond to different $\delta^{\text{max}, \overline Z}$ values for the adjacency matrix used in the neighborhood exposure. The colors correspond to the proposed approach (blue) and OLS (orange). The multiple blue lines per panel correspond to different choices of $\delta^{\text{max}, G}$ and $\delta^{\text{max}, H}$.}
    \label{app_fig:sensitivity_A}
\end{figure}

\cng{We show the estimated exposure-response curves in \cref{app_fig:sensitivity_A}. The different lines in the same panel correspond to different choices of the neighborhood structure in the definition of $G$ and $H$, which, as we see, has no impact on effect estimates. (For OLS, the choice of $G$ and $H$ does not matter, as the exposure model is not incorporated, hence the figures show one line for OLS per panel.) The different panels correspond to different neighborhood structure in the definition of neighborhood exposure. Any differences are minimal, and the qualitative results are unchanged. Results appear not-statistically significant, and the results from the proposed approach and OLS are overlapping.}

\begin{figure}[!ht]
    \centering
    \includegraphics[width=0.8\linewidth]{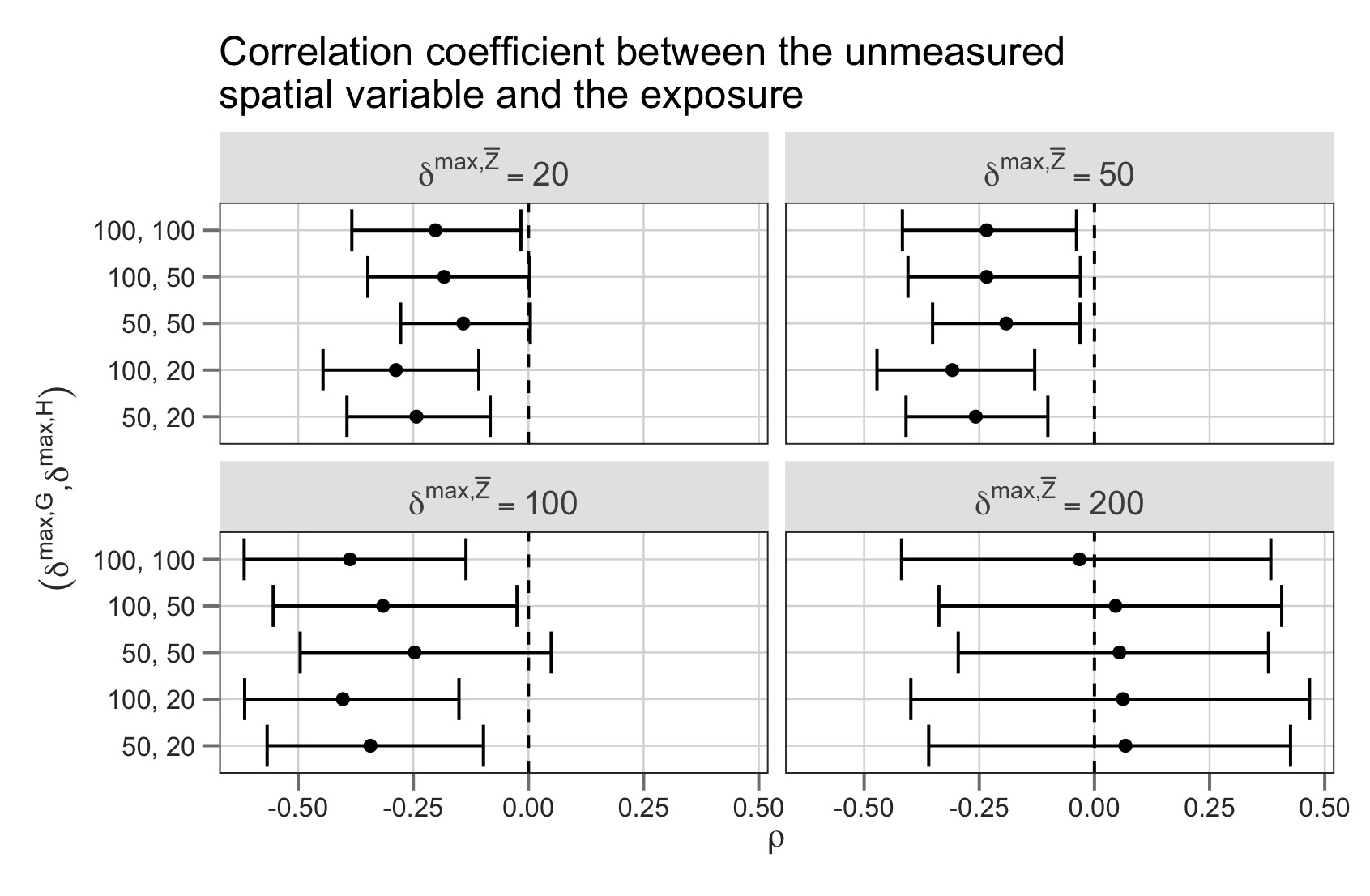}
    \caption{Estimates and 95\% credible intervals for the correlation coefficient of the unmeasured spatial variable and the exposure when we vary the neighborhood structure in the definition of neighborhood exposure (different panels) and the definition of spatial structure in the unmeasured spatial variable and the exposure (y-axis).}
    \label{app_fig:sensitivity_A_rho}
\end{figure}

\cng{Lastly in \cref{app_fig:sensitivity_A_rho} we show the estimated value and 95\% credible interval for $\rho$, the correlation between the local value of the unmeasured spatial variable and the exposure. We see that the values used for the neighborhood structure have minimal impact the value of $\rho$, except for long-range neighborhood exposure $(\delta^{\text{max}, \overline Z} = 200)$. We also find that for larger values of $\delta^{\text{max}, \overline Z}$, the credible intervals for $\rho$ are wider.}  

\subsubsection{Including the log-transformed exposure directly in the outcome model}

\cng{Using the semiparametric model based on the B-splines reduces the sensitivity of estimates on parametric specifications. However, it is more difficult to test the presence of an effect of the local or neighborhood exposure in that setting. We considered an alternative analysis that includes the log-transformed emissions directly in the outcome model, and uses the estimated corresponding coefficients as estimates of the local and interference effects for an increase in the log-exposure by one standard deviation.}

\begin{figure}[!ht]
    \centering
    \includegraphics[width=\linewidth]{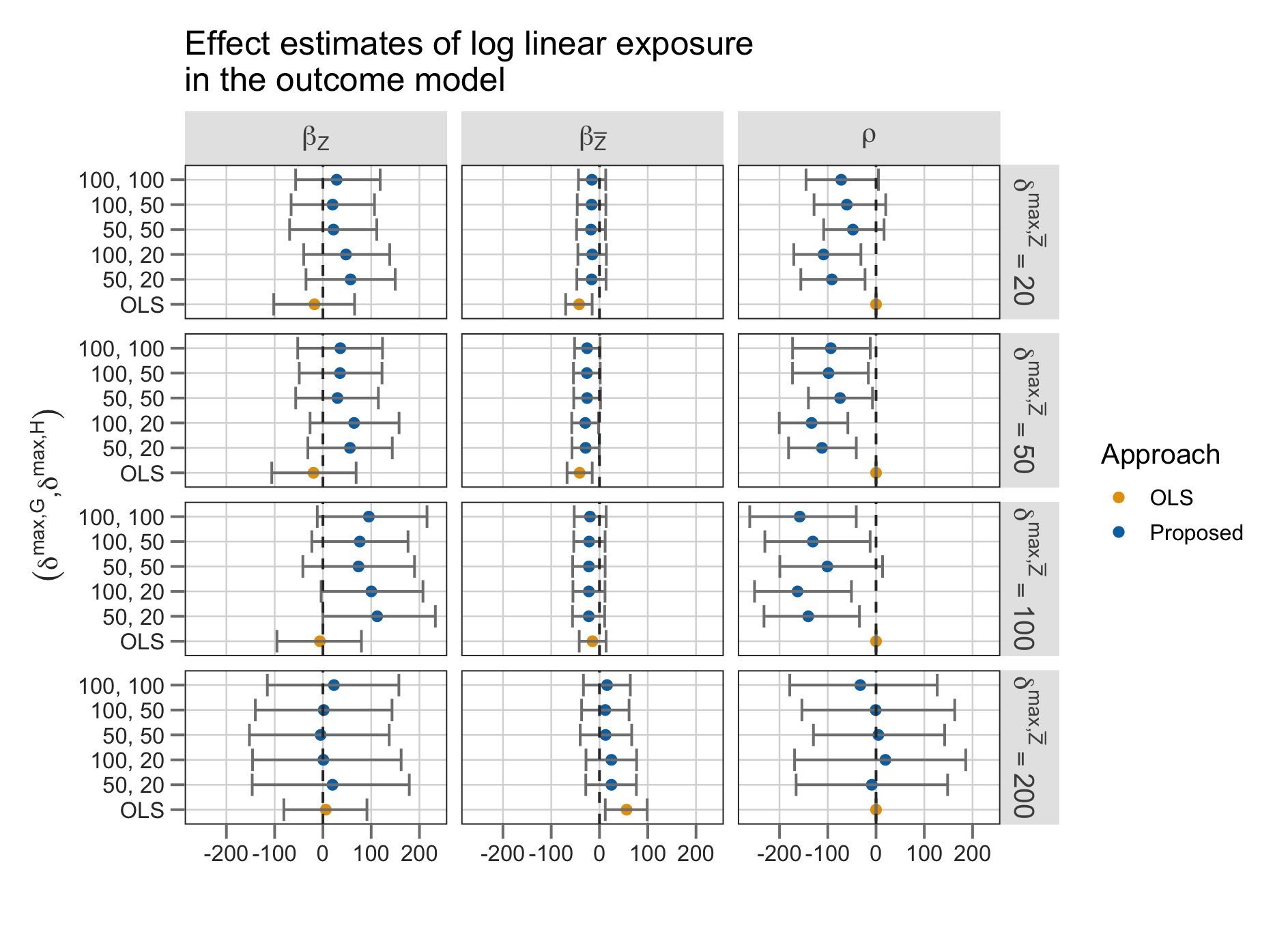}
    \caption{Local and interference effect estimates for a one standard deviation increase in the log-transformed local and neighborhood emissions on cardiovascular mortality rate. We show results base on different definitions of the neighborhood structure for the neighborhood exposure (different panels along the rows) and the definition of spatial structure in the unmeasured spatial variable and the exposure (y-axis).}
    \label{app_fig:sensitivity_log_exposure}
\end{figure}

\cng{The results are shown in \cref{app_fig:sensitivity_log_exposure} where we consider all values of $\delta^{\text{max}, \bar Z}, \delta^{\text{max}, G}$ and $\delta^{\text{max}, H}$, and both the proposed approach and OLS regression. First, we find the same pattern for the correlation coefficient $\rho$, which is consistently negative in all settings, except when $\delta^{\text{max}, \bar Z} = 200$ that is close to 0. For the local effects, almost all analyses considered return estimates with credible intervals that include 0. When $\delta^{\text{max}, \bar Z} < 200$, effect estimates from the proposed approach are always towards the positives reflecting an impact of SO$_2$ emissions on cardiovascular mortality, and indicating that {\it some} unmeasured spatial confounding might be present. However, we cannot rule out that this is an artifact of the parametric specification. For the neighborhood effect, even though OLS returns statistically significant negative estimates in some cases, the proposed approach returns credible intervals that overlap 0, which is more realistic.}

\subsubsection{Excluding weather variables}

\cng{We evaluate the sensitivity of our estimated exposure-response curves to the exclusion of weather variables. Since these variables are expected to be spatial, our methodology should be able to capture them to some extent, and return estimates that are similar with and without the local and neighborhood weather variables explicitly included.}

\begin{figure}[!ht]
\centering
\includegraphics[width = 0.75\textwidth, trim = 20 50 20 10, clip]{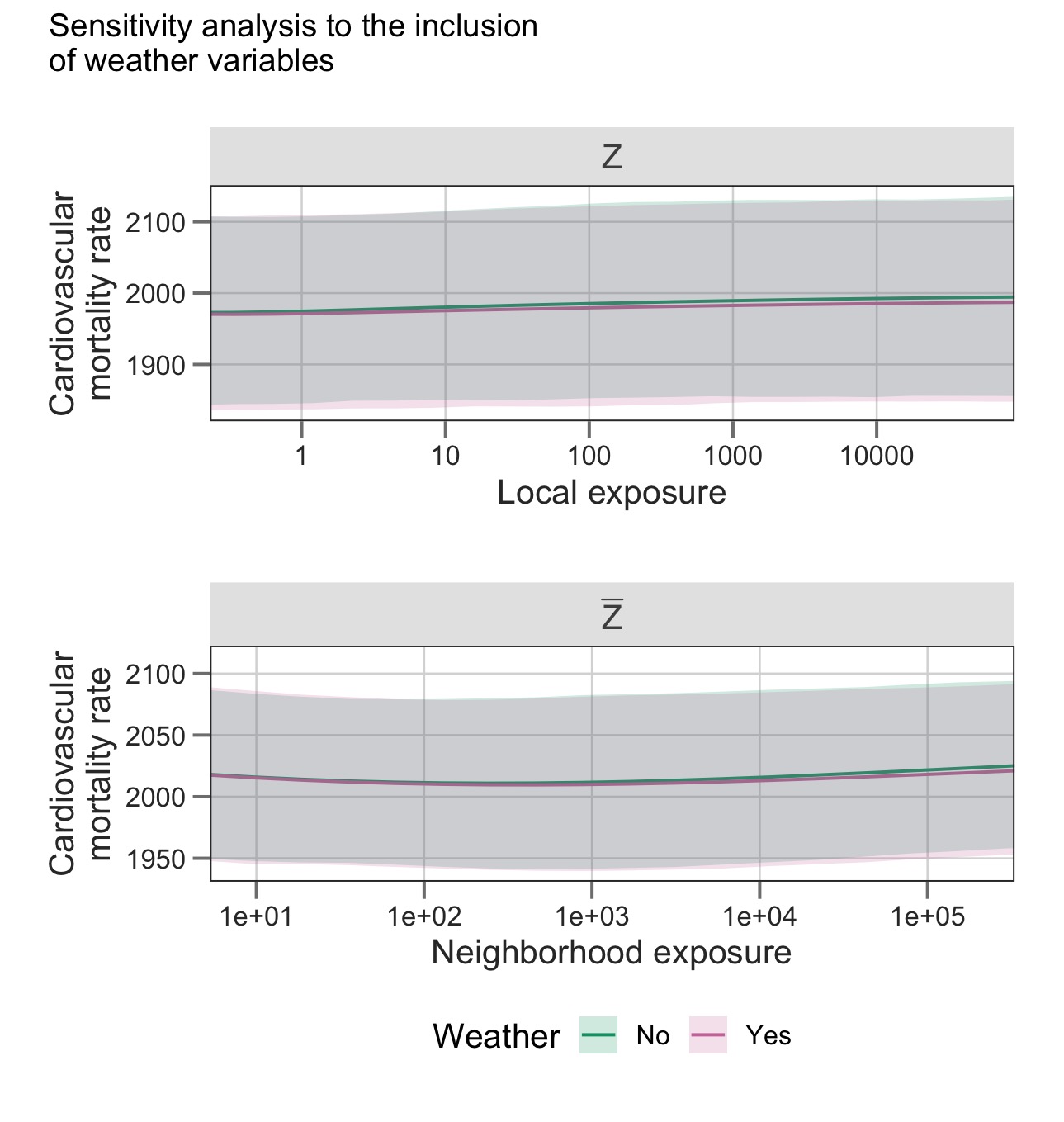}
\caption{Exposure-response curve for local and neighborhood SO$_2$ emissions on cardiovascular mortality rate including and excluding local and neighborhood weather variables from the adjustment set.}
\label{app_fig:sensitivity_weather}
\end{figure}

\cng{\cref{app_fig:sensitivity_weather} shows the estimated curve for local and neighborhood exposure, with and without explicit adjustment for the local and neighborhood values of weather variables from the analysis that uses B-splines for the exposure variables. We find that estimated quantities and uncertainty bands are essentially indistinguishable. We found that, similarly, the analysis that uses the log-transformed exposure directly in the outcome model shows no sensitivity to the inclusion or exclusion of the measured weather variables.}

\subsubsection{Choice of hyperparameters for the prior distributions}

\cng{Lastly, we evaluated the sensitivity of our results to the choice of hyperparameters. We considered two additional sets of hyperparameters that represent moderately and strongly more uninformative prior distributions (compared to the standard values in \cref{subsec:priors}). Specifically, we introduced two scaling parameters $\kappa_\beta$, and $\kappa_{\sigma^2}$ which scale the prior variance of the coefficients and the outcome model residual variance, respectively. In our sensitivity analysis, we specify independent normal prior distributions with variance $\kappa_\beta \sigma^2_\text{prior}$ for all regression coefficients, and a normal prior distribution with variance $\kappa_\beta \sigma^2_{\text{prior}, \overline U}$ for the coefficient of $\overline U$. The values $\sigma^2_\text{prior}$ and $\sigma^2_{\text{prior}, \overline U}$ are equal to $2$ and $0.35^2$, respectively. We also specify that $\sigma^2_Y \sim IG(2 \kappa_{\sigma^2} + 1, \kappa_{\sigma^2} \widetilde \sigma^2_Y / 2)$. Therefore, $\kappa_\beta = \kappa_{\sigma^2} = 1$ represents the case of the recommended hyperparameters, and larger values of $\kappa_\beta$ and smaller values of $\kappa_{\sigma^2}$ imply a less informative prior distribution.}

\begin{figure}[!ht]
\centering
\includegraphics[width = 0.72\textwidth, trim = 20 50 20 10, clip]{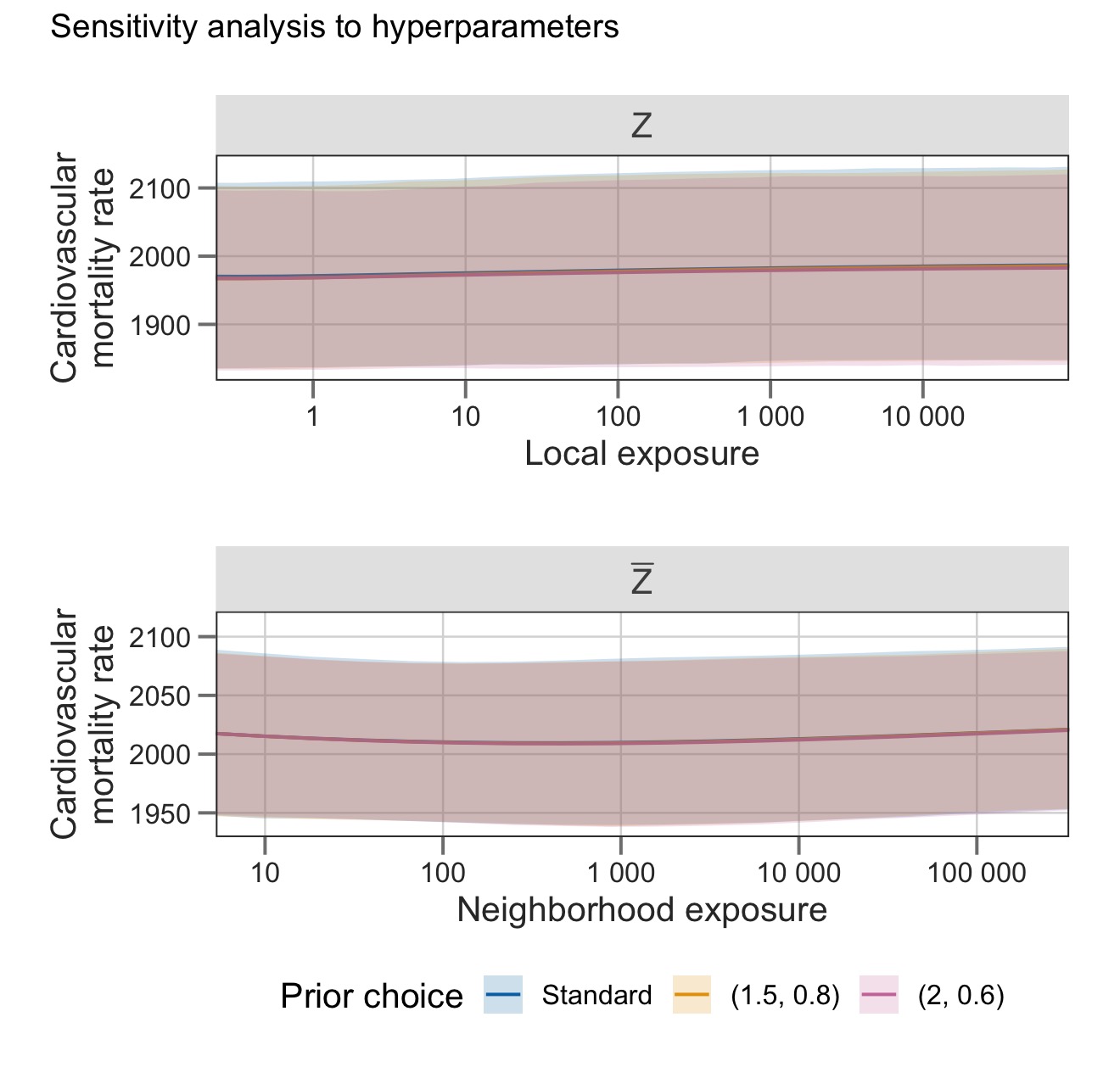}
\caption{Exposure-response curve for local and neighborhood SO$_2$ emissions on cardiovascular mortality rate under different choices of the hyperparameters according to varying $(\kappa_\beta, \kappa_{\sigma^2})$. The standard hyperparameter values correspond to $\kappa_\beta = \kappa_{\sigma^2} = 1$.}
\label{app_fig:sensitivity_hyperparameters}
\end{figure}

\cng{We set $(\kappa_\beta, \kappa_{\sigma^2}) \in \{(1.5, 0.8), (2, 0.6)\}$. The results are shown in \cref{app_fig:sensitivity_hyperparameters}. We see that the choice of hyperparameters has essentially no impact on estimates or uncertainty of the exposure response functions.}

\section{Spatial causal inference when some exposures are undefined}
\label{app_sec:undefined_exposures}

\cng{As we see in \cref{app_tab:app_sample_size}, some counties might have both exposures well-defined, whereas other counties might only have local or neighborhood exposure, with the other being undefined. Specifically, a county has only neighborhood exposure if there are no power plants with SO$_2$ emissions within the county, but there are adjacent counties with SO$_2$ emissions. In turn, a county has only local exposure if there are power plants with SO$_2$ emissions within the county, but no neighboring county has emissions. We extend the causal framework in our manuscript to cover such settings.}

\subsection{The setup}

\cng{We use $N^*$ to denote the total number of units, and $n$ to denote the number of those units with local exposure. We use $i = 1, 2, \dots, N^*$ to denote the $N^*$ units, with the first $n$ having local exposure. Let $\bm Z = (Z_1, Z_2, \dots, Z_n)$ denote the local exposure for these units, with realization $\bm z$. Potential outcomes are of the form $Y_i(\bm z)$ for $i = 1, 2, \dots, N^*$ and $\bm z \in \mathcal{Z}^n$.}

\cng{We formalize the exposure mapping assumption in this setting in the following manner. We assume that there exists an $N^* \times n$ adjacency matrix $A^*$ which reflects whether each of the $N^*$ units is connected with the $n$ units that have local exposure. The entries in $A^*$ can be binary or continuous, and $A^*_{ii} = 0$ for $i = 1, 2, \dots, n$.}

\cng{For units $i$ with $A^*_{ij} = 0$ for all $j = 1, 2, \dots, n$ the neighborhood exposure is undefined, as they have no neighbors with local exposure. For the remaining units, we define the neighborhood exposure as $\overline Z_i = \sum_{j = 1}^n A_{ij}Z_j / \sum_{j = 1}^n A_{ij}$. Out of the $n$ units with local exposure, we assume that the first $n_1$ have both local and neighborhood exposure, whereas the next $n_2 = n - n_1$ units, $i = n_1 + 1, n_1 + 2, \dots, n$, only have local exposure. Furthermore, we assume that units $n + 1, n+2, \dots, N$ have neighborhood exposure, whereas units $N + 1, N+2, \dots, N^*$ have neither local nor neighborhood exposure. This last set of units is excluded from any analysis as they are unaffected by any intervention on the exposures.}

\cng{The exposure mapping assumption in \cref{ass:sutva} can then be restated as
\begin{assumption}
Let $\bm z, \bm z'$ be two treatment vectors in $\mathcal{Z}^n$. For $i = 1, 2, \dots, n_1$, if $z_i = z_i'$ and $\overline z_i = \overline z_i'$, then $Y_i(\bm z) = Y_i(\bm z')$. For units with only local exposure, $i = n_1 + 1, n_1 + 2, \dots, n$, if $z_i = z_i'$, then $Y_i(\bm z) = Y_i(\bm z')$.
For units with only neighborhood exposure, $i = n + 1, n+ 2, \dots, N$, if $\overline z_i = \overline z_i'$, then $Y_i(\bm z) = Y_i(\bm z')$. Therefore, potential outcomes can be denoted as $Y_i(z_i, \overline z_i)$ for $i = 1, 2, \dots, n_1$, as $Y_i(z_i)$ for $i = n_1 + 1, n_1 + 2, \dots, n$, and as $Y_i(\overline z_i)$ for $i = n + 1, n + 2, \dots, N$.
\end{assumption}
}

\cng{We consider hypothesized local exposures $\bm z = (z_1, z_2, \dots, z_n)$ and hypothesized neighborhood exposures $\overline{\bm z} = (\overline z_1, \overline z_2, \dots, \overline z_{n_1}, \overline z_{n + 1}, \overline z_{n+2}, \dots, \overline z_N)$. To define causal effects, we define average potential outcomes within sets of units based on which exposures are well-defined. Specifically, for the units with both exposures, we define $\overline Y^{(b)}(\bm z, \overline{\bm z}) = n_1^{-1} \sum_{i = 1}^{n_1} Y_i(z_i, \overline z_i)$. For the units with only local exposure, we define $\overline Y^{(l)}(\bm z) = n_2^{-1} \sum_{i = n_1 + 1}^n Y_i(z_i)$, and for the units with only neighborhood exposure, we define $\overline Y^{(n)}(\overline{\bm z}) = (N - n)^{-1} \sum_{i = n + 1}^N Y_i(\overline z_i)$.
Then, we define the average potential outcome among those with local exposure (with or without neighborhood exposure) as
$\overline Y^{(b,l)}(\bm z, \overline{\bm z}) = n^{-1} \left( n_1 \overline Y^{(b)}(\bm z, \overline{\bm z}) + n_2 \overline Y^{(l)}(\bm z) \right) = n^{-1}  \sum_{i = 1}^n Y_i(z_i)$.
Similarly we define the average outcome among those with neighborhood exposure (irrespective of whether their local exposure is well-defined or not) as
$\overline Y^{(b,n)}(\bm z, \overline{\bm z}) = (n_1 + N - n)^{-1} \left( n_1 Y^{(b)}(\bm z, \overline{\bm z}) + (N - n) \overline Y^{(n)}(\overline{\bm z}) \right) = (n_1 + N - n)^{-1} \left( \sum_{i = 1}^{n_1} Y_i(z_i) + \sum_{i = n + 1}^N Y_i(z_i) \right).$ We similarly consider average potential outcomes based on $\bm z'$ and $\overline{\bm z}'$.}

\cng{The local effect pertains only to units that have local exposure well-defined. We define it as
\begin{align*}
    \lambda^*(\bm z, \bm z'; \overline{\bm z}) = \E \left[ \overline Y^{(b,l)}(\bm z', \overline{\bm z}) - \overline Y^{(b,l)}(\bm z, \overline{\bm z}) \right].
\end{align*}
Similarly, we define the interference effect based on units that have neighborhood exposure well-defined as
\[
\iota^*(\overline{\bm z}, \overline{\bm z}'; \bm z) = 
\E \left[ \overline Y^{(b,n)}(\bm z, \overline{\bm z}') - \overline Y^{(b,n)}(\bm z, \overline{\bm z}) \right].
\]
}

\subsection{Model on potential outcomes in the presence of undefined exposures}

\cng{In general, there are three sets of units: those with both exposures well-defined, those with only local and those with only neighborhood exposure. In our analysis in the main manuscript, the second set of units is empty. However, we discuss here the general setting with all three sets of units as all three appear in our sensitivity analysis to the neighborhood structure (see \cref{app_tab:app_sample_size}).}

\cng{We use $\mathcal I_Z$ to denote the set of indices with local exposure, $\mathcal I_Z = \{1, 2, \dots, n\}$. We also use $\mathcal I_{\bar Z}$ to denote the set of indices with neighborhood exposure, $\mathcal I_{\bar Z} = \{1, 2, \dots, n_1, n + 1, n + 2, \dots, N\}$. 
We also decompose the covariates $\widetilde C_i$ to three types, those that are available for all units $\widetilde C_i^{(b)}$, those that are available for units with local exposure $\widetilde C_i^{(l)}$, and those that are available for units with neighborhood exposure $\widetilde C_i^{(n)}$. In our study, the first set includes, for example, weather information, the second set includes local power plant information, and the third set includes neighborhood power plant information. Similarly we decompose the coefficients of the covariates as $\bm \beta_C^{(b)}, \bm \beta_C^{(l)}$ and $\bm \beta_C^{(n)}$.}

\cng{Similarly to \cref{eq:linear_sem}, we specify that
\begin{align*}
        Y_i(z, \overline z) &= \beta_0^{11} + \beta_Z^{11} z + \beta_{\bar Z}^{11} \overline z + \\
        & \hspace{40pt}
        +\widetilde C_i^{(b),T} \bm \beta_C^{(b),11} +
        \widetilde C_i^{(l),T} \bm \beta_C^{(l),11} +
        \widetilde C_i^{(n),T} \bm \beta_C^{(n), 11} + \\
        & \hspace{40pt}
        +\beta_U^{11} U_i + \beta_{\bar U}^{11} \overline U_i + \epsilon_i(z, \overline z)
        \hspace{100pt} \text{for } i \in \mathcal I_Z \cap \mathcal I_{\bar Z} \\
        Y_i(z) &= \beta_0^{10} + \beta_Z^{10} z + \\
        & \hspace{40pt}
        +\widetilde C_i^{(b),T} \bm \beta_C^{(b),10} +
        \widetilde C_i^{(l),T} \bm \beta_C^{(l),10} +
        \\
        & \hspace{40pt}
        +\beta_U^{10} U_i + \beta_{\bar U}^{10} \overline U_i + \epsilon_i(z, \overline z)
        \hspace{100pt} \text{for } i \in \mathcal I_Z \cap \mathcal I_{\bar Z}^c \\
        Y_i(\overline z) &= \beta_0^{01} + \beta_{\bar Z}^{01} \overline z + \\
        & \hspace{40pt}
        +\widetilde C_i^{(b),T} \bm \beta_C^{(b),01} +
        \widetilde C_i^{(n),T} \bm \beta_C^{(n), 01} + \\
        & \hspace{40pt}
        +\beta_U^{01} U_i + \beta_{\bar U}^{01} \overline U_i + \epsilon_i(z, \overline z)
        \hspace{100pt} \text{for } i \in \mathcal I_Z^c \cap \mathcal I_{\bar Z}
\end{align*}
with error terms that have corresponding variances $\sigma^{2,11}_Y, \sigma^{2,10}_Y$ and $\sigma^{2,01}_Y$.}

\cng{In principle, we can adopt separate prior distributions for the parameters of the different models and perform estimation separately. However, to gain estimation efficiency, we link the coefficients of the three models. We specify that the three sets of parameters are equal for $\beta_Z, \beta_{\bar Z}, \bm \beta_C, \beta_U, \beta_{\bar U}$ and $\sigma^2_Y$ across the models where the parameter appears. We maintain separate intercepts for the three models.}

\section{Sample average local and interference effects}

\cng{Our definitions of local and interference effects in \cref{eq:local_effect_network} and \cref{eq:interference_effect_network} include an expectation operator. As we discuss there, our estimands can be interpreted as a model-based definition of causal effects where potential outcomes are viewed as random. Alternatively, they can be understood as an expected change in average outcomes over similar networks. Here, we discuss alternative definitions of causal effects that view potential outcomes as fixed quantities and correspond to the specific network only.}

\subsection{Sample average estimands}
\label{app_subsec:sample_estimands}

\cng{This type of estimands are often referred to as {\it sample} average effects, as they pertain to the specific sample at hand. In our setting, these would be defined as
\[
 \lambda^*(\bm z, \bm z'; \overline{\bm z}) = \overline Y(\bm z', \overline{\bm z}) - \overline Y(\bm z, \overline{\bm z})
\]
and
\[
 \iota^*(\overline{\bm z}, \overline{\bm z}'; \bm z) = 
 \overline Y(\bm z, \overline{\bm z}') - \overline Y(\bm z, \overline{\bm z}).
\]
In these definitions, we drop the expectation, and the contrasts reflects differences in the average potential outcomes of the sample only, which are considered as fixed (and not random) quantities.}

\subsection{Estimation of sample average effects}

\cng{Estimation of the effects in Section~\ref{app_subsec:sample_estimands} requires that we know or impute the potential outcomes for all units in the sample under the exposure values that appear in the estimands. In what follows, we use $Y_i^\text{miss}(\cdot)$ to denote the missing potential outcomes for unit $i$ and $\bm Y^\text{miss}(\cdot)$ to denote the collection of missing potential outcomes across all units.}

\cng{Our Bayesian formulation allows for straightforward imputation of such missing potential outcomes using the predictive distribution. Specifically,
\begin{align*}
    p(\bm Y^\text{miss}(\cdot) \mid \bm Y, \bm Z, \bm C) 
    &= 
    \int p(\bm Y^\text{miss}(\cdot) \mid \theta^*, \bm U, \bm Y, \bm Z, \bm C)\ 
    p(\theta^*, \bm U \mid \bm Y, \bm Z, \bm C)\ 
    \mathrm{d}(\theta^*, \bm U).
\end{align*}
Based on \cref{ass:pot_out_independence_units}, 
we have that
\begin{align*}
    p(\bm Y^\text{miss}(\cdot) \mid \bm Y, \bm Z, \bm C) 
    &= 
    \int \prod_i p(Y_i^\text{miss}(\cdot) \mid \theta^*, U_i, \overline U_i, Y_i, Z_i, \overline Z_i, \widetilde C_i)\ 
    p(\theta^*, \bm U \mid \bm Y, \bm Z, \bm C)\ 
    \mathrm{d}(\theta^*, \bm U).
\end{align*}
This implies that the imputation of the missing potential outcomes can be performed separately across units given their own information. 
Then, based on the unconfoundedness assumption, 
this can be re-written as
\begin{equation}
p(\bm Y^\text{miss}(\cdot) \mid \bm Y, \bm Z, \bm C)
= 
\int \prod_i p(Y_i^\text{miss}(\cdot) \mid \theta^*, U_i, \overline U_i, Y_i, \widetilde C_i)\ 
p(\theta^*, \bm U \mid \bm Y, \bm Z, \bm C)\ 
\mathrm{d}(\theta^*, \bm U).
\label{app_eq:impute_correlation}
\end{equation}
Furthermore, based on the assumption of independence of potential outcomes across exposure levels of the same unit, for exposure $(z_i, \overline z_i)$ which are not both equal to the observed, we have
\[
p(Y_i(z_i, \overline z_i) \mid \bm Y, \bm Z, \bm C)
= 
\int p(Y_i(z_i, \overline z_i) \mid \theta^*, U_i, \overline U_i, \widetilde C_i)\ 
p(\theta^*, \bm U \mid \bm Y, \bm Z, \bm C)\ 
\mathrm{d}(\theta^*, \bm U).
\]
}

\cng{This implies the following procedure for imputing missing potential outcomes and estimating sample average causal effects.
For a unit $i$ for which its observed exposures are equal to the exposures in the corresponding estimand, we have that $Y_i(z_i, \overline z_i) = Y_i$, and therefore their observed outcome is used for estimation. Since this potential outcome is known, it contributes no uncertainty to the resulting estimator. For the remaining units, for each sample of $\theta^*$ and $\bm U$ from the posterior distribution, simulate the missing potential outcomes by drawing from the outcome model (including the error term) while setting the local and neighborhood exposure to their values in the corresponding estimand. Then, calculate the average potential outcome for this posterior sample using the observed and the imputed missing potential outcomes. This procedure would provide the posterior distribution for local and interference effects, which can then be summarized by calculating the posterior mean as the estimate, and posterior quantiles to create a credible interval.}

\cng{We note that for the shift-estimands that set $\bm z = \bm Z$ and $\overline{\bm z} = \overline{\bm Z}$, the potential outcomes in $\overline Y(\bm z, \overline{\bm z})$ are all observed. Therefore, the estimation of local and interference effects in this setting requires imputation of potential outcomes that are of the form $Y_i(Z_i + \delta, \overline Z_i)$ or $Y_i(Z_i, \overline Z_i + \overline \delta)$ only.}

\cng{The independence of potential outcomes across exposures for the same unit allows us to impute the different missing potential outcomes for the same unit in the definition of local and interference effects separately and without using the observed outcome. Next, we discuss how to incorporate the possibility of correlation among potential outcomes for the same unit.}

\subsection{Estimating sample-average estimands when potential outcomes for the same unit are correlated}

\cng{When potential outcomes are correlated, the outcome $Y_i$ does not drop out of the conditional distribution of the missing potential outcomes $Y_i^\text{miss}(\cdot)$ in \cref{app_eq:impute_correlation}. At the same time, parameters representing the correlation of potential outcomes are never identifiable since, for each unit, we observe at most one potential outcome \citep{ding2018causal}. Instead, it has been advocated that the correlation of potential outcomes for each unit can be varied as a sensitivity parameter \citep{ding2018causal, li2022bayesian}.}

\cng{Existing work on incorporating the correlation of potential outcomes deals with the case of a binary treatment and potential outcomes of the form $Y(1), Y(0)$. In our setting, the number of potential outcomes $Y_i(z_i, \overline z_i)$ can be up to infinitely many. For continuous exposures, a simplified sensitivity analysis could specify a Gaussian process formulation on the correlation of potential outcomes where for any values $z_i, z_i', \overline z_i, \overline z_i'$, 
\[
\text{Corr}\left(Y_i(z_i, \overline z_i), Y_i(z_i', \overline z_i')\right) = \exp \left\{ - d_Z(z_i, z_i') / \rho_Z - d_{\bar Z}(\overline z_i, \overline z_i') / \rho_{\bar Z} \right\} \quad \text{for all } i
\]
where the correlation is implicitly conditional on all variables, the functions $d_Z, d_{\bar Z}$ are distance metrics that are always greater or equal to 0 such as $d_Z(z, z') = |z - z'|$, and the parameters $\rho_Z, \rho_{\bar Z}$ specify the rate of correlation decay as the distance of exposures grows. Theoretically, different parameters $\rho_Z, \rho_{\bar Z}$ can be specified for each unit, but this would be impractical. Most practically, one can use $\rho_Z = \rho_{\bar Z}$ and vary this value on the positive real line as a sensitivity parameter.}

\end{document}